\documentclass[10pt,journal,compsoc,a4paper]{IEEEtran} 
\ifCLASSOPTIONcompsoc
  \usepackage[nocompress,noadjust]{cite}
\else
  \usepackage[noadjust]{cite}
\fi
\ifCLASSINFOpdf
  \usepackage[pdftex]{graphicx}
\else
  \usepackage[dvips]{graphicx}
\fi
\usepackage{amsmath}
\usepackage{algorithmic}

\usepackage{array}
\usepackage{url}

\usepackage[table]{xcolor}
\usepackage{amssymb}
\usepackage[english]{babel}
\usepackage[utf8]{inputenc}
\usepackage[backref]{hyperref} 
\usepackage{tikz}

\usepackage{pgfplots}
\pgfplotsset{compat=1.14}

\usepackage{textcomp}
\usepackage{enumitem}
\usepackage{multicol}
\usepackage{booktabs} 

\DeclareMathOperator*{\argmax}{\arg\!\max}

\let\oldUrl\url
\renewcommand{\url}[1]{\href{#1}{Source}}

\definecolor{semi_manual_prot_color}{RGB}{89,125,190}
\definecolor{guided_prot_color}{RGB}{118,191,114}
\definecolor{joint_prot_color}{RGB}{199,110,110}

\newlength{\forkmeoffset}
\definecolor{forkmebg}{HTML}{CC0000}
\definecolor{forkmefg}{HTML}{EEEEEE}
\definecolor{forkmetext}{HTML}{f9f3f3}

\newcommand{\forkme}[1][west]{
	\ifthenelse{\equal{#1}{east}}{%
		\tikzset{forkmerot/.style={rotate=-45}}
	}{%
		\tikzset{forkmerot/.style={rotate=45}}
	}
	\begin{tikzpicture}[remember picture, overlay]
	\node[forkmerot, shift={(0, -\forkmeoffset)}] at (current page.north #1) {
		\begin{tikzpicture}[remember picture, overlay]
		\node[fill=forkmebg, text centered, minimum width=50em, minimum height=3.0em, text=forkmefg](fmogh) at (0pt, 0pt) {   \fontfamily{phv}\selectfont\bfseries {%
				\hypersetup{pdfborder=0 0 0}%
				\href{https://github.com/mamrehn/interactive_image_segmentation_evaluation}{\color{forkmetext}Fork me on GitHub}%
		} };
		\draw[forkmefg!60, dashed, line width=.08em, dash pattern=on .5em off 1.5\pgflinewidth] (-25em,1.2em) rectangle (25em,-1.2em);
		\end{tikzpicture}
	};
\end{tikzpicture}
}

\usepackage{rebuttal}
\usepackage[sort]{cleveref} 
\usepackage{pdfpages}
\usepackage{changepage}
\usepackage{layout}
\usepackage{xkeyval}

\newboolean{includereviewanswers}

\newcommand{\additioncaption}[1]{#1}
\newcommand{\deletioncaption}[1]{}
\newcommand{\changecaption}[2]{#1}

\definecolor{mynicegreen}{RGB}{102,252,102}
\hypersetup{
	colorlinks=false,
	linkcolor=red,
	anchorcolor=black,
	citecolor=green,
	filecolor=cyan,
	menucolor=red,
	runcolor=cyan, 
	urlcolor=magenta,
	citebordercolor=mynicegreen, 
	filebordercolor=cyan, 
	linkbordercolor=red, 
	menubordercolor=red, 
}

\begin{document}

\setboolean{includereviewanswers}{false}

%
\title{A Semi-Automated Usability Evaluation Framework for Interactive Image Segmentation Systems}
%
%
%
%

\author{Mario~Amrehn,
        Stefan~Steidl,
        Reinier~Kortekaas,
        Maddalena~Strumia,\\
        Markus~Weingarten,
        Markus~Kowarschik,
		Andreas~Maier
\IEEEcompsocitemizethanks{\IEEEcompsocthanksitem M.\ Amrehn, S.\ Steidl, and A.\ Maier are with the Pattern Recognition Lab, Computer Science Department, Friedrich-Alexander University Erlangen-Nuremberg, Germany\protect\\
E-mail: \href{https://www5.cs.fau.de/~amrehn}{mario.amrehn@fau.de} 
\IEEEcompsocthanksitem R.\ Kortekaas, M.\ Strumia, M.\ Weingarten, and M.\ Kowarschik are with Siemens Healthineers \mbox{AG}, Forchheim, Germany}
}

%
%

\markboth{Accepted as Research Article at the International Journal of Biomedical Imaging, Hindawi}%
{Amrehn \MakeLowercase{\textit{et al.}}: Usability Evaluation of Interactive Image Segmentation Systems}
%



\IEEEtitleabstractindextext{%
\begin{abstract}
For complex segmentation tasks, the achievable accuracy of fully automated systems is inherently limited.
Specifically, 
when a precise segmentation result is desired for a small amount of given data sets, semi-automatic methods exhibit a clear benefit for the user.
The optimization of human computer interaction (\mbox{HCI}) is an essential part of interactive image segmentation. 
Nevertheless, publications introducing novel interactive segmentation systems (\mbox{ISS}) often lack an objective comparison of \mbox{HCI} aspects. 
It is demonstrated, that even when the underlying segmentation algorithm is the same throughout interactive prototypes, 
their user experience may vary substantially. 
As a result, users prefer simple interfaces as well as a considerable degree of freedom to control each iterative step of the segmentation. 

In this article, an objective method for the comparison of \mbox{ISS} is proposed, based on extensive user studies.
A summative qualitative content analysis is conducted via abstraction of visual and verbal feedback given by the participants.
A direct assessment of the segmentation system is executed by the users via the system usability scale (\mbox{SUS}) and \mbox{AttrakDiff-2} questionnaires.
Furthermore, an approximation of the findings regarding usability aspects in those studies is introduced, conducted solely from the system-measurable user actions during their usage of interactive segmentation prototypes.
The prediction of all questionnaire results has an average relative error of $8.9$\,\%, which is close to the expected precision of the questionnaire results themselves.
This automated evaluation scheme may significantly reduce the resources necessary to investigate each variation of a prototype's user interface (\mbox{UI}) features and segmentation methodologies.
\end{abstract}

}

\maketitle


\forkme[east]


\IEEEdisplaynontitleabstractindextext

%
\IEEEpeerreviewmaketitle

\IEEEraisesectionheading{\section{Introduction}\label{sec:introduction}}

\IEEEPARstart{T}{o} the best of our knowledge, there is not one publication in which user based scribbles are combined with standardized questionnaires in order to assess an interactive image segmentation system's quality.
This type of synergetic usability measure is a contribution of this work. 
In order to provide a guideline for an objective comparison of interactive image segmentation approaches,
a prototype providing a \mbox{semi-manual} pictorial user input, introduced in \textbf{Sec.}\,\ref{sec:semi-manual_prototype}, is compared to a prototype with a guiding
menu-driven {UI}, described in \textbf{Sec.}\,\ref{sec:guided_prototype}.
Both evaluation results are analyzed with respect to a joint prototype, defined in \textbf{Sec.}\,\ref{sec:joint_prototype}, incorporating aspects of both interface techniques.
All three prototypes are built utilizing modern web technologies. 
An evaluation of the interactive prototypes is performed utilizing pragmatic usability aspects described in \textbf{Sec.}\,\ref{sec:results_pragmatic}, as well as hedonic usability aspects analyzed in \textbf{Sec.}\,\ref{sec:results_hedonic}.
\addition[label=c:a121,ref=c:c12]{%
These aspects are evaluated via two standardized questionnaires (System Usability Scale and \mbox{AttrakDiff-2}) which form the ground truth for a subsequent prediction of the questionnaires' findings via a regression analysis outlined in \textbf{Sec.}\,\ref{sec:prediction_of_questionnaire_results}.
The outcome of questionnaire result prediction from interaction log data only is detailed in \textbf{Sec.}\,\ref{sec:prediction_of_questionnaire_results_from_log_data}.%
This novel automatic assessment of pragmatic as well as hedonic usability aspects is a contribution of this work. %
}
\addition[label=c:a211,ref=c:c21]{%
Our source code release for the automatic usability evaluation from user interaction log data can be found at \oldUrl{https://github.com/mamrehn/interactive_image_segmentation_evaluation}. %
}

\subsection{Image Segmentation Systems}
Image segmentation can be defined as the partitioning of an image into a finite number of semantically non-overlapping regions.
A semantic label can be assigned to each region.
In medical imaging, each individual region of a patients' abdominal tissue might be regarded as healthy or cancerous.
Segmentation systems can be grouped into three principal categories,
each differing in the degree of involvement of an operating person (user): manual, automatic, and interactive.
(1) During manual tumor segmentation, a user provides all elements $i$ in the image grid which have neighboring elements $N(i)$ of different labels than $i$.
The system then utilizes this closed curve contour line information to infer the labels for remaining image elements via simple region growing. 
This minimal assistance by the system causes the overall segmentation process of one lesion to take up to several minutes of user interaction time.
However, reaching an appropriate or even perfect segmentation result (despite noteworthy \mbox{inter-observer} difference~\cite{becker2017increased}) is feasible~\cite{kim2016interobserver,hong2014interobserver}. 
In practice, few \mbox{time-consuming} manual segmentations are performed by domain experts, in order to utilize the results as a reference standard in radiotherapy planning~\cite{moltz2011analysis}.
(2) A fully automated approach does not involve a user's interference with the system.
The introduced deficiency in domain knowledge for accurately labeling regions may be restored partially by automated segmentation approaches. 
The maximum accuracy of the segmentation result is therefore highly dependent on the individual set of rules or amount of training data available. 
If the segmentation task is sufficiently complex, a perfect result may not be reachable. 
(3) Interactive approaches aim at a fast and exact segmentation by combining substantial assistance by the system with knowledge about a very good estimate of the true tumor 
extent provided by trained physicians during the segmentation process~\cite{olabarriaga1997setting}. 
In contrast to fully automated solutions, prior knowledge is (also) provided during the segmentation process.
Although, interactive approaches are also costly in terms of manual labor to some extent, 
they can supersede fully automated techniques in terms of accuracy.
Due to their exact segmentation capabilities, interactive segmentation techniques are frequently chosen to outline pathologies during imaging assisted medical procedures, like hepatocellular carcinomata during trans-catheter arterial chemoembolization 
(see \textbf{Sec.}\,\ref{sec:tace}).

\subsection{Evaluation of Image Segmentation Systems} 
Performance evaluation is one of the most important aspects during the continuous improvement of systems and methodologies.
With non-interactive computer vision and machine learning systems for image segmentation, an objective comparison of systems  
can be achieved by evaluating \mbox{pre-selected} data sets for training and testing. 
Similarity measures between segmentation outcome and ground truth images are utilized to quantify the quality of the segmentation result. 

With \addition[label=c:a23,ref=c:c23]{interactive segmentation systems} (\mbox{ISS}), a complete ground truth data set would also consist of the adaptive user interactions which advance the segmentation process.
Therefore, when comparing \mbox{ISS}, the user needs to be involved in the evaluation process.
User interaction data however is highly dependent on 
(1) the users' domain knowledge and the unique learning effect of the human throughout a period of exposure to the problem domain, 
(2) the system's underlying segmentation method and the users' preferences toward this technique, as well as 
(3) the design and usability (the user experience~\cite{hassenzahl2006user,law2009understanding}) of the interface which is presented to the user during the interactive segmentation procedure~\cite{caro1979inter,hong2014interobserver}.
This includes users' differing preferences towards diverse interaction systems and tolerances for unexpected system behavior.
Considering \mbox{(1--3)}, an analytically expressed objective function for an interactive system is hard to define.
Intuitively, the user wants to achieve a satisfying result in a short amount of time with ease~\cite{kohli2012user}. 
A direct assessment of a system's usability is enabled via standardized questionnaires, as described in \textbf{Sec.}\,\ref{sec:questionnaires}.
Individual usage of \mbox{ISS} can be evaluated via the segmentation result's similarity to the ground truth labeling according to the S{\o}rensen-Dice coefficient (\mbox{Dice})~\cite{dice1945measures} after each interaction.
The interaction data utilized for these segmentations has to be representative in order to generalize the evaluation results.

\subsection{Types of User Interaction}

As described by Olabarriaga et al.~\cite{olabarriaga2001interaction} as well as Zhao and Xie~\cite{zhao2012interactive}, user interactions can be categorized with 
regards to the type of interface an \mbox{ISS} provides.
The following categories are emphasized.
(1) A pictorial mask image is the most intuitive form of user input. 
Humans use this technique when transferring knowledge via a visual medium~\cite{puranik2011scribbles}.
The mask overlayed on the visualization of the image \mbox{$\mathbf{I}\in\mathbb{R}^{w,h}$} to segment consists of structures called scribbles, where $w$ is the width and $h$ is the height of the \mbox{2-D} image $\mathbf{I}$ in pixels.
Scribbles are seed points, lines, and complex shapes, each represented as a set of individual seed points.
One seed point is a tuple \mbox{$\mathbf{s}_i=\left(\mathbf{p}_i,\mathbf{\ell}_i\right)$}, where \mbox{$\mathbf{p}_i\in\mathbb{R}^2$} describes the position of the seed in image space.
The class label of a scribble in a binary segmentation system is represented by \mbox{$\mathbf{\ell}_i\in\left\{\text{background},\text{foreground}\right\}$}.
Scribbles need to be defined by the user in order to act as a representative subset $\mathbf{S}$ of the ground truth segmentation \mbox{$\mathbf{G}=\left\{\mathbf{s}_1,\mathbf{s}_2,\dots\right\}$}. 

(2) A menu-driven user input scheme as in~\cite{rupprecht2015image, udupa1997multiple} limits the user's scope of action.
Users trade distinct control over the segmentation outcome for more guidance provided by the system.
The locations or the shapes of newly created scribbles are fixed before presentation to the user. 
It is challenging to achieve an exact segmentation result using a method from this category.
Rupprecht et al.~\cite{rupprecht2015image} describe significant deficits in finding small objects and outline a tendency of the system to automatically choose 
seed point locations near the object border,
which cannot be labeled by most users' visual inspection and would therefore not have been selected by the users themselves.
Advantages of \mbox{menu-driven} user input are the high level of abstraction of the process, enabling efficient guidance for inexperienced users in their 
decision which action to perform for an optimal segmentation outcome (regarding accuracy over time or 
number of interactions)~\cite{olabarriaga1999human,olabarriaga2001interaction}.

\subsection{Generation of Representative User Input}

Nickisch et al.~\cite{nickisch2010learning} describe crowd sourcing and user studies as two methods to generate plausible user input data.
The cost efficient crowd sourcing method often lacks control and knowledge of the users' motivation.
Missing context information for crucial aspects of the data acquisition procedure creates a challenging task objectifying the evaluation results.
Specialized fraud detection methods are commonly used in an attempt to pre-filter the recorded corpus and extract a usable subset of data. 
McGuinness and O'Connor~\cite{mcguinness2010comparative} proposed an evaluation of \mbox{ISS} via extensive user experiments. 
In these experiments, users are shown images with descriptions of the objects they are required to extract. 
Then, users mark foreground and background pixels utilizing a platform designed for this purpose. 
These acquisitions are more time-consuming and cost intensive than \mbox{crowd-sourcing}, since they require a constant involvement of users. 
However, the study's creators are able to control many aspects of the data recording process, which enables detailed observations of user reactions.
The data samples recorded are a representative subset of the focus group of the finalized system. 
A user study aims at maximizing repeatability of its results.
In order to increase the objectivity of the evaluation in this work, a user study is chosen to be conducted.
The study is described in \textbf{Sec.}\,\ref{sec:usability_test_setup}.

\subsection{State-of-the-art Evaluation of Interactive Segmentation Systems}

\subsubsection{Segmentation Challenges}

In segmentation challenges like \mbox{SLIVER07}~\cite{van20073d} (mainly) fully automated approaches are competing for the highest score regarding a predefined image quality metric.
Semi-automatic methods are allowed for submission if the manual interaction with the test data is strictly limited to pre-processing and (single seed point) initialization of an otherwise fully automated process. 
\mbox{ISS} may be included into the contests' final ranking, but are regarded as non-competing, 
since the structure of the challenges is solely designed for automated approaches.
The \mbox{PROMISE12} challenge~\cite{litjens2014evaluation} had a separate category for proposed interactive approaches, where the user (in this case, the person also describing the algorithm) may add an unlimited number of hints during segmentation, without observing the experts' ground truth for the test set.
No group of experts was provided to operate the interactive method for comparative results.
The submitted interactive methods' scores in the challenge's ranking are therefore highly dependent on the domain knowledge of single operating users and can not be regarded as an objective measure.

\subsubsection{Comparisons for Novel Segmentation Approaches}

In principle, with every new proposal of an interactive segmentation algorithm or interface, the authors have to demonstrate the new method's capabilities in an objective 
comparison with already established techniques.
The effort spent for these comparisons by the original authors varies substantially. 
According to~\cite{kohli2012user}, many evaluation methods only consider a fixed input.
This approach is especially unsuited for evaluation, without simultaneously defining an appropriate interface, which actually validates that a real person 
utilizing this {UI} is capable of generating similar input patterns to the ones provided.
Although, there are some overview publications, which compare several approaches~\cite{zhao2013overview,olabarriaga2001interaction,mcguinness2010comparative,mcguinness2011toward,amrehn2016comparative}, the number of publications outlining new methods is disproportionately greater, 
leaving comparisons insufficiently covered.
\addition[label=c:a122,ref=c:c12]{%
Olabarriaga et al.~\cite{olabarriaga2001interaction} main contribution is the proposition of criteria to evaluate interactive segmentation methods: accuracy, repeatability, and efficiency. 
McGuinness et al.~\cite{mcguinness2010comparative} utilized a unified user interface with multiple underlying segmentation methods for the survey they conducted. 
They recorded the current segmentation masks after each interaction to gauge segmentation accuracy over time.
Instead of utilizing a standardized questionnaire, users were asked to rate the difficulty and perceived accuracy of the segmentation tasks on a scale of 1 to 5. 
Their main contribution is an empirical study by $20$ subjects segmenting with four different segmentation methods in order to conclude that one of the four methods is best, given their data and participants.
Their ranking is primarily based on the mean accuracy over time achieved per segmentation method.
McGuinness et al.~\cite{mcguinness2011toward} define a robot user in order to simulate user interactions during an automated interactive segmentation system evaluation. 
However, they do not investigate the similarity of their rule-based robot user to seed input pattern by individual human subjects. 
} %
\addition[label=c:a123,ref=c:c12]{%
Zhao et al.~\cite{zhao2013overview} concluded in their overview over interactive medical image segmentation techniques, that there is a clear need of well-defined performance evaluation protocols for interactive systems.
} %

In \textbf{Tab.}\,\ref{tab:interactiveSegmentationEvaluationComparison}, a clustering of popular publications describing novel interactive segmentation techniques is depicted.
The evaluation methods can be compared by the type of data utilized as user input. 
Note that there is a trend towards more elaborate evaluations in more recent publications.
\addition[label=c:a124,ref=c:c12]{%
The intent and perception of the interacting user are a valuable resource worth considering when comparing interactive segmentation systems~\cite{yang2010user}.
However, only two of the $42$ related publications listed in \textbf{Tab.}\,\ref{tab:interactiveSegmentationEvaluationComparison} make use of the insights about complex thought processes of a human utilizing an interactive segmentation system for the ranking of novel interactive segmentation methods.
Ramkumar et al.~\cite{ramkumar2016using,ramkumar2016user} acquire these data by well-designed questionnaires, but do not automate their evaluation method.
We propose an automated, i.\,e.\ scalable, system to approximate pragmatic as well as hedonic usability aspects of a given interactive segmentation system.
} %

\begin{table*}[thp]%
	\caption{Overview of seed point location selection methods for a set of influential publications in the field of interactive image segmentation.
	Additional \additioncaption{unordered} 
	seed information can be retrieved \deletioncaption{in arbitrary order} 
	by 
	a) manually drawn seeds or 
	b) randomly generated seeds.
	Seeds can be inferred rule-based from the ground truth segmentation by 
	c) sampling the binary mask image, 
	d) from provided bounding box mask images, 
	e) random sampling from tri-maps generated by erosion and dilation, or 
	f) by a robot user i.\,e.\ user simulation.
	\additioncaption{A tri-map specifies background, foreground, and mixed areas.} 
	Seeds can also be provided by real users via the 
	g) final seed masks after all interactions on one input image, or 
	h) the \changecaption{ordered}{actual} iterative scribbles. 
	i) Questionnaire data from \emph{Goals, Operators, Methods, and Selection rules} 
	(\mbox{GO}) as well as \emph{National Aeronautics and Space Administration Task Load Index} (\mbox{TL}) 
	may be retrieved by interviewing users after the segmentation process.
	\additioncaption{%
		Check marks indicate the usage of seeds in the publications listed.
		Publications with check marks in brackets display these seeds but do not utilize them for evaluation.
	} %
	}
	\label{tab:interactiveSegmentationEvaluationComparison}
	\rowcolors{3}{gray!1}{gray!4}
	\setlength\extrarowheight{1pt}
	\resizebox{\textwidth}{!}{%
		\begin{tabular}{r|r|p{0.09\linewidth}p{0.09\linewidth}|p{0.09\linewidth}p{0.09\linewidth}p{0.09\linewidth}p{0.09\linewidth}|p{0.09\linewidth}p{0.09\linewidth}p{0.09\linewidth}}
			\multicolumn{2}{r}{} & \multicolumn{2}{c}{Arbitrary Seeds} & \multicolumn{4}{c}{Seeds Derived from GT} & \multicolumn{3}{c}{Multiple User Data based Seeds} \\[1pt]\hline
			& & (a) & (b) & (c) & (d) & (e) & (f) & (g) & (h) & (i) \\[1pt]
			Year & Publication & Manual & Random & Binary Mask & Box & Tri-maps & Robot & Final Seeds &  Scribbles & Questionnaire \\[1pt]\hline 
			\additioncaption{2019} & \additioncaption{Amrehn~\cite{amrehn2019interactive}} & & & & & & \additioncaption{$\checkmark$~\cite{kohli2012user,xu2016deep,wang2017deepigeos}} & & & \\[1pt]
			2018 & Chen~\cite{chen2018swipecut} & ($\checkmark$) & & & & & $\checkmark$~\cite{rupprecht2015image} & & $\checkmark$ ($N=10$) & \\[1pt]
			& Amrehn~\cite{amrehn2018ideal} & & & & & & $\checkmark$ & & & \\[1pt]
			2017 & Liew~\cite{liew2017regional} & ($\checkmark$) & ($\checkmark$) & & & & $\checkmark$~\cite{kohli2012user} & & & \\[1pt] 
			& Wang~\cite{wang2017interactive} & & & & & & & & $\checkmark$ ($N=2$) & \\[1pt]
			& Wang~\cite{wang2017deepigeos} & & & & & & $\checkmark$~\cite{amrehn2017uinet} & & $\checkmark$ ($N=2$) & \\[1pt]
			& Amrehn~\cite{amrehn2017uinet} & & & & & $\checkmark$ & $\checkmark$~\cite{wang2017deepigeos} & & & \\[1pt]
			& Amrehn~\cite{amrehn2017robust} & & & & & & $\checkmark$ & & & \\[1pt]
			2016 & Ramkumar~\cite{ramkumar2016using} & & & & & & & & & $\checkmark$(GO, TL) \\[1pt]
			& Ramkumar~\cite{ramkumar2016user} & & & & & & & & & $\checkmark$(TL) \\[1pt] 
			& Jiang~\cite{jiang2016automatic} & & & $\checkmark$~\cite{martin2001database} & & & & & $\checkmark$ ($N=5$) & \\[1pt]
			& Xu~\cite{xu2016deep} & ($\checkmark$) & ($\checkmark$) & & & & $\checkmark$ & & & \\[1pt]
			& Chen~\cite{chen2016interactive} & & & & & & & $\checkmark$ & & \\[1pt] 
			2015 & Andrade~\cite{andrade2015supervised} & & & & & & & & \href{https://github.com/flandrade/dataset-interactive-algorithms}{$\checkmark$} & \\[1pt] 
			& Rupprecht~\cite{rupprecht2015image} & & & & & $\checkmark$ & & $\checkmark$ & & \\[1pt] 
			2014 & Bai~\cite{bai2014error} & $\checkmark$ & $\checkmark$ & & & & & & & \\[1pt]
			2013 & Jain~\cite{jain2013predicting} & & & $\checkmark$ & & & & & $\checkmark$ & \\[1pt]
			& He~\cite{he2013interactive} & $\checkmark$ & & & & & & & & \\[1pt]
			2012 & Kohli~\cite{kohli2012user} & $\checkmark$ & & & & $\checkmark$ & $\checkmark$ & ($\checkmark$) & $\checkmark$ & \\[1pt]
			2011 & Zhao~\cite{zhao2011benchmark} & & $\checkmark$ & & & $\checkmark$ & & & & \\[1pt]
			& Top~\cite{top2011active} & ($\checkmark$) & & & & & $\checkmark$ & & $\checkmark$ ($N=4$) & \\[1pt]
			& McGuinness~\cite{mcguinness2011toward} & ($\checkmark$) & & & & & $\checkmark$ & $\checkmark$ & & \\[1pt]
			2010 & Nickisch~\cite{nickisch2010learning} & $\checkmark$ & & & & $\checkmark$ & & ($\checkmark$) & $\checkmark$ & \\[1pt]  
			& Gulshan\cite{gulshan2010geodesic} & $\checkmark$ & & & & & & ($\checkmark$) & & \\[1pt]
			& Batra~\cite{batra2010icoseg} & $\checkmark$ & & & & & & $\checkmark$ & & \\[1pt]
			& Ning~\cite{ning2010interactive} & $\checkmark$ & & & & & & & & \\[1pt]
			& Price~\cite{price2010geodesic} & $\checkmark$ & & & $\checkmark$~\cite{singaraju2009p} & $\checkmark$~\cite{rother2004grabcut} & & & & \\[1pt]
			& Moschidis~\cite{moschidis2010systematic} & & $\checkmark$ & & & & & & & \\[1pt]
			2009 & Moschidis~\cite{moschidis2009simulation} & & $\checkmark$ & & & $\checkmark$ & & & & \\[1pt]
			& Singaraju~\cite{singaraju2009p} & & & & $\checkmark$ & $\checkmark$~\cite{rother2004grabcut} & & & & \\[1pt]
			2008 & Duchenne~\cite{duchenne2008segmentation} & $\checkmark$ & & & & $\checkmark$~\cite{rother2004grabcut} & & & & \\[1pt]
			& Levin~\cite{levin2008closed} & $\checkmark$ & & & & & & & & \\[1pt]
			& Vicente~\cite{vicente2008graph} & $\checkmark$ & & & & & & & & \\[1pt]
			2007 & Protiere~\cite{protiere2007interactive} & $\checkmark$ & & & & & & & & \\[1pt]
			2006 & Boykov~\cite{boykov2006graph} & $\checkmark$ & & & & & & & & \\[1pt]
			& Grady~\cite{grady2006random} & $\checkmark$ & & & & & & & & \\[1pt]
			2005 & Vezhnevets~\cite{vezhnevets2005growcut} & $\checkmark$ & & & & & & & & \\[1pt]
			& Cates,\cite{cates2005case} & ($\checkmark$) & & & & & & & $\checkmark$ ($N=8+3$) & \\[1pt]
			2004 & Li~\cite{li2004lazy} & & & & & & & & $\checkmark$ & \\[1pt]  
			& Rother~\cite{rother2004grabcut} & $\checkmark$ & & ($\checkmark$) & ($\checkmark$) & \href{https://web.archive.org/web/20161203110733/http://research.microsoft.com/en-us/um/cambridge/projects/visionimagevideoediting/segmentation/grabcut.htm}{$\checkmark$} & & & & \\[1pt]
			& Blake~\cite{blake2004interactive} & & & $\checkmark$ & & $\checkmark$~\cite{martin2001database} & & & & \\[1pt]
			2001 & Martin~\cite{martin2001database} & & & $\checkmark$ & &  \href{https://www2.eecs.berkeley.edu/Research/Projects/CS/vision/grouping/segbench/}{$\checkmark$} & & & & \\[1pt] 
		\end{tabular}%
	}%
\end{table*}

\subsection{Clinical Application for Interactive Segmentation}\label{sec:tace}

\begin{figure}
	\centering
	\resizebox{0.85\columnwidth}{!}{%
		{\def\arraystretch{1.1}\tabcolsep=2pt
			\begin{tabular}{lll}
				\includegraphics[trim={10 10 10 10},clip,height=0.255\textheight,width=0.255\textheight
				]{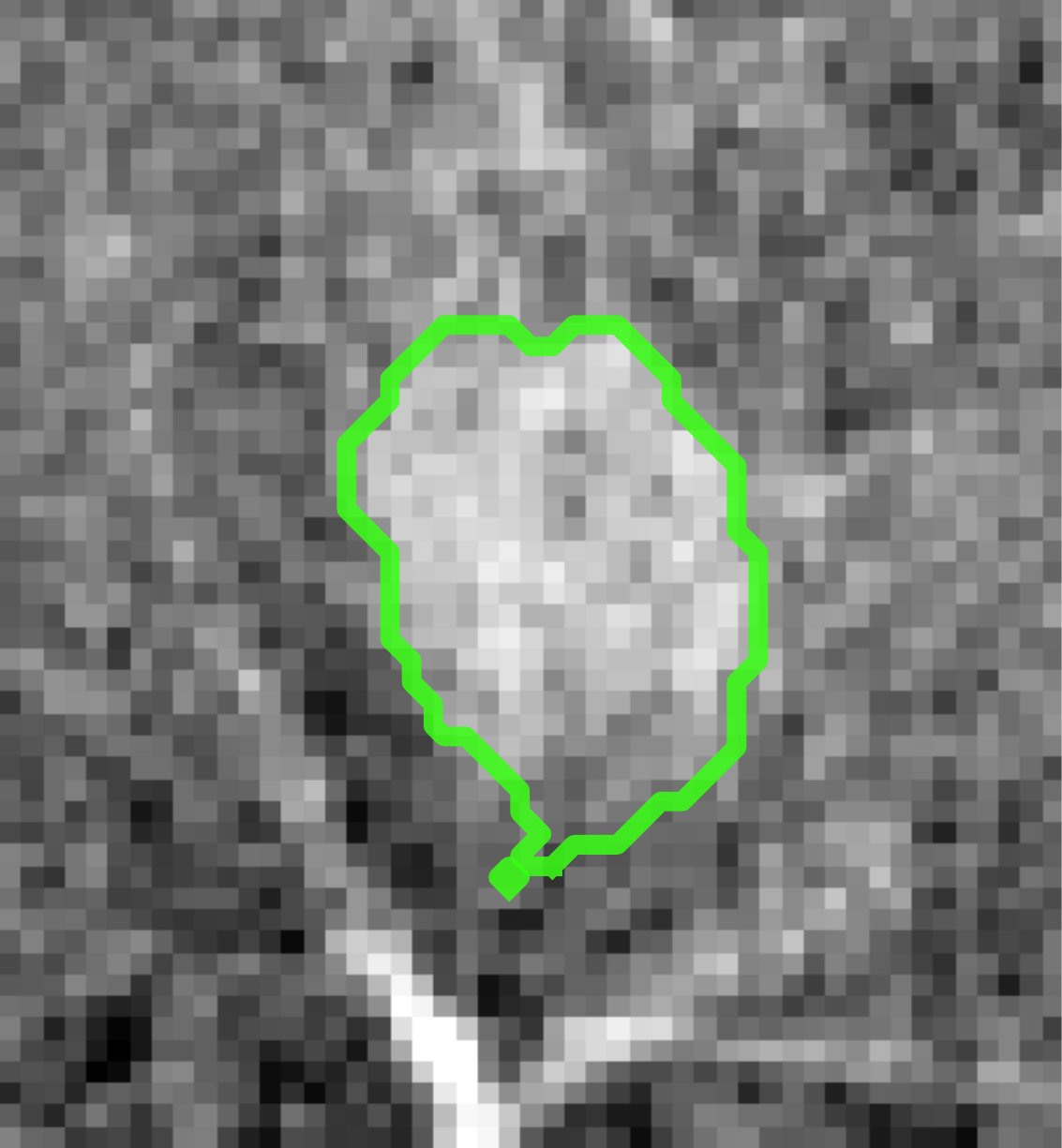} & %
				\includegraphics[trim={10 10 10 10},clip,height=0.255\textheight,width=0.255\textheight
				]{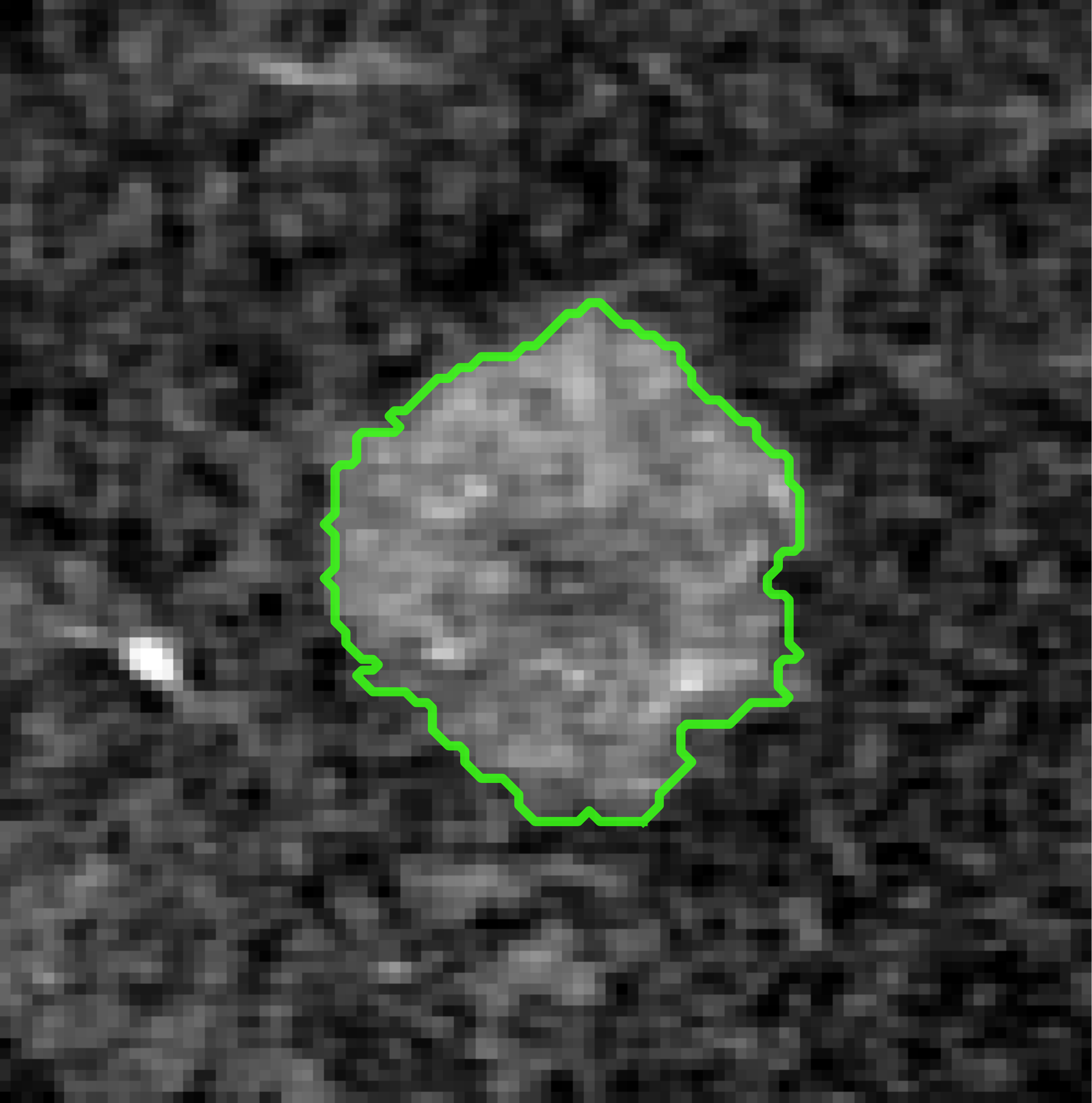} & %
				\includegraphics[trim={10 10 10 10},clip,height=0.255\textheight,width=0.255\textheight
				]{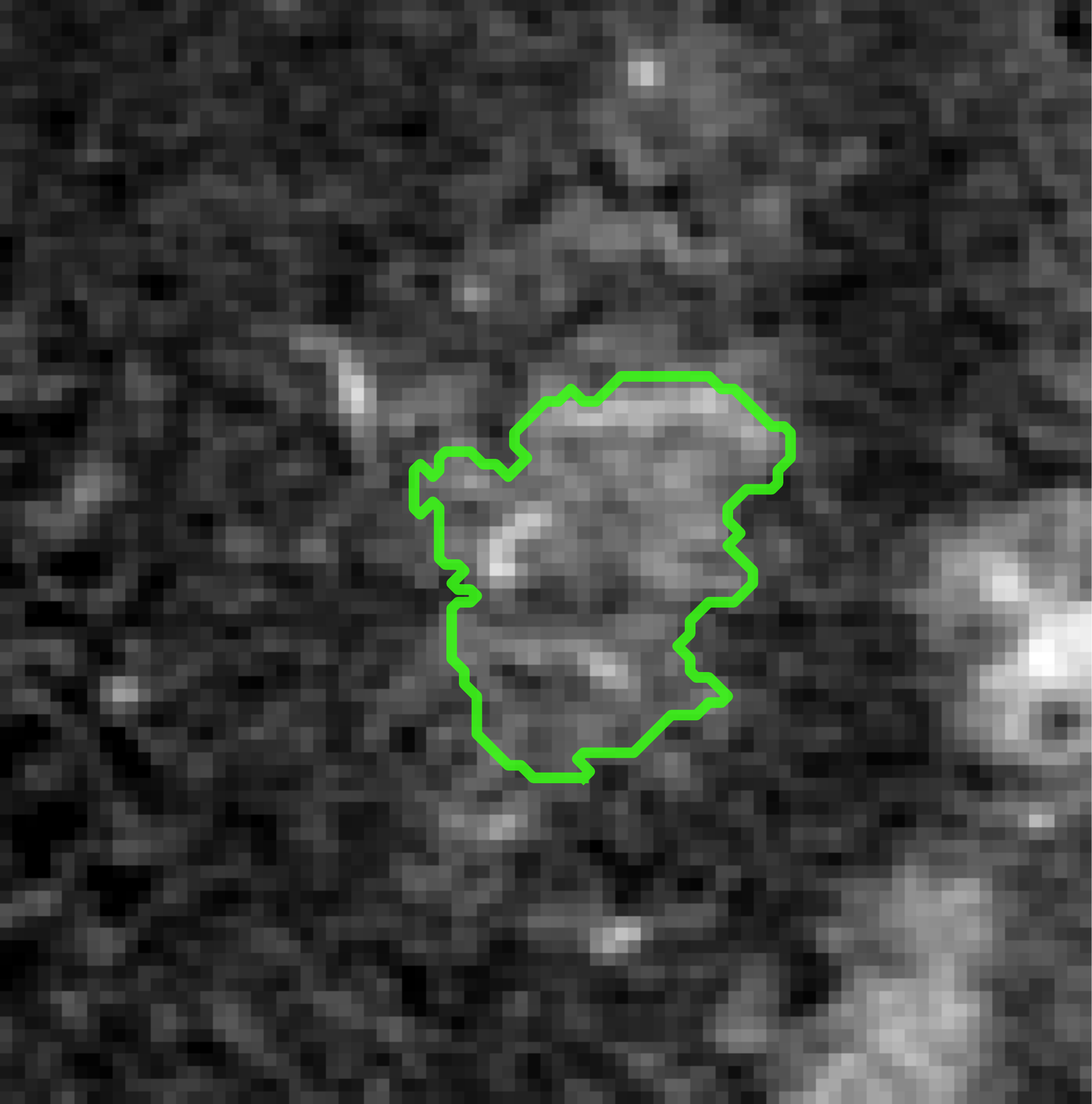}\\
				\includegraphics[trim={10 10 10 10},clip,height=0.255\textheight,width=0.255\textheight
				]{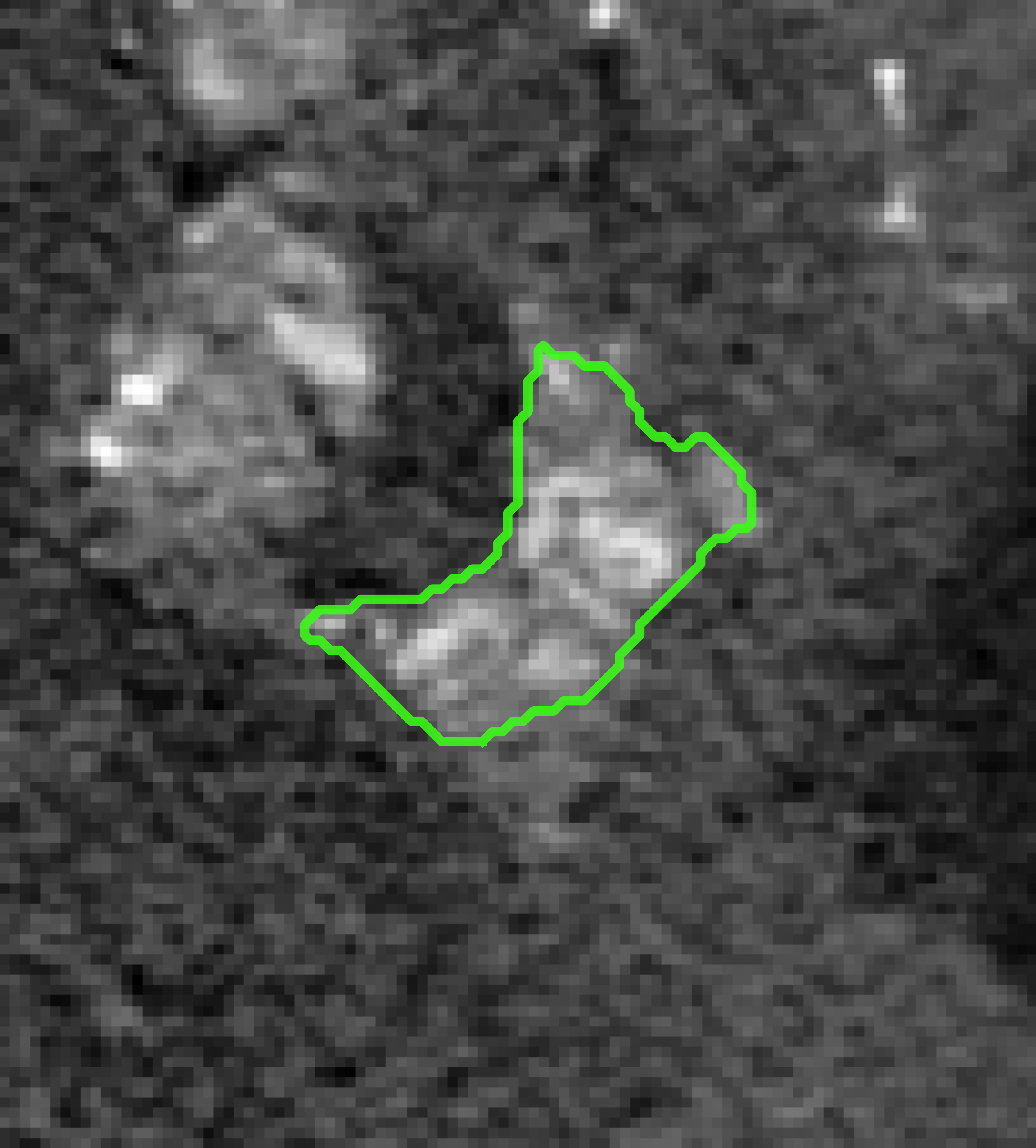} & %
				\includegraphics[trim={10 10 10 10},clip,height=0.255\textheight,width=0.255\textheight
				]{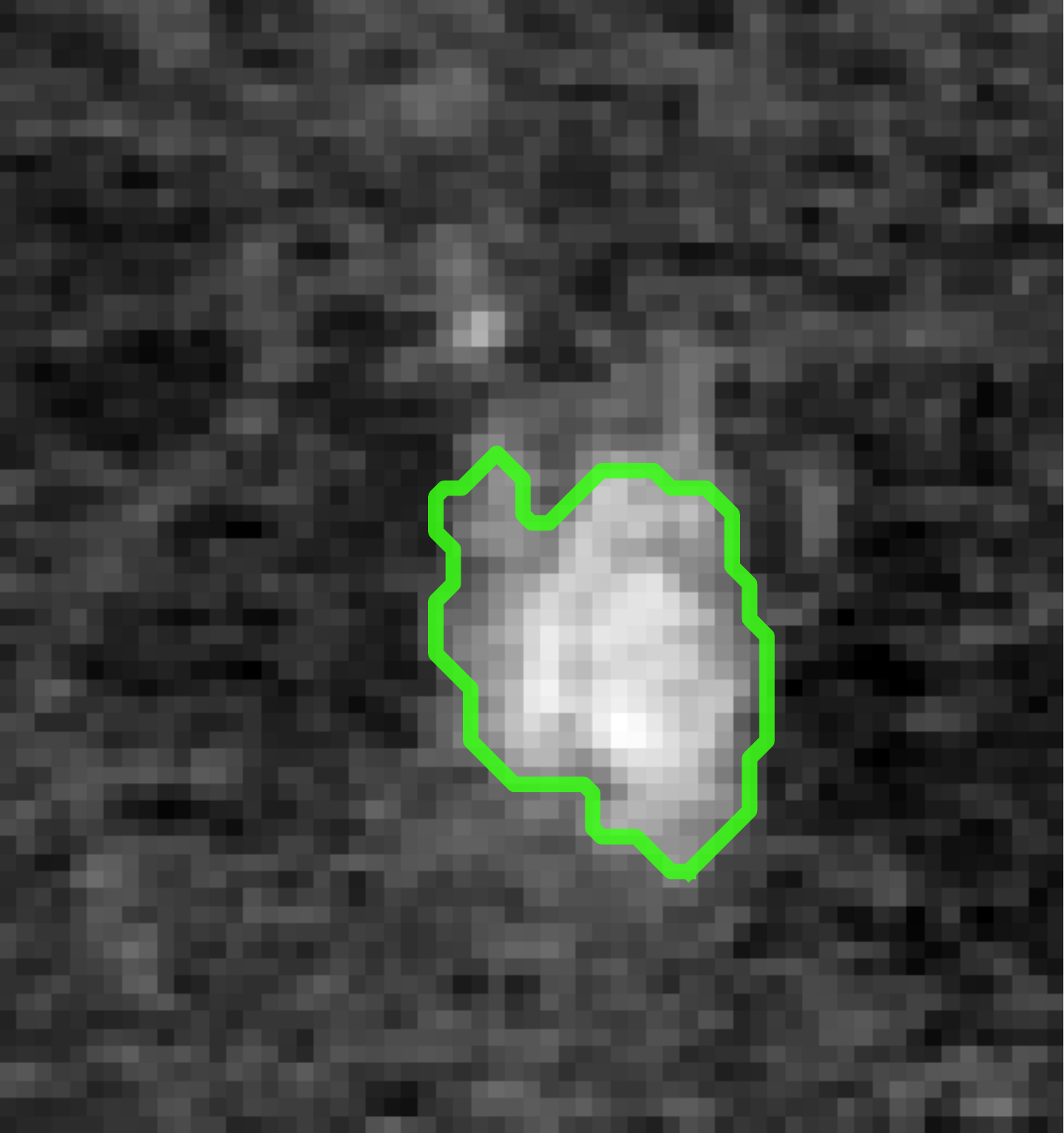} & %
				\includegraphics[trim={10 10 10 10},clip,height=0.255\textheight,width=0.255\textheight
				]{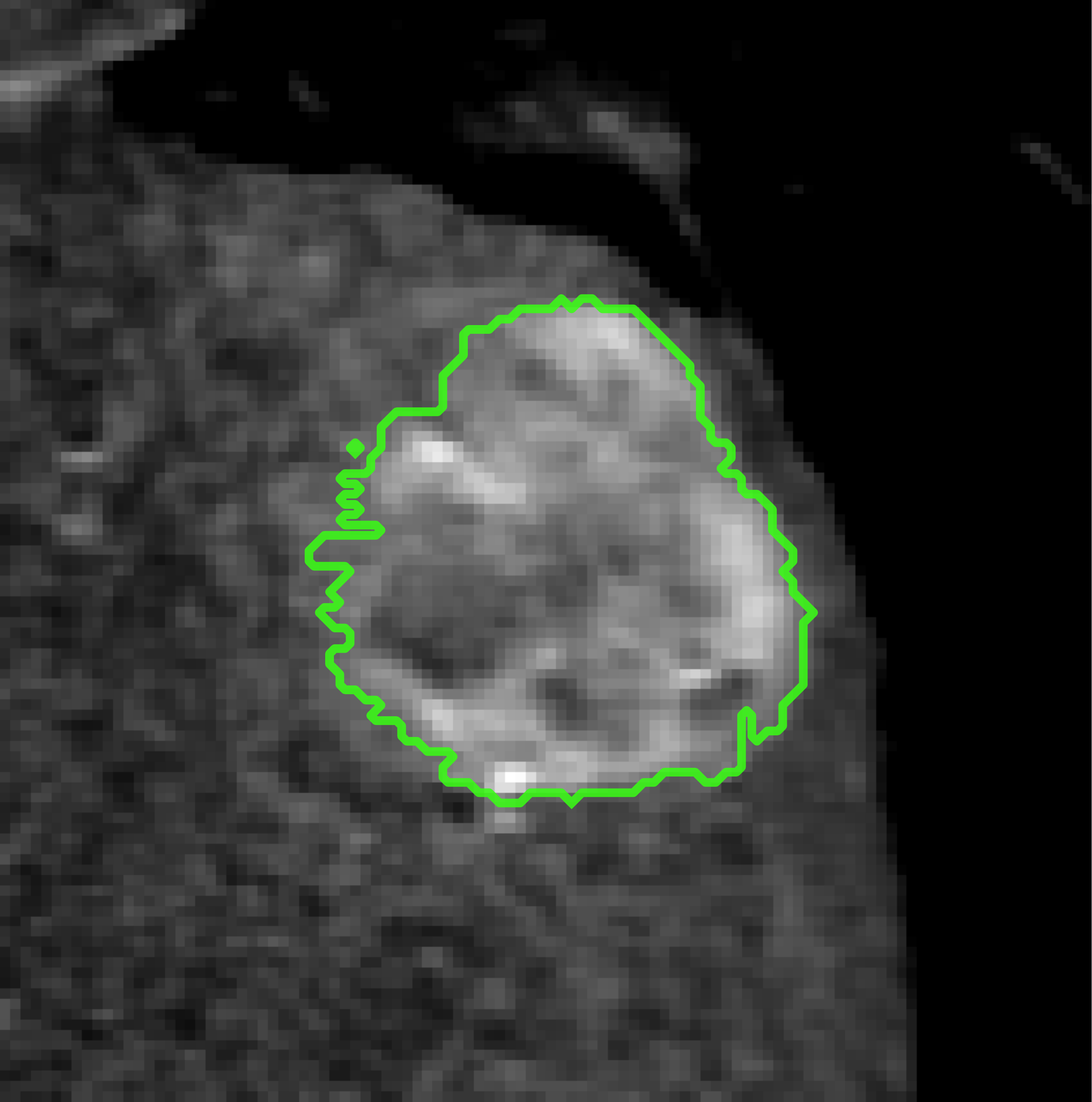}\\
				\includegraphics[trim={10 10 10 10},clip,height=0.255\textheight,width=0.255\textheight
				]{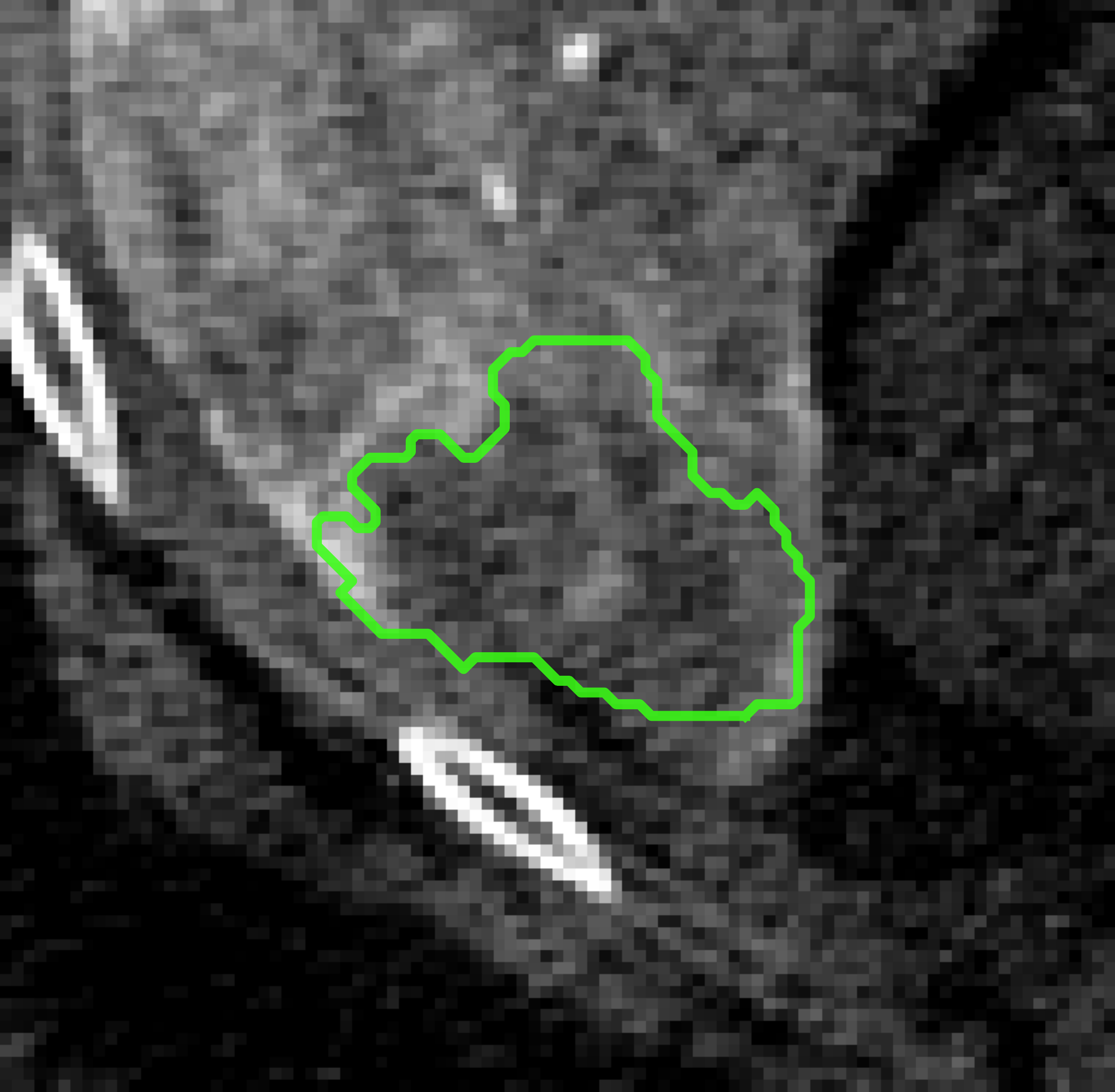} & %
				\includegraphics[trim={10 10 10 10},clip,height=0.255\textheight,width=0.255\textheight
				]{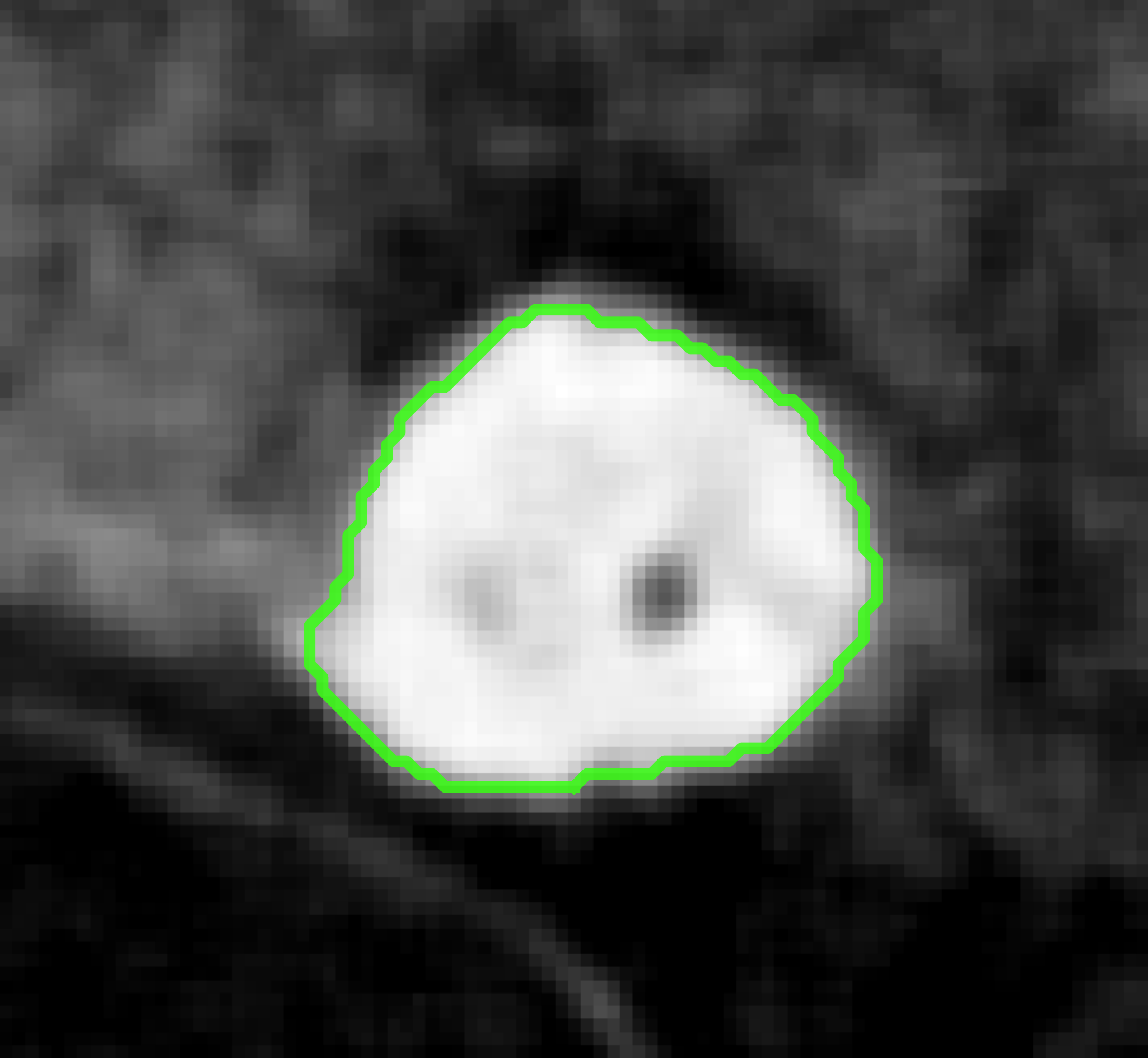} & %
				\includegraphics[trim={10 10 10 10},clip,height=0.255\textheight,width=0.255\textheight
				]{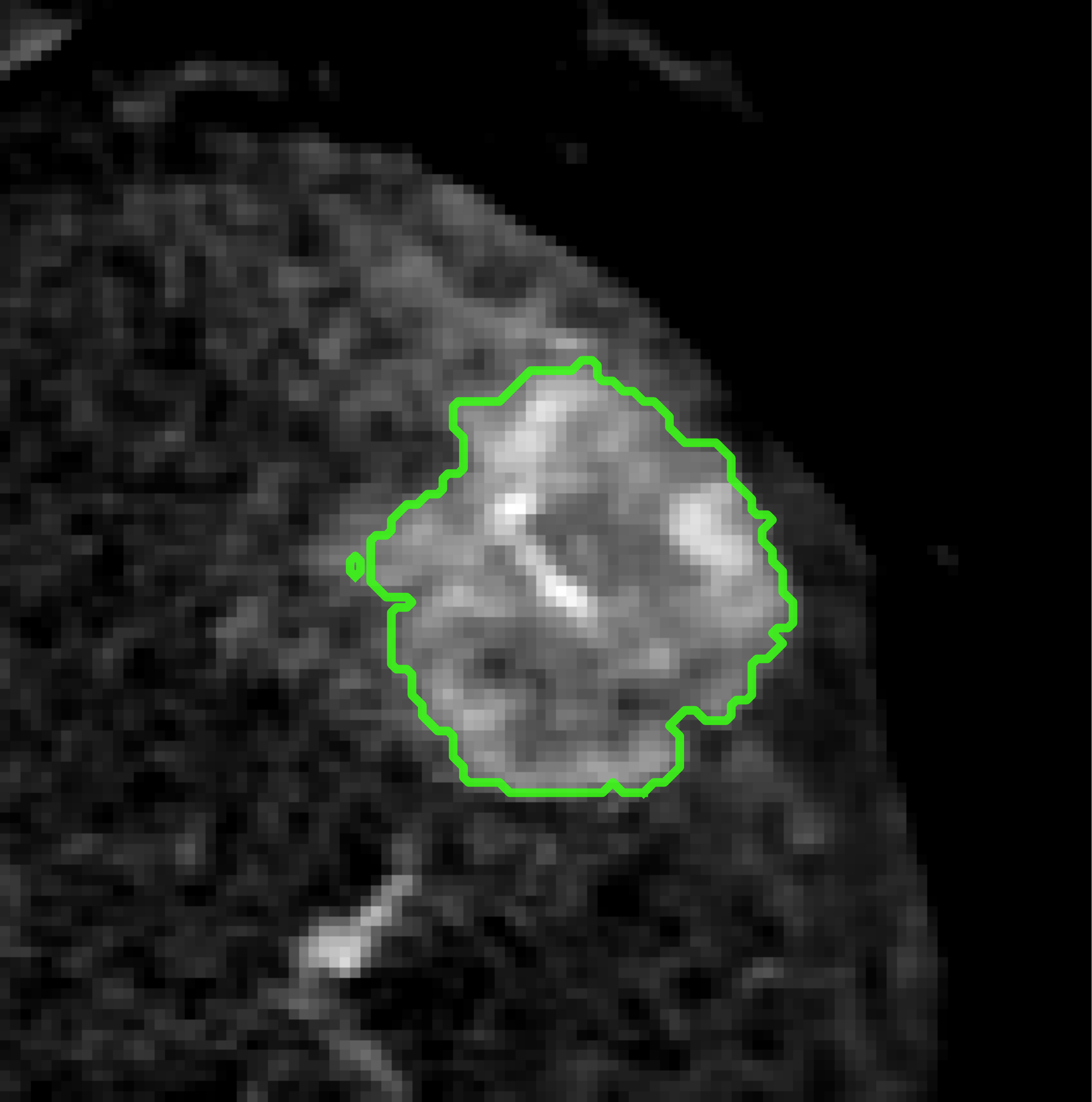}%
			\end{tabular}%
		}
	}%
	\caption{Liver lesion segmentations. 
		Depicted are central slices through the volumes of interest of reconstructed images acquired by a C-arm {\additioncaption{CB}CT} scanner.  
		The manually annotated ground truth segmentation is displayed as an overlay contour line in green.}%
	\label{fig:hepatic_tumor_segmentation_outcome}%
\end{figure}

Hepatocellular carcinoma (\mbox{HCC}) is among the most prevalent malignant tumors worldwide~\cite{chung2006transcatheter, mcglynn2011global}.
Only \mbox{$20$\,--\,$30$\,\%} of cases are curable via surgery.
Both, a patient's \mbox{HCC} and hepatic cirrhosis in advanced stages may lead on to the necessity of alternative treatment methods.
For these inoperable cases, trans-catheter arterial chemoembolization (\mbox{TACE})~\cite{lewandowski2011transcatheter} is a promising and widely used minimally invasive intervention technique~\cite{bruix2005management,bruix2011management}.
During \mbox{TACE}, \mbox{extra-hepatic} collateral vessels are occluded, which previously supplied the {HCC} with oxygenated blood.
To locate these vessels, it is crucial to find the exact shape as well as the position of the tumor inside the liver.
Interventional radiology is utilized to generate a volumetric cone-beam C-arm computed tomography (\mbox{CBCT})~\cite{strobel20093d} image of the patient's abdomen, which is processed to precisely outline and label the lesion.
The toxicity of \mbox{TACE} decreases, the less healthy tissue is labeled as pathologic.
The efficacy of the therapy increases, the less cancerous tissue is falsely labeled as healthy~\cite{lo2002randomized}.
However, precisely outlining the tumor is challenging, especially due to its variations in size and shape, as well as a high diversity in X-ray attenuation coefficient values representing the lesion as illustrated in \textbf{Fig.}\,\ref{fig:hepatic_tumor_segmentation_outcome}. %
While fully automated systems may yield insufficiently accurate segmentation results, \mbox{ISS} tend to be well suited for an application during \mbox{TACE}.

\section{Methods}\label{sec:methods}

\addition[label=c:a241,ref=c:c24]{In the following Section, the segmentation method underlying the user interface prototypes is described in \textbf{Sec.}\,\ref{sec:segmentation_method} in order to 
subsequently adequately outline the different characteristics of each novel interface prototype in \textbf{Sec.}\,\ref{sec:sgmentation_prototypes}.
Usability evaluation methods utilized are detailed regarding questionnaires in \textbf{Sec.}\,\ref{sec:questionnaires}, semi-structured feedback in \textbf{Sec.}\,\ref{sec:qualitative_measures}, as well as the test environment in \textbf{Sec.}\,\ref{sec:hci_evaluation}.
}

\subsection{Segmentation Method}\label{sec:segmentation_method}

\mbox{GrowCut}~\cite{vezhnevets2005growcut} is a seeded image segmentation algorithm based on cellular automaton theory.
The automaton is a tuple \mbox{$(\mathbf{G}_\mathbf{I},\mathbf{Q},\delta)$}, where $\mathbf{G}_\mathbf{I}$ is the \change[label=c:c241,ref=c:c24]{data the automaton operates on. In this case $\mathbf{G}_\mathbf{I}$ is the graph of image}{graph of} $\mathbf{I}$, where the pixels/voxels act as nodes $\mathbf{v}_e$.
The nodes are connected by edges on a grid defined by the Moore neighborhood system.
\addition[label=c:a242,ref=c:c24]{$\mathbf{Q}$ defines the automaton's possible states and $\delta$ the state transition function utilized.}
\begin{equation}
\mbox{$\mathbf{Q}\ni\mathbf{Q}_e^t=\left((\mathbf{p}_e, \,\mathbf{\ell}_e^t), \,\mathbf{\Theta}_e^t, \,\mathbf{c}_e, \,\mathbf{h}_e^t\right)$}
\label{eq:growcutgraph}
\end{equation}
\change[label=c:c242,ref=c:c24]{As detailed in \textbf{Eq.}\,\ref{eq:growcutgraph}, $\mathbf{Q}$ is the set of each node's state, where 
$\mathbf{p}_e$ is the node's position in image space and $\mathbf{\ell}_e^t$ is the class label}{is a state set, where \mbox{$\mathbf{\Theta}_e^t \in [0.0, 1.0]\subset \mathbb{R}$} is the strength} of node $e$ at \mbox{GrowCut} iteration $t$. 
\change[label=c:c243,ref=c:c24]{\mbox{$0 \le \mathbf{\Theta}_e^t \le 1$} is the strength of $e$ at iteration $t$.
The feature vector $\mathbf{c}_e$ describes}{and $c_e$ is the feature vector describing} the node's characteristics.
\addition[label=c:a2410,ref=c:c24]{ %
	The pixel value $\mathbf{I}\left(\mathbf{p}_e\right)$ at image location $\mathbf{p}_e$ is typically utilized as feature vector $\mathbf{c}_e$~\cite{vezhnevets2005growcut}.
} 
Here, we additionally define $\mathbf{h}_e^t \in \mathbb{N}^{0}$ as a counter for accumulated label changes of $e$ during the \mbox{GrowCut} iteration, as described in~\cite{amrehn2018ideal}, with 
\change[label=c:c244,ref=c:c24]{%
	\mbox{$\mathbf{h}_e^{t=0}=0$}}{%
	\mbox{$\mathbf{h}_e^0=0$}}. 
\addition[label=c:a243,ref=c:c24]{ %
	Note that this extension of GrowCut is later utilized for seed location suggestion in two of the three prototypes tested.} %
\change[label=c:c245,ref=c:c24]{A node's strength \mbox{$\mathbf{\Theta}_e^{t=0}$}}{\mbox{$\mathbf{\Theta}_e^0$}} 
is initialized with $1$ for scribbles, i.\,e.\ \change[label=c:c246,ref=c:c24]{%
\mbox{$(\mathbf{p}_e, \,\mathbf{\ell}_e^{t=0})\in\mathbf{S^{t=0}}$}}{%
\mbox{$(\mathbf{p}_e, \,\mathbf{\ell}_e^0)\in\mathbf{S^0}$}}%
, and $0$ otherwise.

Iterations \mbox{$\operatorname{\delta}\left(\mathbf{Q}_e^t\right)=\mathbf{Q}_e^{t+1}$} are performed utilizing local state transition rule $\delta$:
starting from initial seeds, labels are propagated based on local intensity features $\mathbf{c}$.
At each discrete time step $t$, each node $f$ attempts to conquer its direct neighbors.
A node $e$ is conquered if \addition[label=c:a244,ref=c:c24]{the condition in \textbf{Eq.}\,\ref{eq:growcutisconquered} is true.} %
\begin{align}
	\mathbf{\Theta}_f^t\cdot\operatorname{g}(\mathbf{c}_e,\mathbf{c}_f)&>\mathbf{\Theta}_e^t\,,\ \text{where}\label{eq:growcutisconquered}\\
	\operatorname{g}(\mathbf{c}_e,\mathbf{c}_f) &= 1 - \frac{\Vert\mathbf{c}_e-\mathbf{c}_f\Vert_2}{\max_{j,k}\Vert\mathbf{c}_j-\mathbf{c}_k\Vert_2}
\end{align}
If node $e$ is conquered, the automaton's state set is updated \addition[label=c:a245,ref=c:c24]{ %
	according to \textbf{Eq.}\,\ref{eq:growcutupdatestate}.
	If $e$ is not conquered, the node's state remains unchanged, i.\,e.\ \mbox{$\mathbf{Q}_e^{t+1}=\mathbf{Q}_e^t$}.
} %
\begin{equation} \mbox{$\mathbf{Q}_e^{t+1}=((\mathbf{p}_e,\mathbf{\ell}_f^t),\mathbf{\Theta}_f^t\cdot\operatorname{g}(c_e,c_f),\mathbf{c}_e,\mathbf{h}_e^t+1)$},
\label{eq:growcutupdatestate}
\end{equation}
The process is guaranteed to converge with positive and bounded node strengths \addition[label=c:a246,ref=c:c24]{ %
	($\forall_{e,t} \ \mathbf{\Theta}_e^t \le 1$) %
} monotonously decreasing \addition[label=c:a247,ref=c:c24]{ %
	(since $\operatorname{g}(.) \le 1$). 
	The image's final segmentation mask after convergence is encoded as part of state $\mathbf{Q}^{t=\infty}$, specifically in $(\mathbf{p}_e, \,\mathbf{\ell}_e^{t=\infty})$ for each node $e$. 
}

\subsection{Interactive Segmentation Prototypes}\label{sec:sgmentation_prototypes}

Three interactive segmentation prototypes with different \mbox{UIs} were implemented for usability testing.
The segmentation technique applied in all prototypes is based on the \mbox{GrowCut} approach as described in \textbf{Sec.}\,\ref{sec:segmentation_method}. 
\mbox{GrowCut} allows for efficient and parallelizable computation of image segmentations while providing an acceptable accuracy from only few initial seed points.
\addition[label=c:a248,ref=c:c24]{The method is also chosen due to its tendency to benefit from careful placement of large quantities of seed points.}
It is therefore well suited for an integration into a highly interactive system.
\addition[label=c:a249,ref=c:c24]{A learning-based segmentation system was not utilized for usability testing due to its inherent dependence of segmentation quality on the characteristics of prior training data, which potentially adds a significant bias to the test results, given only a small data set as utilized in the scope of this work.}

All three user interfaces provided include an \emph{undo} button to reverse the effects of the user's latest action.
A \emph{finish} button is used to define the stopping criterion for the interactive image partitioning.
The transparency of both, the contour line and seed mask displayed, is adjustable to one of five fixed values via the \emph{opacity} toggle button.
The image contrast and brightness (windowing) can be adapted with standard control sliders for the window width and the window center operating on the image intensity value range~\cite{jin2001contrast}.
All protoypes incorporate a \emph{help} button used to provide additional guidance for the prototype's usage during the segmentation task.
The segmentation process starts with a set of pre-defined background-labels $\mathbf{S}^0$ along the edges of the image, 
since an object is assumed to be located in its entirety inside the displayed region of the image.

\subsubsection{\mbox{Semi-manual} Segmentation Prototype}\label{sec:semi-manual_prototype}

\begin{figure}
	\includegraphics[width=\columnwidth,height=0.61296534017\columnwidth]{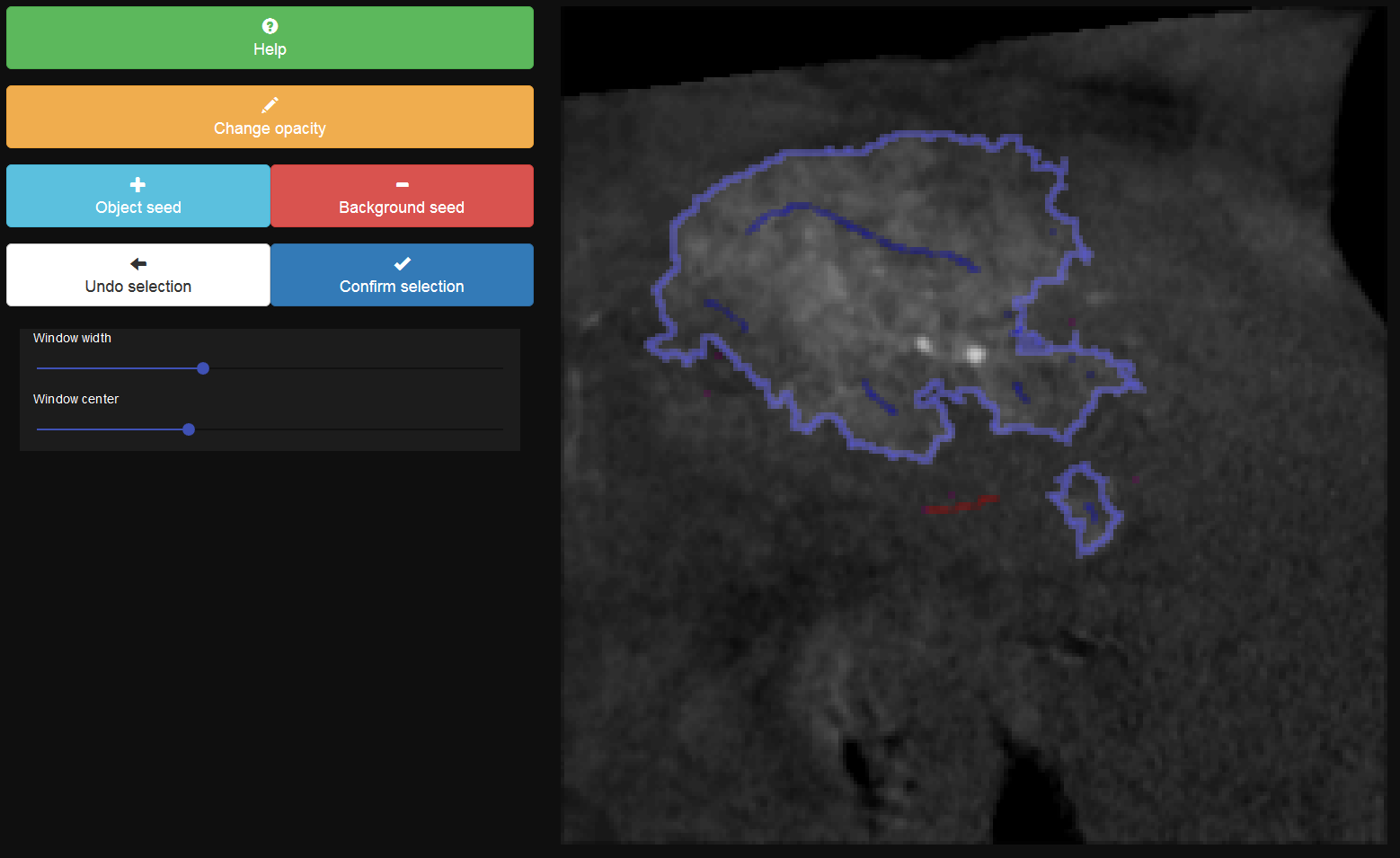}
	\caption{\mbox{Semi-manual} segmentation prototype user interface.
		The current segmentation's contour line (light blue) is \changecaption{adjusted towards the user's estimate of the ground truth segmentation}{adapted} by manually adding foreground (blue) or background (red) seed points.}
	\label{fig:semi-manual_prototype}
\end{figure}

The \mbox{UI} of the \mbox{semi-manual} prototype, depicted in \textbf{Fig.}\,\ref{fig:semi-manual_prototype}, provides several interaction elements.
A user can add seed points as an overlay mask displayed on top of the image.
These seed points have a pre-defined label of either \mbox{\emph{foreground}} for the object or \mbox{\emph{background}} used for all other image elements.
The label of the next brush strokes (scribbles) can be altered via the buttons named \mbox{\emph{object seed}} and \mbox{\emph{background seed}}.
After each interaction \mbox{$n\in\mathbb{N}$}, a new iteration of the seeded segmentation is started given the image $\mathbf{I}$ as well as the 
updated set of seeds \mbox{$\mathbf{S}^n=\mathbf{S}^{n-1}\cup\{\mathbf{s}^n_1,\mathbf{s}^n_2,\dots\}$} as input.

\subsubsection{Guided Segmentation Prototype}\label{sec:guided_prototype}

\begin{figure}
	\includegraphics[width=\columnwidth,height=0.61296534017\columnwidth]{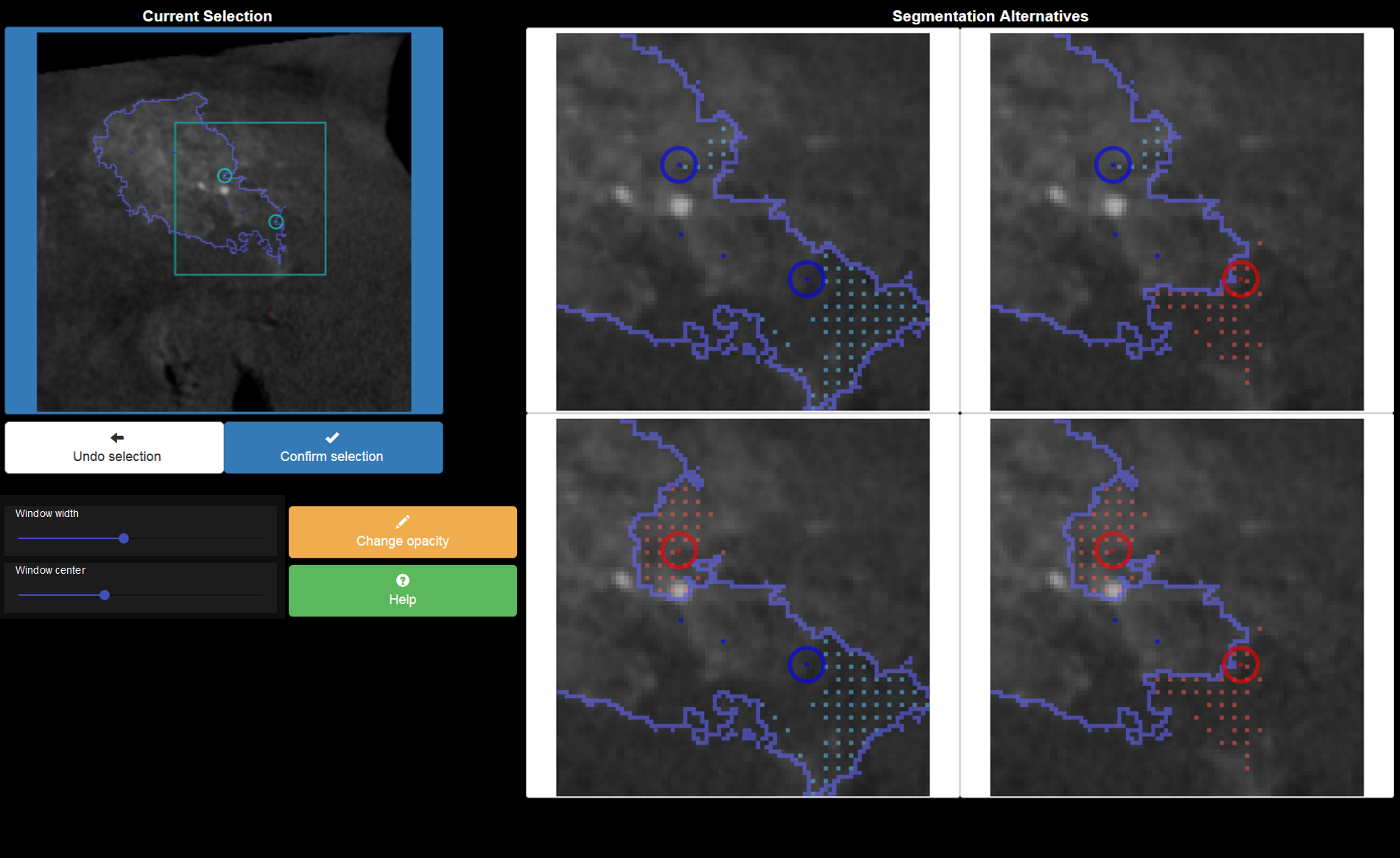}
	\caption{Guided segmentation prototype user interface.
		The current segmentation displayed on the upper left can be improved by choosing one of the four segmentation alternatives displayed on the right.
		The user is expected to choose the upper-right option in this configuration, 
		\additioncaption{due to the two new seeds' matching background and foreground labels}.} %
	\label{fig:guided_prototype}
\end{figure}

The system selects two seed point locations \mbox{$\mathbf{p}^n_1$ and $\mathbf{p}^n_2$}, each with the lowest label certainty values assigned by the previous segmentation process. 
The seed point locations are shown to the user in each iteration $n$, as depicted in \textbf{Fig.}\,\ref{fig:guided_prototype}.
There are four possible labeling schemes for those points in the underlying \mbox{two-class} classification problem, since each seed point 
\mbox{$\mathbf{s}^n_i=(\mathbf{p}^n_i,\mathbf{\ell}^n_i)$} has a label \mbox{$\mathbf{\ell}^n_i\in\{{background},{foreground}\}$}.
The interface providing advanced user guidance displays the four alternative segmentation contour lines, which are a result of the four possible next steps 
during the iterative interactive segmentation with respect to the labeling of the new seed points $\mathbf{s}^n_1$ and $\mathbf{s}^n_2$.
The user selects the only correct labeling, where all displayed object and background seeds are inside the object of interest and the image background, respectively.
The alternative views on the right act as four buttons to define a selection.
To further assist the user in their decision making, the region of interest, defined by $\mathbf{p}^n_1$ and $\mathbf{p}^n_2$, is zoomed in for the option view 
on the right and displayed as a cyan rectangle in the overview image on the left of the \mbox{UI}.
The differences regarding the previous iteration's contour line and one of the four new options each are highlighted by dotted areas in the four overlay mask images.
After the user selects one of the labelings, the two new seed points are added to the current set of scribbles $\mathbf{S}^n$.
The scribbles \mbox{$\mathbf{S}^n:=\mathbf{S}^{n-1}\cup\left\{\mathbf{s}^n_1,\mathbf{s}^n_2\right\}$} are utilized as input for the next iteration, on which basis two new 
locations \mbox{$\mathbf{p}^{n+1}_1$ and $\mathbf{p}^{n+1}_2$} are computed.

The system-defined locations of the additional seed points can be determined by \mbox{$\argmax_e\mathbf{h}_e^{t=\infty,{n-1}}$}, the location(s) with maximum number of label changes during \mbox{GrowCut} segmentation.
Frequent changes define specific image elements and areas in which the \mbox{GrowCut} algorithm indicates uncertainty in finding the correct labels.  %
Two locations in $\mathbf{h}^{t=\infty,{n-1}}$ are then selected as $\mathbf{p}^n_1$ and $\mathbf{p}^n_2$, which stated the most changes in labeling during the previous 
segmentation with input image $\mathbf{I}$ and seeds $\mathbf{S}^{n - 1}$. %

\subsubsection{Joint Segmentation Prototype}\label{sec:joint_prototype} 

\begin{figure}
	\includegraphics[width=\columnwidth,height=0.61296534017\columnwidth]{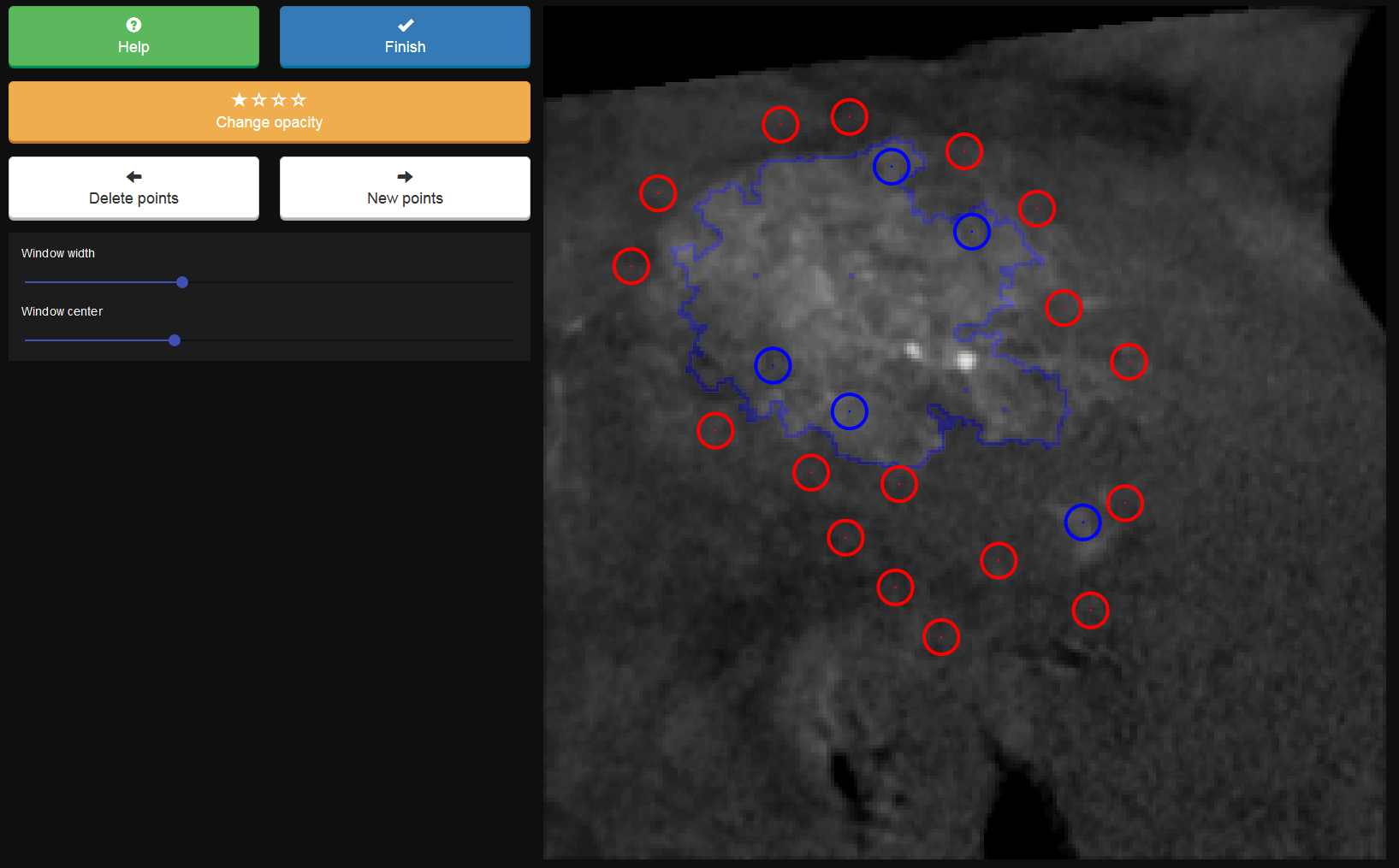}
	\caption{Joint segmentation prototype user interface.
	The user toggles the labels of pre-positioned seed points\additioncaption{, which positions are displayed to them as colored circles,} to properly indicate their inclusion into the set of object or background representatives.
	New seeds can be added at the position of \additioncaption{current} interaction via a long-press on the overlay image.
	The segmentation result as well as \additioncaption{the} displayed contour line adapt accordingly after each interaction.}
	\label{fig:joint_prototype}
\end{figure}

\begin{figure}
	\centering
	\includegraphics[height=0.61296534017\columnwidth]{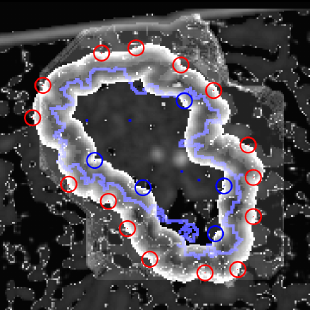}%
	\caption{The approximated influence map for new seed point locations \additioncaption{for the joint segmentation prototype}.
	The map is generated by a weighted sum of gradient magnitude image, number of cell changes \additioncaption{$h_e^{t=\infty}$ per cell $e$} 
	obtained from \additioncaption{the} previous \mbox{GrowCut} segmentation, \changecaption{as well as the}{and} distance to the contour line of the current segmentation.
	}%
	\label{fig:joint_prototype_prob_map}
\end{figure}

The joint prototype depicted in \textbf{Fig.}\,\ref{fig:joint_prototype} is a combination of a pictorial interaction scheme and a menu-driven approach.
(1) A set of \mbox{$J\in\mathbb{N}$} pre-selected new seeds is displayed in each iteration.
The seeds' initial labels are set automatically, based on whether their position is inside (foreground) or outside (background) the current segmentation mask.
The user may toggle the label of each of the new seeds, which also provides an intuitive \mbox{\emph{undo}} functionality. 
The automated suggestion process for new seed point locations is depicted in \textbf{Fig.}\,\ref{fig:joint_prototype_prob_map}.
The seed points are suggested deterministically based on the indices of the maximum values in an element-wise sum of three approximated influence maps.
These maps are 
the gradient magnitude image of $\mathbf{I}$, 
the previous label changes \mbox{$\mathbf{h}^{t=\infty,{n-1}}$} per element in $\mathbf{G}_\mathbf{I}$ weighted by an empirically determined factor of $17/12$, 
and an influence map based on the distance of each element in $\mathbf{I}$ to the current contour line.
Note that for the guided prototype (see \textbf{Sec.}\,\ref{sec:guided_prototype}), only $\mathbf{h}$ was used for the selection of suggested seed point locations.
This scheme was extended for the joint prototype, since extracting \mbox{$J\approx20$} instead of only the top two points solely from $\mathbf{h}$ potentially introduces suggested point locations forming impractical local clusters instead of spreading out with higher variance in the image domain. 
This process approximates the true influence or entropy (information gain) of each possible location for a new seed.

When all seed points \mbox{$\left\{\mathbf{s}^n_1,\mathbf{s}^n_2,\dots,\mathbf{s}^n_J\right\}$} presented to the user are toggled to their correct label, 
the user may click on the \emph{new points} button to initiate the next iteration with an updated set of seed points 
\mbox{$\mathbf{S}^n=\mathbf{S}^{n-1}\cup\{\mathbf{s}^n_1,\mathbf{s}^n_2,\dots,\mathbf{s}^n_J\}$}.
Another set of seed points \mbox{$\{\mathbf{s}^{n+1}_1,\mathbf{s}^{n+1}_2,\dots,\mathbf{s}^{n+1}_J\}$} is generated and displayed.

(2) In addition to pre-selected seeds, a single new seed point $\mathbf{s}^n_0$ can be added manually via a user's long-press on any location in the image.
A desired change in the current labeling of this region is interpreted given this user action.
Therefore, the new seed point's initial label is set by inverting the current label of the given location.
A new segmentation is initiated by this interaction based on \mbox{$\mathbf{S}^n=\mathbf{S}^{n-1}\cup\left\{\mathbf{s}^{n}_0,\mathbf{s}^{n}_1,\dots,\mathbf{s}^n_J\right\}$}.
Note that the labels of \mbox{$\mathbf{s}^n_i$} are still subject to change via toggle interactions until the \mbox{\emph{new points}} button is pressed.

\subsection{Questionnaires}\label{sec:questionnaires}

\subsubsection{System Usability Scale (\mbox{SUS})}\label{sec:questionnaires_sus}

\begin{figure}%
	\includegraphics[width=\linewidth]{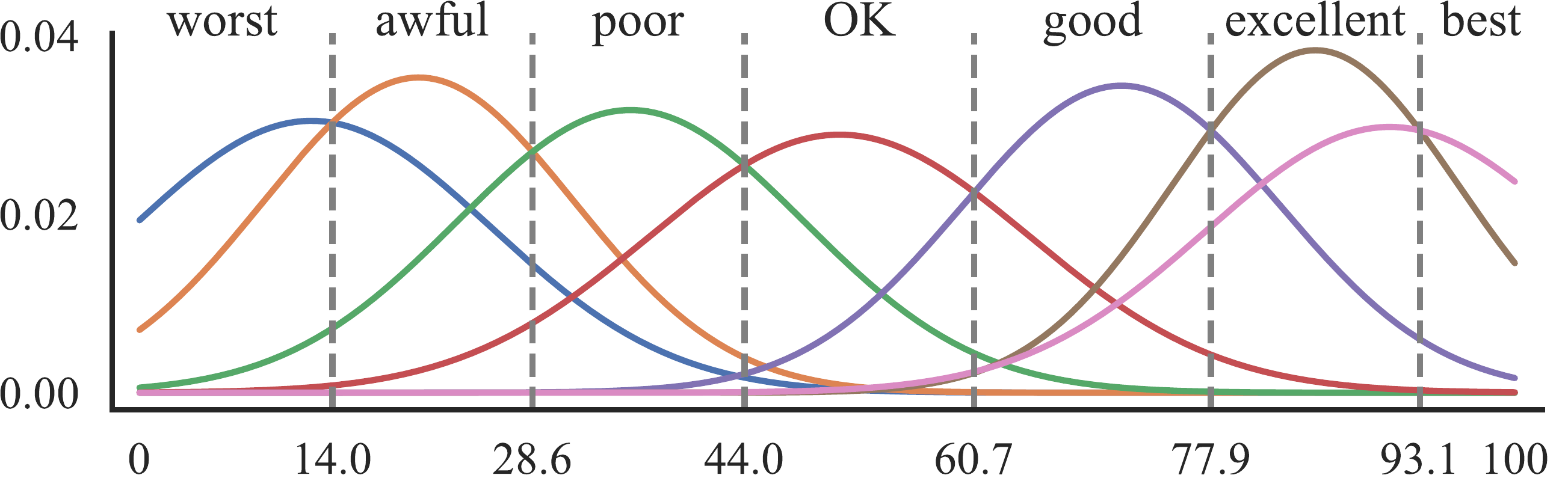}\\
	\centerline{\small System usability scale (\mbox{SUS}) rating}%
	\caption{Mapping from a \mbox{SUS} score to an adjective rating scheme proposed by Bangor et al.~\cite{bangor2009determining}\additioncaption{. %
	Given a \mbox{SUS} rating, the relative height of the Gaussian distributions approximate the probabilities for each adjective. %
	Distributions' $\mu$ and $\sigma$ were extracted evaluating} $959$ surveys 
	\additioncaption{with added adjective rating as an 11th question}.
	}%
	\label{fig:sus_adjective}%
\end{figure}

The \mbox{SUS}~\cite{brooke1996sus,lewis2009factor} is a widely used, reliable, and low-cost survey to assess the overall usability of a prototype, product, or service~\cite{kortum2013usability}.
Its focus is on pragmatic quality evaluation~\cite{ISO92411998,ISO92412018}.
The survey is technology agnostic, which enables a utilization of the usability of many types of user interfaces and \mbox{ISS}~\cite{bangor2009determining}.
The questionnaire consists of ten statements and an unipolar five-point Likert scale~\cite{likert1932technique}. %
This allows for an assessment in a time span of about three minutes per participant.
The statements are as follows:
\begin{enumerate}
	\item I think that I would like to use this system frequently.
	\item I found the system unnecessarily complex.
	\item I thought the system was easy to use.
	\item I think that I would need the support of a technical person to be able to use this system.
	\item I found the various functions in this system were well integrated.
	\item I thought there was too much inconsistency in this system.
	\item I would imagine that most people would learn to use this system very quickly.
	\item I found the system very cumbersome to use.
	\item I felt very confident using the system.
	\item I needed to learn a lot of things before I could get going with this system.
\end{enumerate}
The Likert scale provides a fixed choice response format to these expressions.
The \mbox{$(N-1)/2$\,th} choice in an \mbox{$N$-point} Likert scale always is the neutral element.
Using the scale, subjects are asked to define their degree of consent to each given statement.
The fixed choices for the five-point scale are named \emph{strongly disagree}, \emph{disagree}, \emph{undecided}, \emph{agree}, and \emph{strongly agree}.
During the evaluation of the survey, these names are assigned values 
\change[label=c:c171,ref=c:c17]{%
\mbox{$\mathbf{x}^\text{SUS}_{s,i} \in \left\{0, 1, \dots, 4\right\}$} per subject $s$}%
{$\mathbf{x}_i$ from zero to four} 
in the order presented, for statements with index \change[label=c:c131,ref=c:c13]{\mbox{$i \in \left\{1, 2, \dots, 10\right\}$}}{\mbox{$i \in \left[1, 10\right]$}}. 
\mbox{SUS} scores enable simple interpretation schemes, understandable also in multi-disciplinary project teams.
The result of the \mbox{SUS} survey is a single scalar value, in the range of zero to $100$ as a composite measure of the overall usability.
The score is computed according to 
\addition[label=c:a131,ref=c:c13]{\textbf{Eq.}\,\ref{eq:sus_score}}, 
\addition[label=c:a151,ref=c:c15]{as outlined in~\cite{brooke1996sus},} 
given $S$ participants, where 
\change[label=c:c172,ref=c:c17]{$\mathbf{x}^\text{SUS}_{s,i}$}{$\mathbf{x}_s$} 
is the response to 
\change[label=c:c173,ref=c:c17]{the statement}{all statements} $i$ by subject $s$.
\begin{equation}
	\operatorname{sus}(\mathbf{x}) = \frac{2.5}{S} \sum_{s}\left[\, \sum_{\text{odd } i} \mathbf{x}^\text{SUS}_{s,i} + \sum_{\text{even } i} (4 - \mathbf{x}^\text{SUS}_{s,i})\, \right]
	\label{eq:sus_score}
\end{equation}
\addition[label=c:a156,ref=c:c15]{A neutral participant (\mbox{$\forall_{i} \ \mathbf{x}^\text{SUS}_{s,i} = 2$}) would produce a \mbox{SUS} score of $50$.} 
Although the \mbox{SUS} score allows for straightforward comparison of the usability throughout different systems, there is no simple intuition associated with the resulting scalar value.
\addition[label=c:a141,ref=c:c14]{ %
	\mbox{SUS} scores do not provide a linear mapping of a system's quality in terms of overall usability. %
}
In practice, a \mbox{SUS} of less than $80$ is often interpreted as an indicator of a substantial usability problem with the system.
Bangor et al.~\cite{bangor2008empirical,bangor2009determining} %
proposed an interpretation of the score in a seven-point scale.
\addition[label=c:a142,ref=c:c14]{ %
	They added an eleventh question to $959$ surveys they conducted.
	Here, participants were asked to describe the overall system as one of these seven items of an adjective rating scale%
}: 
\emph{worst imaginable}, \emph{awful}, \emph{poor}, \emph{OK}, \emph{good}, \emph{excellent}, and \emph{best imaginable}.
\addition[label=c:a143,ref=c:c14]{ %
	The resulting \mbox{SUS} scores could then be correlated with the adjectives.
	The mapping from scores to adjectives resulting from their evaluation is depicted in \textbf{Fig.}\,\ref{fig:sus_adjective}. %
}
This mapping also enables an absolute interpretation of a single \mbox{SUS} score. %

\subsubsection{Semantic Differential \mbox{AttrakDiff-2}}\label{sec:questionnaires_attrakdiff}

A semantic differential is a technique for the measurement of meaning as defined by Osgood et al.~\cite{osgood1952nature,osgood1957measurement}.
Semantic differentials are based on the theory, that the implicit anticipatory response of a person to a stimulus object is regarded as the object's meaning.
Since these implicit responses themselves cannot be recorded directly, more apparent responses like verbal expressions have to be considered~\cite{mehrabian1974approach,fishbein1975belief}.
These verbal responses have to be sensitive to and maximally dependent on meaningful states while independent from each other~\cite{osgood1957measurement}.
Hassenzahl et al.~\cite{hassenzahl2003attrakdiff, hassenzahl2000hedonic} defined a set of $28$ pairs of verbal expressions suitable to represent a subject's opinion on the hedonic  as well as pragmatic quality (both aspects of perception) and attractiveness (an aspect of assessment) of a given interactive system separately~\cite{hassenzahl2001effect}.
During evaluation, the pairs of complementary adjectives are clustered into four groups, each associated with a different aspect of quality.
Pragmatic quality (\mbox{PQ}) is defined as the perceived usability of the interactive system, which is the ability to assist users to reach their goals by providing utile and usable functions~\cite{hassenzahl2008user}.
The attractiveness (\mbox{ATT}) quantizes the overall appeal of the system~\addition[label=c:a281,ref=c:c28]{\cite{hassenzahl2002importance}}.
The hedonic quality (\mbox{HQ})~\cite{diefenbach2008give} is separable into hedonic identity (\mbox{HQ-I}) and hedonic stimulus (\mbox{HQ-S}).
\mbox{HQ-I} focuses on a user's identification with the system and describes the ability of a product to communicate with other persons benefiting the user's self-esteem~\cite{hassenzahl2007hedonic}. 
\mbox{HQ-S} describes the perceived novelty of the system. \mbox{HQ-S} is associated with the desire to advance ones knowledge and proficiencies.  %
The clustering into these four groups for the $28$ word pairs are defined as depicted in \textbf{Tab.}\,\ref{tab:attrakdiff_statements}.
\begin{table*}%
	\caption{\mbox{AttrakDiff-2} statement pairs \deletioncaption{ for each category}. 
		\additioncaption{The pairs of complementary adjectives are clustered into four groups, each associated with a different aspect of quality.}
		All $28$ pairs are presented to participants in randomized order. %
	}%
	\label{tab:attrakdiff_statements}%
	\resizebox{\textwidth}{!}{%
	\begin{tabular}{llll}%
		Pragmatic quality ({PQ}) & Attractiveness ({ATT}) & Hedonic identity ({HQ-I}) & Hedonic stimulus ({HQ-S})\\\hline
		complicated, simple & bad, good & alienating, integrating & cautious, bold \\
		confusing, clearly structured & disagreeable, likeable & cheap, premium & conservative, innovative\\
		cumbersome, straightforward & discouraging, motivating & isolating, connective & conventional, inventive\\
		impractical, practical & rejecting, inviting & separates me from, brings me closer to people & dull, captivating\\
		technical, human & repelling, appealing & tacky, stylish & ordinary, novel\\
		unpredictable, predictable & ugly, attractive & unpresentable, presentable & undemanding, challenging\\
		unruly, manageable & unpleasant, pleasant & unprofessional, professional & unimaginative, creative
	\end{tabular}%
	}%
\end{table*}

For each participant, the order of word pairs and order of the two elements of each pair are randomized prior to the survey's execution.
A bipolar~\cite{mccroskey1989bipolar} seven-point Likert scale is presented to the subjects to express their relative tendencies toward one of the two opposing statements (\mbox{poles}) of each expression pair, where index three denotes the neutral element.
For the questionnaire's evaluation for subject \change[label=c:c132,ref=c:c13]{\mbox{$s \in \left\{0, 1, \dots, S-1\right\}$}}{\mbox{$s \in [0, S)$}}, each of the seven adjective pairs \change[label=c:c133,ref=c:c13]{\mbox{$i \in \left\{0, 1, \dots, 6\right\}$}}{\mbox{$i \in [0, 6]$}} per group 
\mbox{$g \in \{\text{PQ}, \,\text{ATT}, \,\text{HQ-I}, \,\text{HQ-S}\}$} is assigned a score \mbox{$\mathbf{x}^g_{s,i} \in \left\{1, 2, \dots, 7\right\}$} by each participant, reflecting their tendency towards the positive of the two adjectives.
The overall ratings per group are \addition[label=c:a132,ref=c:c13]{defined in \cite{hassenzahl2003attrakdiff} as} the mean scores computed over all subjects $s$ and statements $i$, 
\addition[label=c:a133,ref=c:c13]{as depicted in \textbf{Eq.}\,\ref{eq:attrakdiff_score}}%
. 
Here, $S$ is the number of participants in the survey.
\begin{equation}
	\operatorname{attrakdiff}(\mathbf{x}, \,g) = \frac{1}{7 \cdot S} \sum_{s} \sum_{i} \mathbf{x}^g_{s,i}
	\label{eq:attrakdiff_score}
\end{equation}
Therefore, a neutral participant would produce an \mbox{AttrakDiff-2} score of four.
The final averaged score of each group $g$ ranges from one (worst) to seven (best rating).

An overall evaluation of the \mbox{AttrakDiff-2} results can be conducted in the form of a portfolio representation~\cite{hassenzahl2008user}.
\mbox{HQ} is the mean of a system's \mbox{HQ-I} and \mbox{HQ-S} scores.
{PQ} and {HQ} scores of a specific system and user are visualized as a point in a two-dimensional graph.
The $95$\,\% confidence interval is an estimate of plausible values for rating scores from additional study participants, 
and determines the extension of the rectangle around the described data point in each dimension.
A small rectangle area represents a more homogeneous rating among the participants than a larger area.
If a rectangle completely lies inside one of the seven fields with associated adjectives defined in~\cite{hassenzahl2008user}, this adjective is regarded as the dominant descriptor of the system.
Otherwise, systems can be particularized by overlapping fields' adjectives.
If the confidence rectangles of two systems overlap in their one-dimensional projection on either \mbox{HQ} or \mbox{PQ}, their difference in \mbox{AttrakDiff-2} scores in regards to this dimension is not significant.

\subsection{Qualitative Measures}\label{sec:qualitative_measures}

In order to collect, normalize, and analyze visual and verbal feedback given by the participants, a summative qualitative content analysis is conducted via abstraction~\cite{hsieh2005three,elo2008qualitative}.
The abstraction method reduces the overall transcript material while preserving its substantial contents by summarization.
The corpus retains a valid mapping of the recording.
An essential part of abstraction is the formulation of macro operators like elimination, generalization, construction, integration, selection and
bundling.
The abstraction of statements is increased iteratively by the use of macro operators, which map statements of the current level of abstraction to the next, while clustering items based on their similarity~\cite{mayring2014qualitative}.

\subsection{HCI Evaluation}\label{sec:hci_evaluation}  %

A user study is the most precise method for the evaluation of the quality of different interactive segmentation approaches~\cite{nickisch2010learning}.  %
Analytical measures as well as subjective measures can be derived from standardized user tests~\cite{gao2013mental}.
From interaction data recorded during the study, the reproducibility of segmentation results as well as the achievable accuracy with a given system per time can be estimated.
The complexity and novelty of the system can be expressed via the observed convergence to the ground truth over time spent by the participants segmenting multiple images each.
The user's satisfaction with the interactive approaches is expressed by the analysis of questionnaires, which the study participant fills out %
immediately %
after their tests are conducted and before any discussion or debriefing has started. %
The respondent is asked to fill in the questionnaire as spontaneously as possible.
Intuitive answers are desired as user feedback instead of well-thought-out responses for each item in the questionnaire~\cite{brooke1996sus}.

For the randomized A/B study, individuals are selected to approximate a representative sample of the intended users of the final system~\cite{siroker2013b}.
During the study, subjects are given multiple interactive segmentation tasks to fulfill each in a limit time frame.
The user segments all $m$ images provided with two different methods (A and B).
All subjects are given $2 \cdot m$ tasks in a randomized order to prevent a learning effect bias, which would allow for higher quality outcomes for the later tasks.
Video and audio data of the subjects are recorded.
Every user interaction recognized by the system and its time of occurrence are logged.

\section{Experiments}\label{sec:experiments}

\subsection{Data Set for the Segmentation Tasks}\label{sec:study_data_sets}

\begin{figure}%
	\resizebox{\columnwidth}{!}{%
		{\def\arraystretch{1.1}\tabcolsep=2pt
		\begin{tabular}{cccc}%
			\includegraphics[height=0.2357\textwidth]{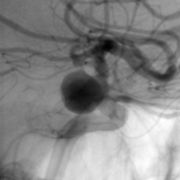} &
			\includegraphics[height=0.2357\textwidth]{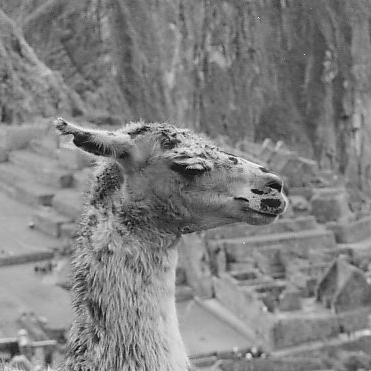} &
			\includegraphics[height=0.2357\textwidth]{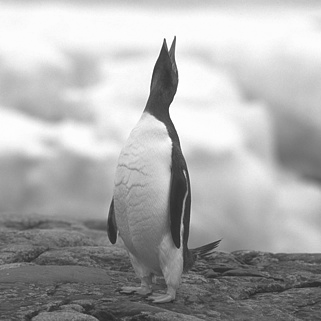} &
			\includegraphics[height=0.2357\textwidth]{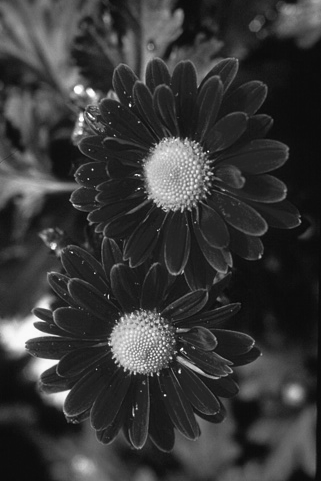} \\
			\includegraphics[height=0.2357\textwidth]{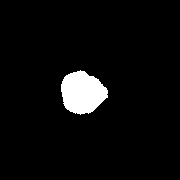} &
			\includegraphics[height=0.2357\textwidth]{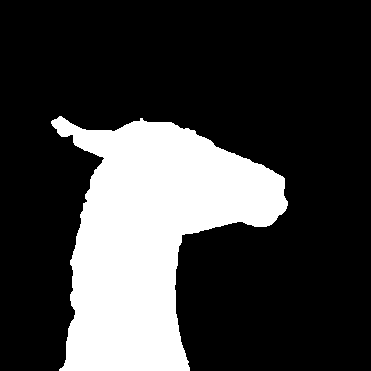} &
			\includegraphics[height=0.2357\textwidth]{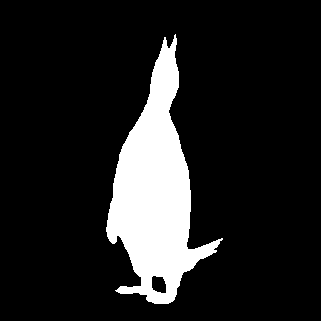} &
			\includegraphics[height=0.2357\textwidth]{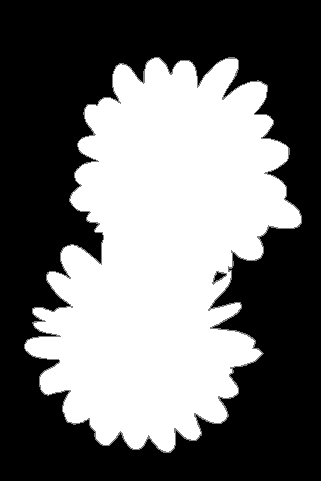} \\
			\Large{(a)} & \Large{(b)} & \Large{(c)} & \Large{(d)}
		\end{tabular}%
	}
	}%
	\caption{In the top row, image data utilized in the usability tests are depicted.
		In the bottom row, the ground truth segmentations of the images are illustrates.
		The image of a contrast enhanced aneurysm (a) and its ground truth \additioncaption{annotation by a medical expert} were composed for this study.
		Images (b\,--\,d) are selected from the \href{https://web.archive.org/web/20161203110733/http://research.microsoft.com/en-us/um/cambridge/projects/visionimagevideoediting/segmentation/grabcut.htm}{GrabCut image database} initially created for~\cite{rother2004grabcut}.%
	}
	\label{fig:study_data_sets}%
\end{figure}

In \textbf{Fig.}\,\ref{fig:study_data_sets} the data set used for the usability test is depicted.
For this evaluation, the \mbox{RGB} colored images are converted to grayscale in order to increase similarity to the segmentation process of medical images acquired from \mbox{CBCT}.
The conversion is performed in accordance with the \href{https://www.itu.int/rec/R-REC-BT.709/en}{\mbox{ITU--R BT.709-6}}  recommendation~\cite{recommendation1990basic} for the extraction of true luminance 
\change[label=c:c161,ref=c:c16]{%
\mbox{$\mathbf{I}\in\mathbb{R}^{w,h}$}}{%
\mbox{$\mathbf{Y}\in\mathbb{R}^{w,h}$}} 
defined by the International Commission on Illumination (\mbox{CIE}) from contemporary cathode ray tube (\mbox{CRT}) phosphors via \change[label=c:c134,ref=c:c13]{\textbf{Eq.}\,\ref{eq:rgbtograyscale}, 
where \mbox{$\mathbf{I}'_R\in\mathbb{R}^{w,h}$}, \mbox{$\mathbf{I}'_G\in\mathbb{R}^{w,h}$}, and \mbox{$\mathbf{I}'_B\in\mathbb{R}^{w,h}$}}{where $\mathbf{R}$, $\mathbf{G}$, and $\mathbf{B}$} are the linear red, green, and blue color channels \addition[label=c:a165,ref=c:c16]{of \mbox{$\mathbf{I}'\in\mathbb{R}^{w,h,3}$}} respectively.
\begin{equation}
	\mathbf{I} = 0.2126 \cdot \mathbf{I}'_R + 0.7152 \cdot \mathbf{I}'_G + 0.0722 \cdot \mathbf{I}'_B 
	\label{eq:rgbtograyscale}
\end{equation}
\addition[label=c:a161,ref=c:c16]{Image \textbf{Fig.}\,\ref{fig:study_data_sets}}(b) is initially presented to the \change[label=c:c162,ref=c:c16]{study participants}{users} in order to familiarize themselves with the upcoming segmentation process.
The segmentation tasks associated with images \addition[label=c:a162,ref=c:c16]{\textbf{Fig.}\,\ref{fig:study_data_sets}}(a,\,c,\,d) are then displayed sequentially to the subjects in randomized order.
The images are chosen to fulfill two goals of the study.
(1) Ambiguity of the ground truth has to be minimized in order to suppress noise in the quantitative data.
Each test person should have the same understanding and consent about the correct outline of the object to segment.
Therefore, clinical images can only be utilized with groups of specialized domain experts.
(2) The degree of complexity should vary between the images displayed to the users.
Image (b), depicted in \textbf{Fig.}\,\ref{fig:study_data_sets}, of moderate complexity with regards to its disagreement coefficient~\cite{hanneke2007bound}, is displayed first to learn the process of segmentation with the given prototype.
\addition[label=c:a163,ref=c:c16]{ %
	Users are asked for an initial testing of a prototype's features utilizing this image without any time pressure.
	The subsequent interactions during the segmentations of the remaining three images are recorded for each prototype and participant.
}
The complexity increases from (a) to (d), \addition[label=c:a164,ref=c:c16]{according to the \mbox{GTs'} Minkowski-Bouligand dimensions~\cite{mandelbrot1967long}}. %
The varying complexity enables a more objective and extended differentiation of subjects' performances with given prototypes. %

\subsection{Usability Test Setup}\label{sec:usability_test_setup}

\begin{figure}
	\includegraphics[width=\columnwidth]{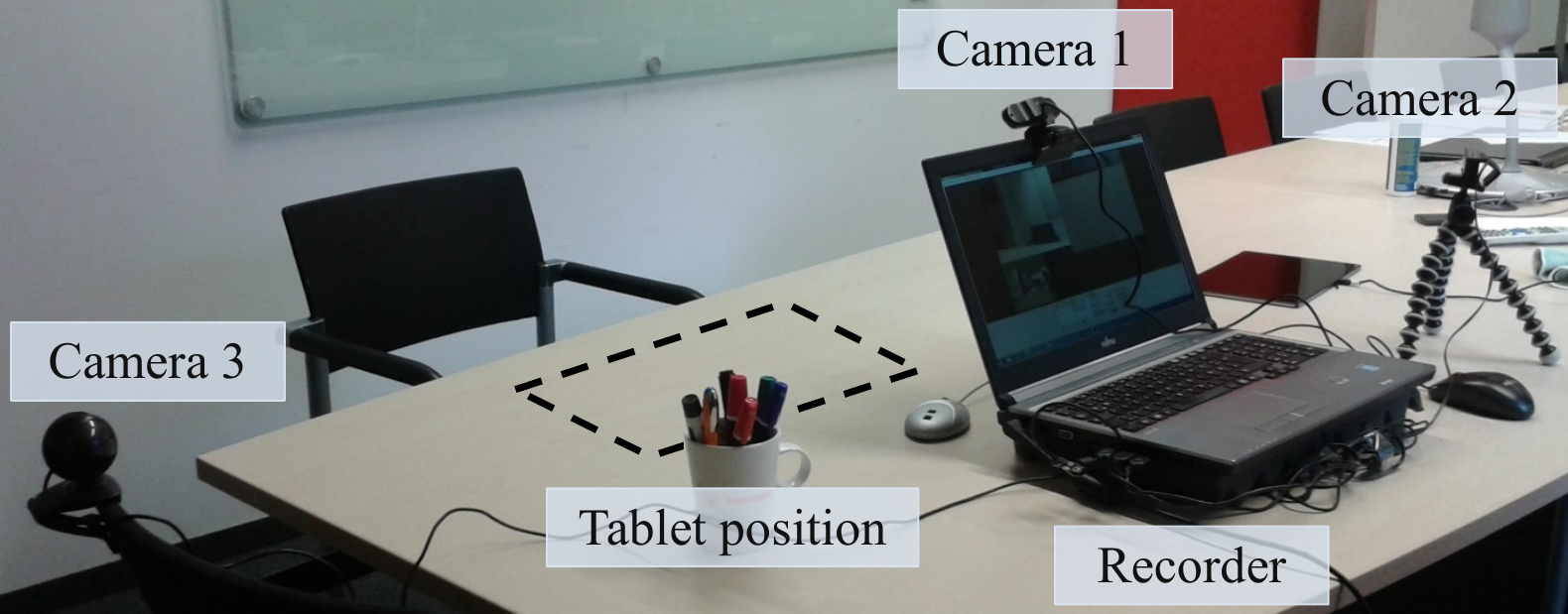}
	\caption{User testing setup for the usability evaluation of the prototypes.
		In this environment, a user performs an interactive segmentation on a mobile tablet computer while sitting.
		\mbox{RGB} cameras record the hand motions on the input device and facial expressions of the participant.
		\additioncaption{In addition, each recognized input is recorded on the tablet device (the interaction log).}
	}
	\label{fig:study_setup}
\end{figure}

Two separate user studies are conducted to test all prototypes described in \textbf{Sec.}\,\ref{sec:sgmentation_prototypes}, 
in order to keep the time for each test short (less than \change[label=c:c261,ref=c:c26]{$10$ minutes per prototype}{$20$ minutes}), thus retaining the focus of the participants, while minimizing the occurrence of learning effect artifacts in the acquired data.
\addition[label=c:a261,ref=c:c26]{Note that the participants use this time not only to finish the segmentation tasks, but also to familiarize themselves with the novel interaction system, as well as to form opinions about the system while testing their provided interaction features.}
(1) The first user test is a randomized A/B test of the \mbox{semi-manual} prototype (\textbf{Sec.}\,\ref{sec:semi-manual_prototype}) and the guided prototype (\textbf{Sec.}\,\ref{sec:guided_prototype}).
Ten individuals are selected as test subjects due to their advanced domain knowledge in the fields of medical image processing and mobile input devices.
The subjects are given the task to segment \mbox{$m=3$} different images with varying complexity, which are described in \textbf{Sec.}\,\ref{sec:study_data_sets}, in random order.
A fourth input image of medium complexity is provided for the users to familiarize themselves with the \mbox{ISS} before the tests.
As an interaction device, a mobile tablet computer is utilized, since the final segmentation method is intended for usage via such a medium.
The small $10.1$ inch \mbox{($13.60\,\text{cm}\cdot21.75\,\text{cm}$)} \mbox{WUXGA} display and fingers utilized as a multi-touch pointing device further exacerbate the challenge to fabricate an exact segmentation for the participants~\cite{norman2010gestural}.
The user study environment is depicted in \textbf{Fig.}\,\ref{fig:study_setup}.
Audio and video recordings are evaluated via a qualitative content analysis, described in \textbf{Sec.}\,\ref{sec:qualitative_measures}, 
in order to detect possible improvements for the tested prototypes and their interfaces.
After segmentation, each participant fills out the \mbox{SUS} (\textbf{Sec.}\,\ref{sec:questionnaires_sus}) and \mbox{AttrakDiff-2} (\textbf{Sec.}\,\ref{sec:questionnaires_attrakdiff}) questionnaires.

(2) The second user test is conducted for %
the joint segmentation prototype (\textbf{Sec.}\,\ref{sec:joint_prototype}). %
The data set and test setup are the same as in the first user study and all test persons of study (1) also participated in study (2). One additional subject participated only in study (2).
Two months passed between the conduction of the two studies, in which the former participants were not exposed to any of the prototypes.
Therefore, the learning effect bias for the second test is neglectable.

\subsection{Prediction of Questionnaire Results}\label{sec:prediction_of_questionnaire_results} %

The questionnaires' \mbox{PQ}, \mbox{HQ}, \mbox{HQ-I}, \mbox{HQ-S}, \mbox{ATT}, and \mbox{SUS} results are predicted, based on features extracted from the interaction log data.
For the prediction, a regression analysis is performed.
Stochastic Gradient Boosting Regression Forests (\mbox{GBRF}) are an additive model for regression analysis~\cite{friedman2001greedy,friedman2002stochastic,hastie2009boosting}.
In several stages, shallow regression trees are generated.
Such a tree is a weak base learner each resulting in a prediction error \mbox{$\varepsilon = b + v$}, with high bias $b$ and low variance $v$. 
These regression trees are utilized 
to minimize an arbitrarily differentiable loss function each on the negative gradient of the previous stage's outcome, thus reducing the overall bias via boosting~\cite{breiman1999using}.
The Huber loss function~\cite{huber1964robust} is utilized for this evaluation due to its increased robustness to outliers in the data with respect to the squared error loss. %

The collected data set of user logs is split randomly in a ratio of \mbox{$4:1$} for training and testing. %
An exhaustive grid search over $20,480$ parameter combinations is performed for each of the six \mbox{GBRF} estimators (one for each questionnaire result) with scorings based on an eight-fold cross-validation on the training set.
\subsubsection{Feature Definition}\label{sec:feature_definition}
The collected data contains $31$ samples with $216$ possible features each.
The $31$ questionnaire results (\mbox{PQ}, \mbox{HQ}, \mbox{HQ-S}, \mbox{HQ-I}, \mbox{ATT}, \mbox{SUS}), are predicted based on 
features extracted from the interaction log data of the four images segmented with the system\deletion[label=c:d271,ref=c:c27]{(see \textbf{Fig.}\,\ref{fig:study_data_sets})}.
Four features are the relative median seed positions per user and their standard deviation in two dimensions.
$22$ additional features, like the number of undo operations (\mbox{\emph{\#Undos}}) and number of interactions (\mbox{\emph{\#Interactions}}), the overall computation time (\mbox{\emph{$\Sigma$Computation\_time}}), overall interaction time (\mbox{\emph{$\Sigma$Interaction\_time}}), elapsed real time (\mbox{\emph{$\Sigma$Wall\_time}}), \mbox{\emph{Final\_Rand\_index}}, and \mbox{\emph{Final\_Dice\_score}} are reduced to one scalar value each by the mean and median, over the four segmentations per prototype and user, 
to obtain $48$ base features.
Since these features each only correlate weakly with the questionnaire results, composite features are added in order to assist the model's learning process for feature relations.
Added features are composed of one base feature value divided by (the mean or median of) %
computation time, interaction time, or elapsed real time. %
The relations between those time values themselves are also added.
In total, $216$ features directly related to the interaction log data are used.
In addition, a principal component analysis (\mbox{PCA}) is performed in order to add $10$\,\% ($22$) features with maximized variance to the directly assessed ones to further assist the feature selection step via \mbox{GBRFs}.

\subsubsection{Feature Selection for \mbox{SUS} Prediction}\label{sec:sus_prediction}

For the approximation of \mbox{SUS} results, a feature selection step is added to decrease the prediction error by an additional three percent points:
here, after the described initial grid search, $1$\,\% (205) of the \mbox{GBRF} estimators, with the lowest mean deviance from the ground truth, %
are selected to approximate the most important features.
From those estimators, the most important features for the \mbox{GBRFs} are extracted via a \emph{$1/\text{loss}$}-weighted %
feature importance voting. %
This feature importance voting by $205$ estimators ensures a more robust selection than deciding the feature ranking from only a single trained \mbox{GBRF}.
After the voting, a second grid search over the same $20,480$ parameter combinations, but with a reduction from $238$ to only $25$ of the most important features is performed.

\section{Results}\label{sec:results} 

\subsection{Overall Usability}\label{sec:results_overall_usability}  %

\begin{figure*}%
	\resizebox{\textwidth}{!}{%
		{\def\arraystretch{1.5}\tabcolsep=4pt
			\begin{tabular}{cccc!{\color{gray}\vrule}c}%
				\rotatebox{90}{$\quad$ \mbox{SUS} score per Subject} &
				\includegraphics[height=0.2357\textwidth]{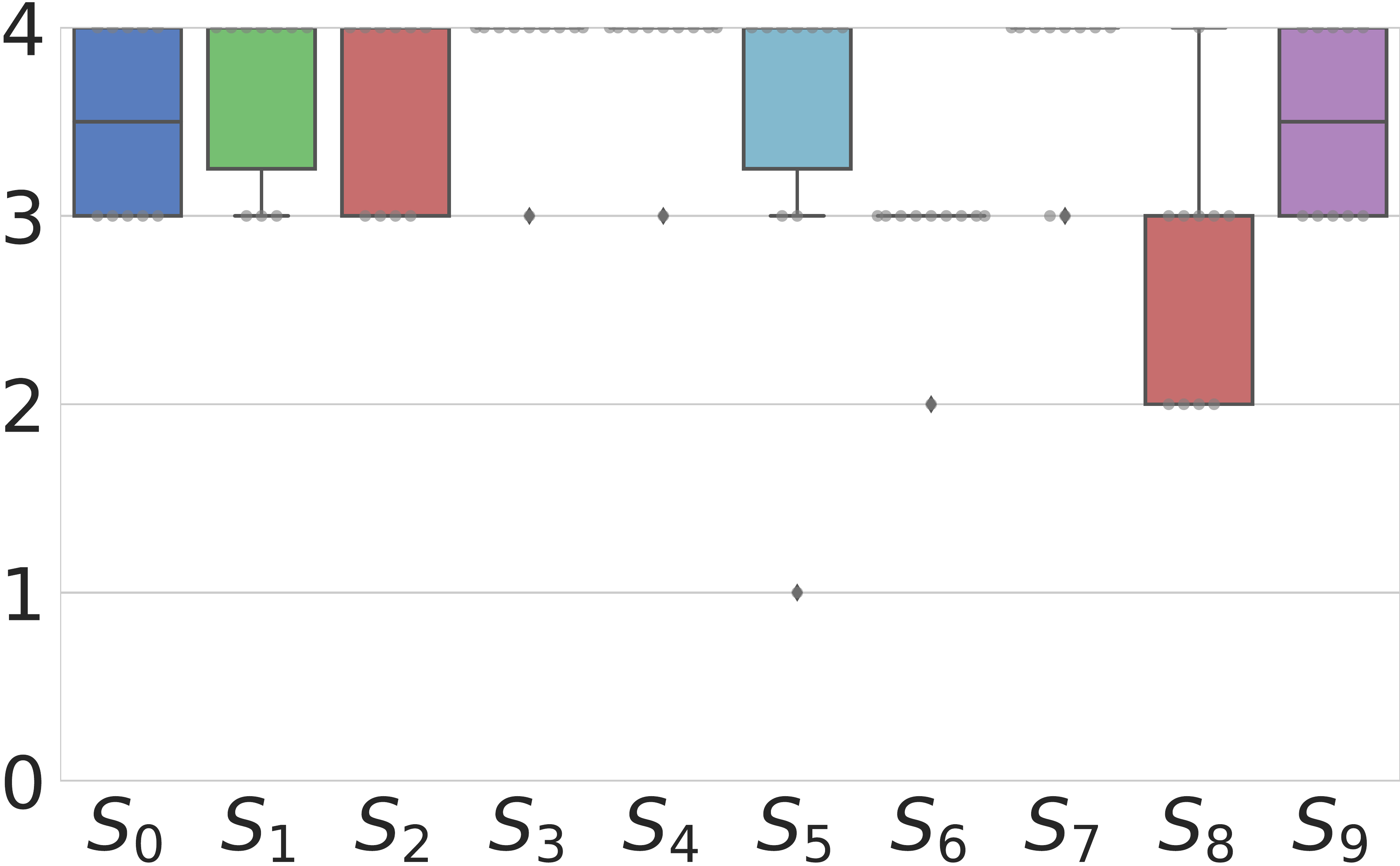} &
				\includegraphics[height=0.2357\textwidth]{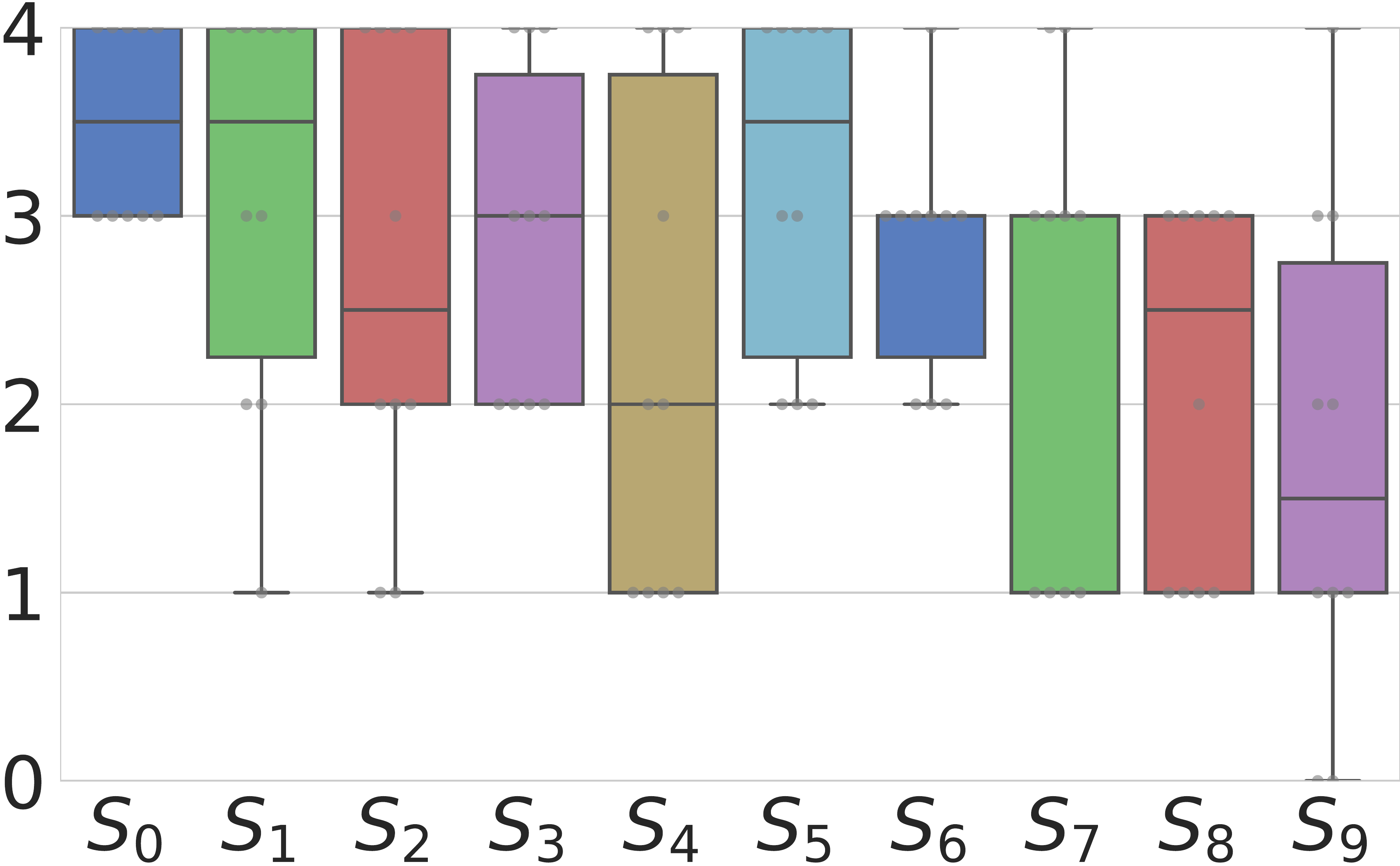} &
				\includegraphics[height=0.2357\textwidth]{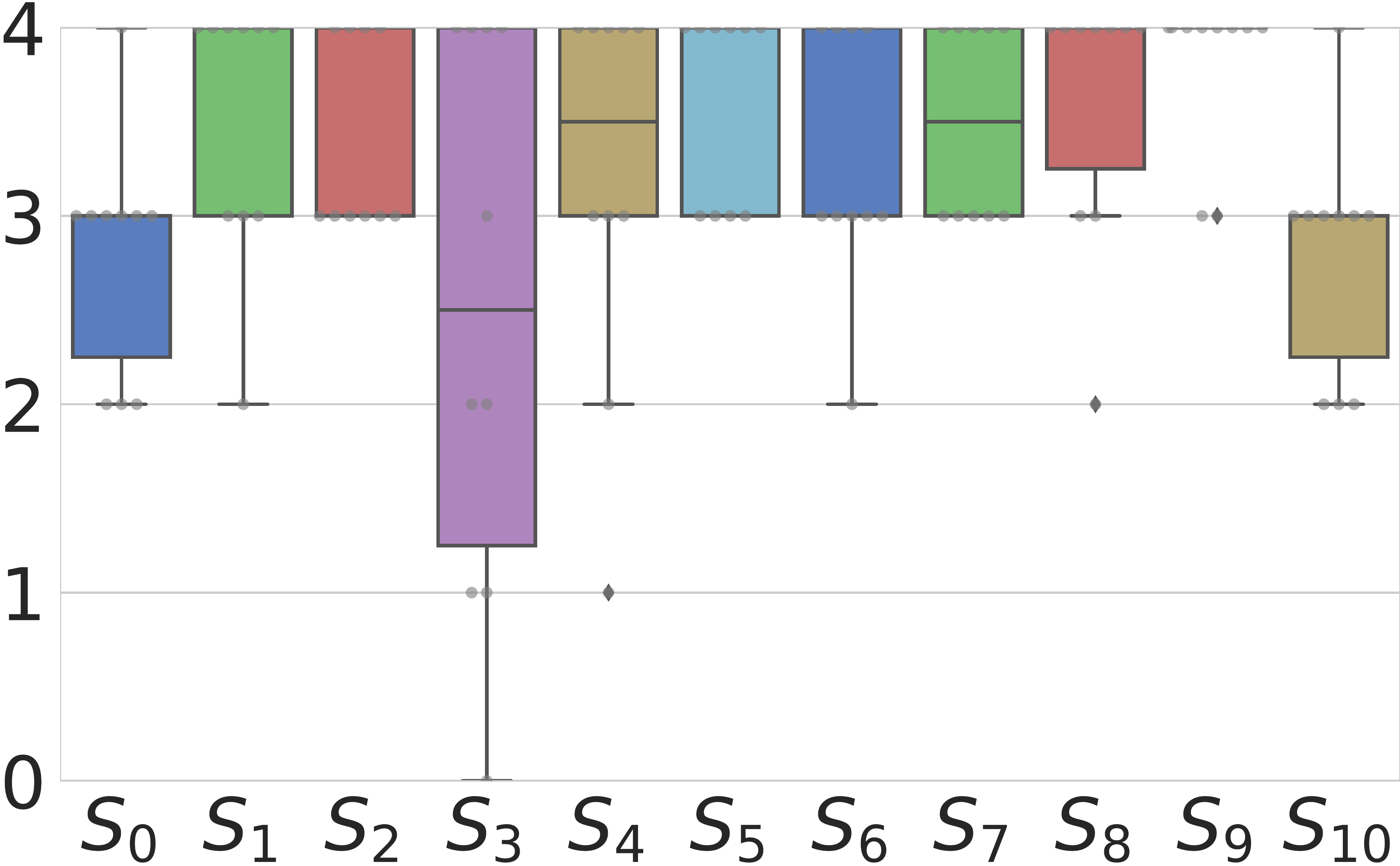} & \\
				\rotatebox{90}{$\quad$ \mbox{SUS} score per Statement} &
				\includegraphics[height=0.2357\textwidth]{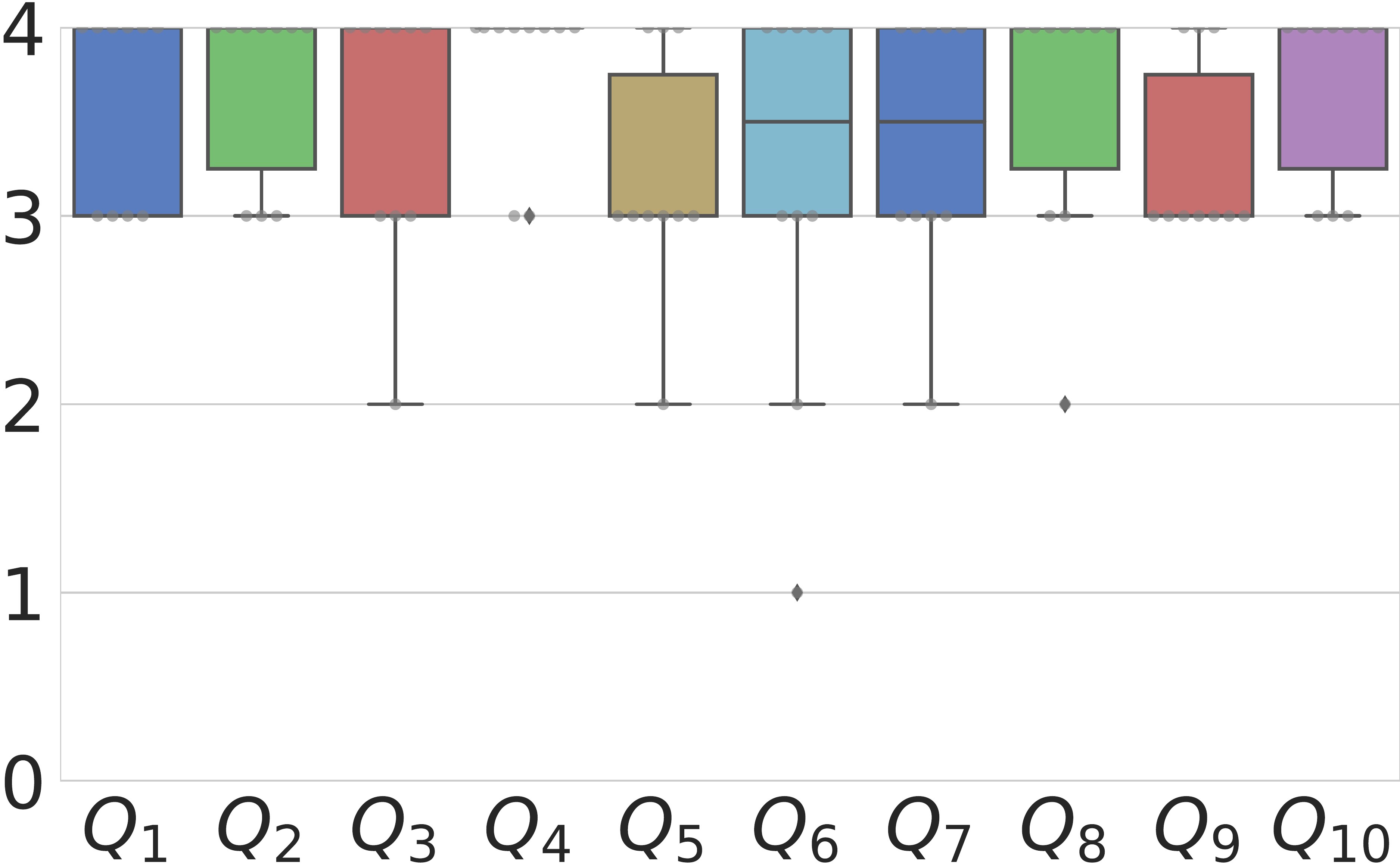} &
				\includegraphics[height=0.2357\textwidth]{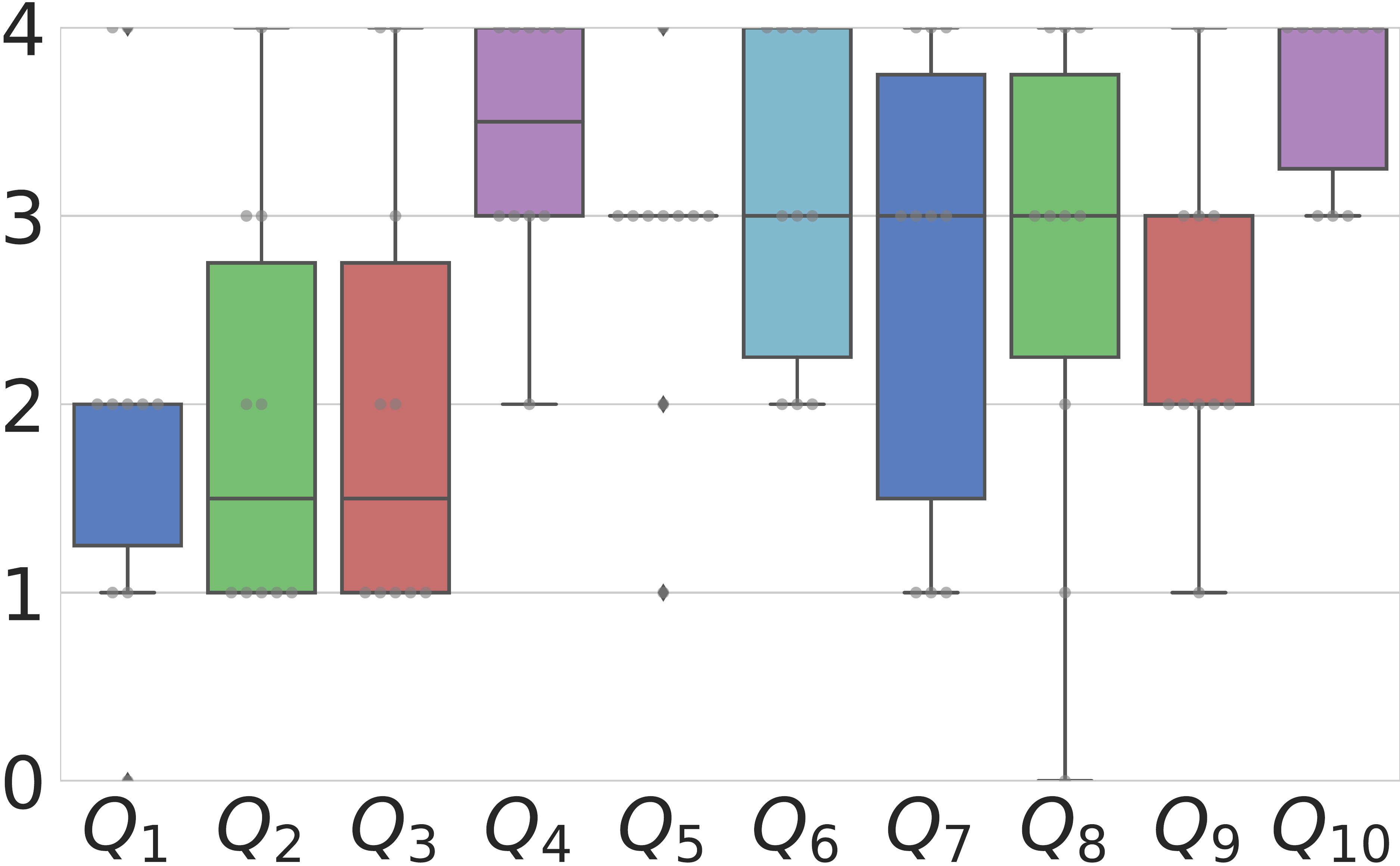} &
				\includegraphics[height=0.2357\textwidth]{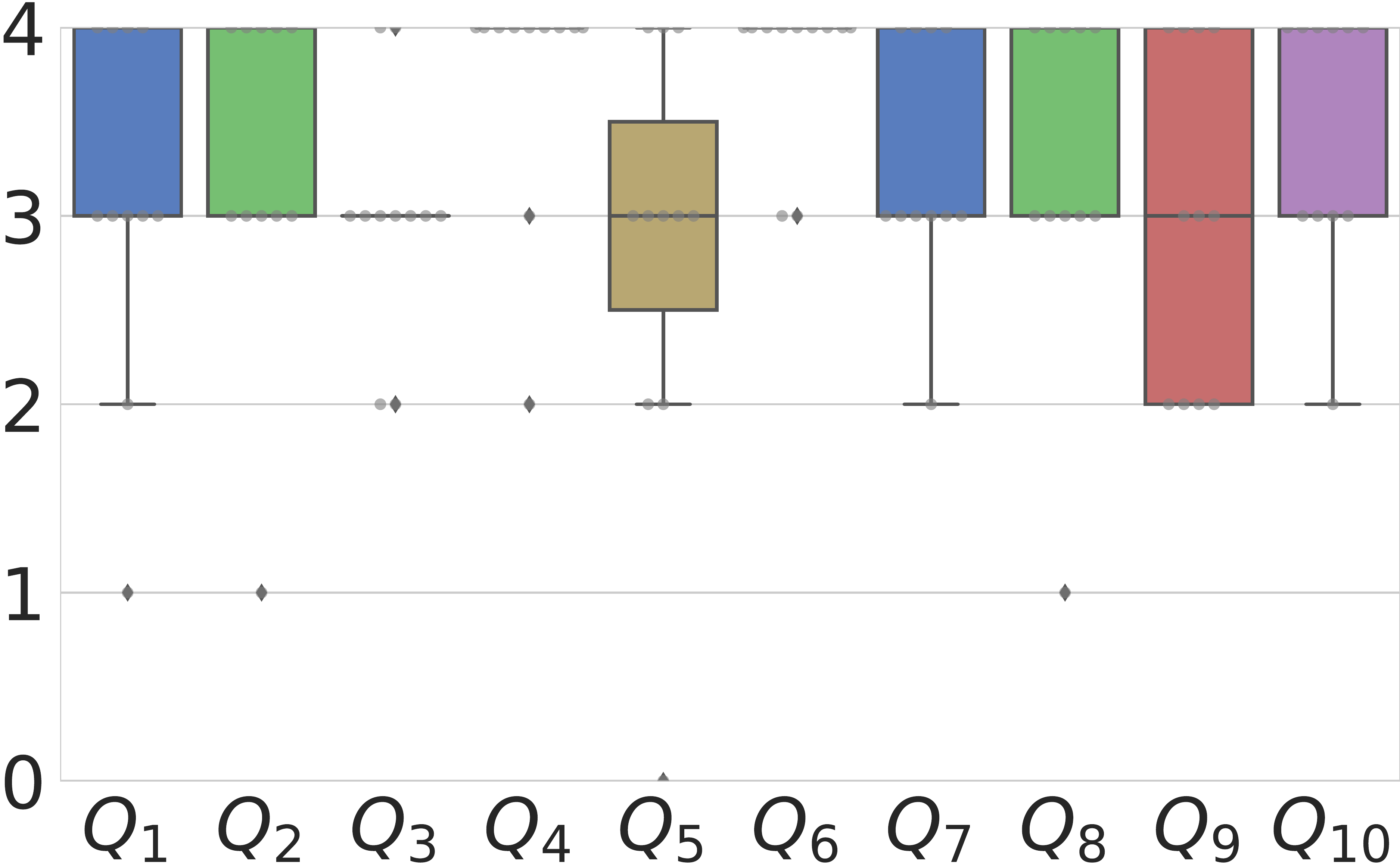} & 
				\includegraphics[height=0.2357\textwidth]{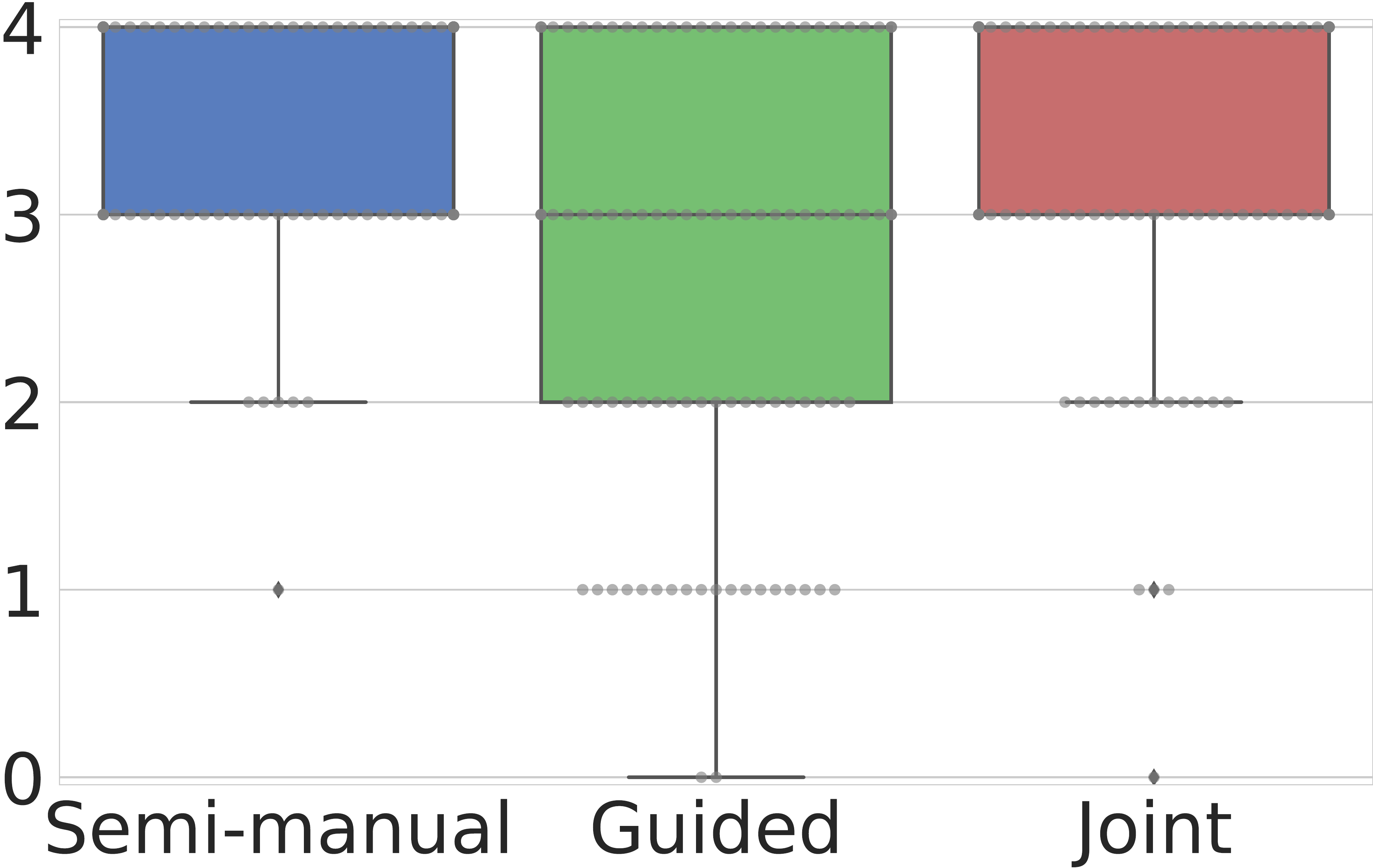} \\
				& SUS \mbox{semi-manual} prototype & \mbox{SUS} guided prototype & \mbox{SUS} joint prototype & \mbox{SUS} overall\\
			\end{tabular}%
		}
	}%
	\caption{Results of the \mbox{SUS} questionnaires per prototype.
		Values are normalized \additioncaption{in accordance with \textbf{Eq.}\,\ref{eq:sus_score}}, such that $4$ is considered the best possible result for each question.
		The \mbox{Semi-manual} prototype's \mbox{SUS} mean is $88$, guided prototype's mean is $67$, and joint prototype's mean \mbox{SUS} score is $82$.
	}
	\label{fig:result_sus}%
\end{figure*}%

The result of the \mbox{SUS} score is depicted in \textbf{Fig.}\,\ref{fig:result_sus}.
According to the mapping (\textbf{Fig.}\,\ref{fig:sus_adjective}) introduced in \textbf{Sec.}\,\ref{sec:questionnaires_sus}, the adjective rating of the \mbox{semi-manual} and joint prototypes are \emph{excellent} ($88$ respective $82$), the adjective associated with the guided prototype is \emph{good} ($67$).

\begin{figure}
	\centering %
	\resizebox{0.75\columnwidth}{!}{%
		{\Large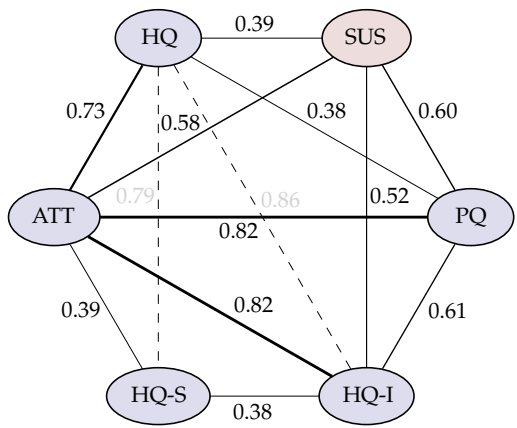}%
	}%
	\caption{Pearson correlation coefficients for the \mbox{AttrakDiff-2} (blue) and \mbox{SUS} (red) questionnaire results, 
		based on the acquired questionnaire data.
		The line thickness is proportionate to correlation strength \additioncaption{of the different aspects of quality measured}.}%
	\label{fig:result_questionnaire_results_correlation}%
\end{figure}

A graph representation of the similarity of individual usability aspects, based on the acquired questionnaire data, is depicted in \textbf{Fig.}\,\ref{fig:result_questionnaire_results_correlation}.
Based on the Pearson correlation coefficients utilized as a metric for similarity, the \mbox{SUS} score has the most similarity to the pragmatic (\mbox{PQ}) and attractiveness (\mbox{ATT}) usability aspects provided by the \mbox{AttrakDiff-2} questionnaire.

\subsection{Pragmatic Quality}\label{sec:results_pragmatic}
\begin{figure*}%
	\resizebox{\textwidth}{!}{%
		{\def\arraystretch{1.45}\tabcolsep=4pt
		\begin{tabular}{cccc!{\color{gray}\vrule}c}%
			\rotatebox{90}{$\quad$ \mbox{AttrakDiff-2} per Subject} &
			\includegraphics[height=0.2357\textwidth]{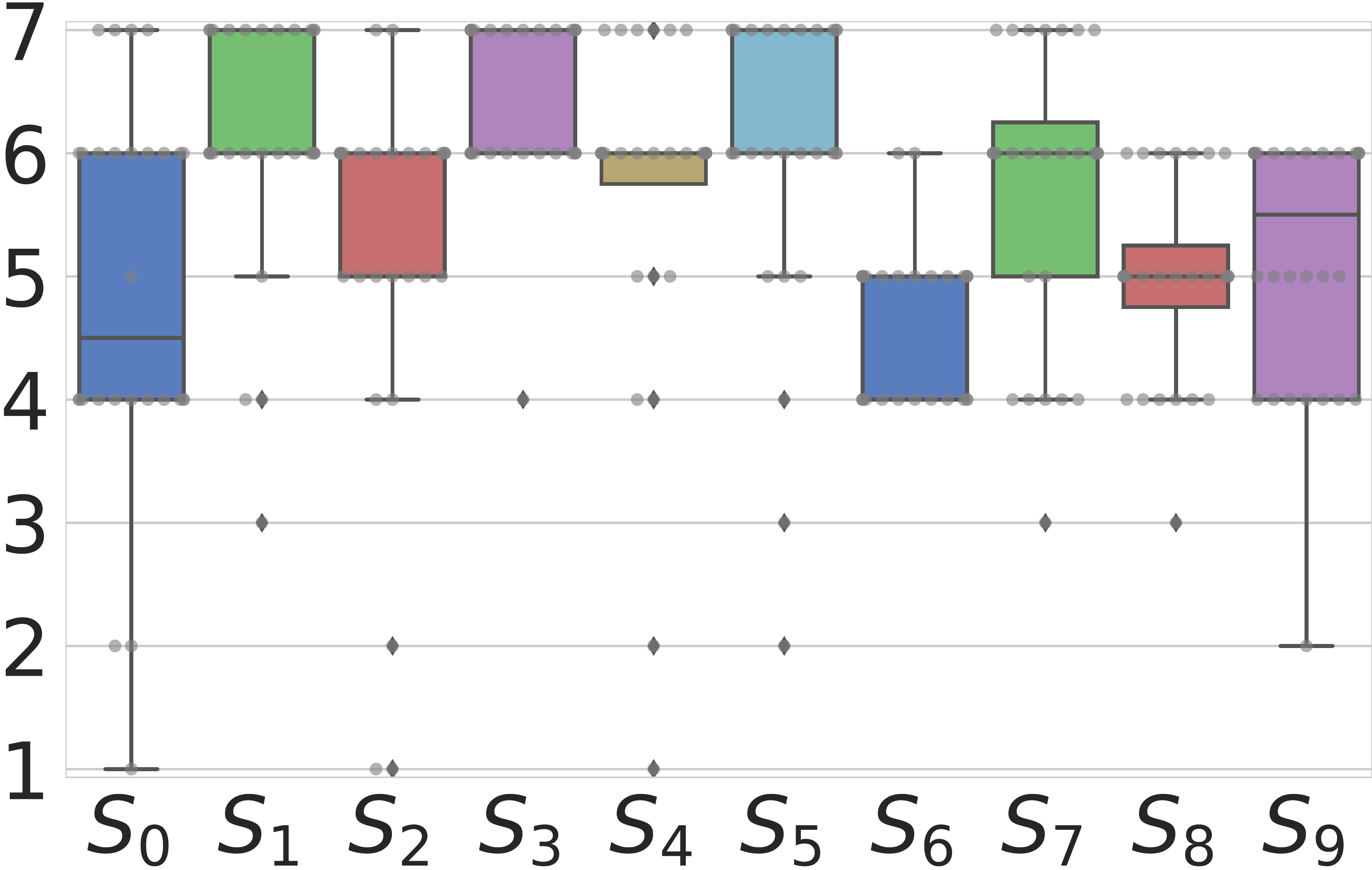} &
			\includegraphics[height=0.2357\textwidth]{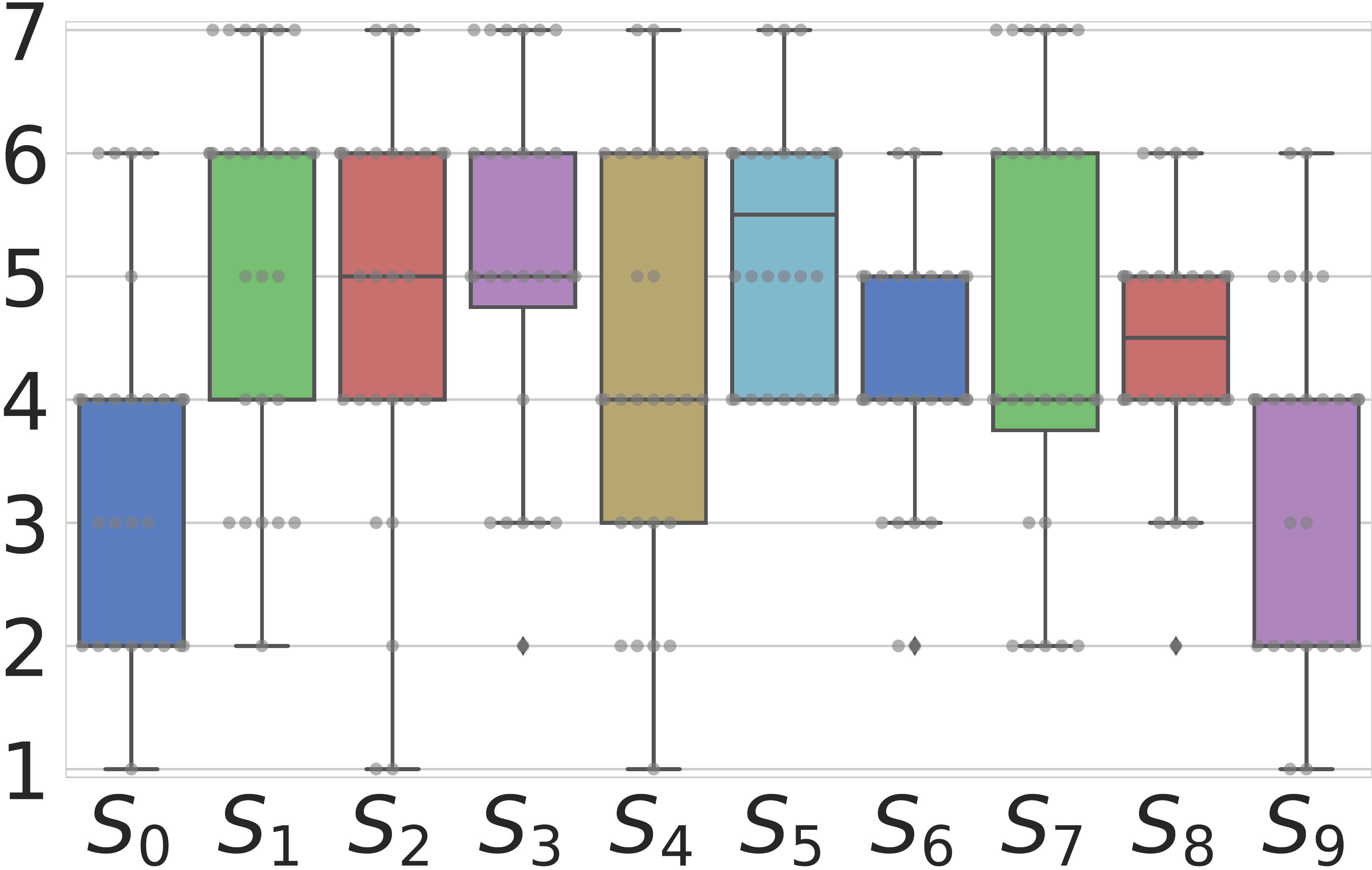} &
			\includegraphics[height=0.2357\textwidth]{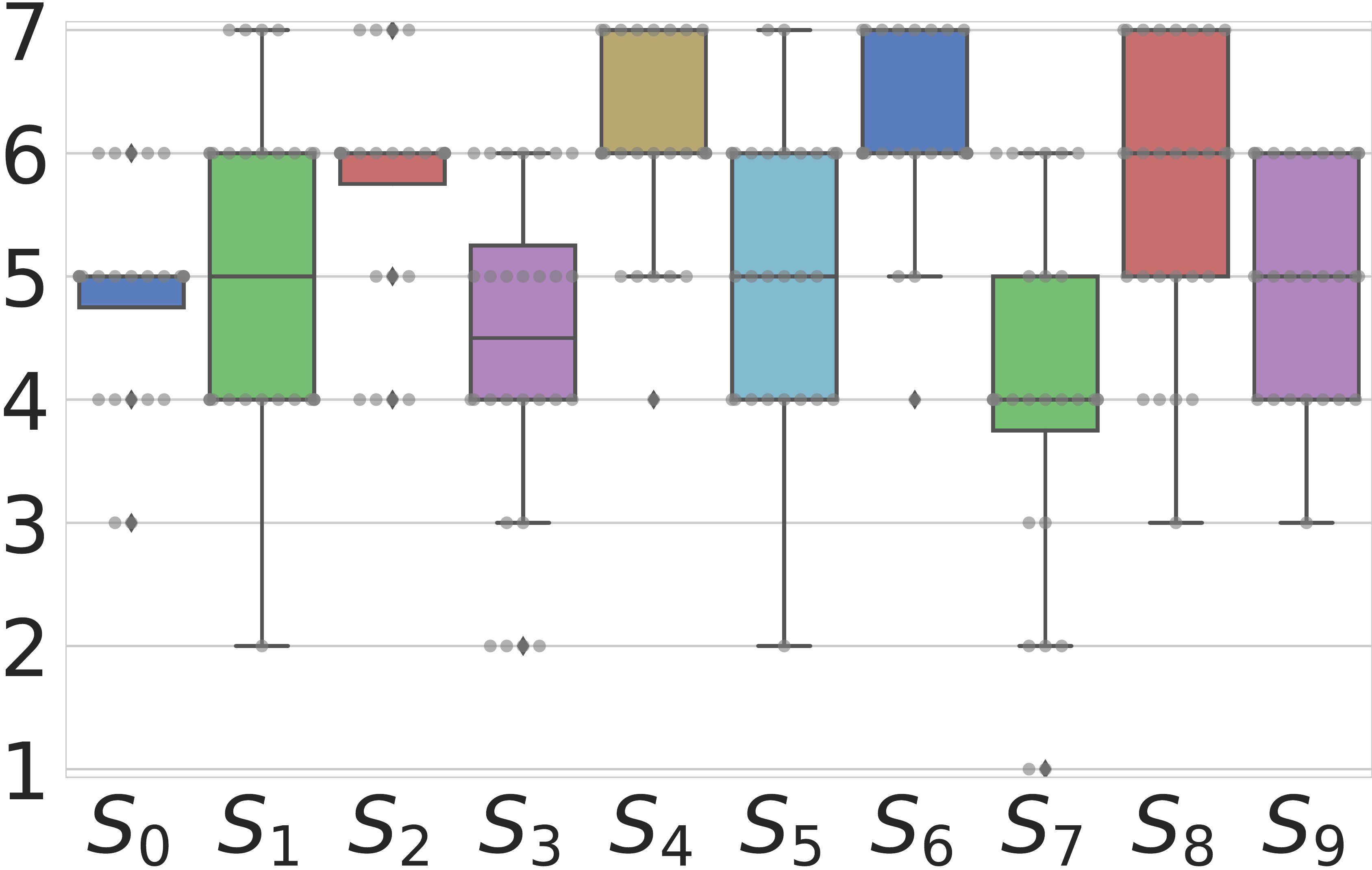} & 
			\includegraphics[height=0.2357\textwidth]{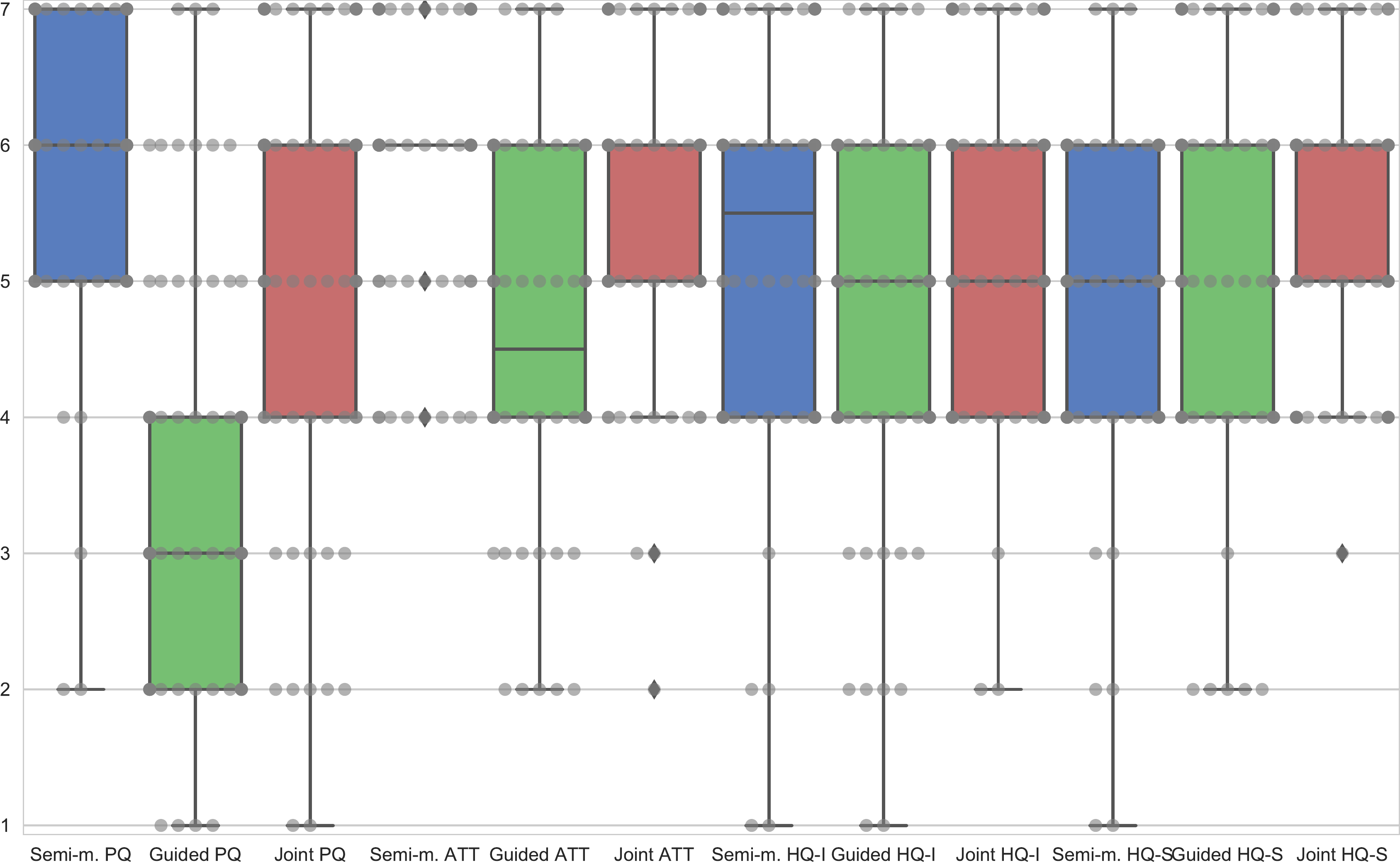} \\
			\rotatebox{90}{$\quad$ \mbox{AttrakDiff-2} per Statement} &
			\includegraphics[height=0.2357\textwidth]{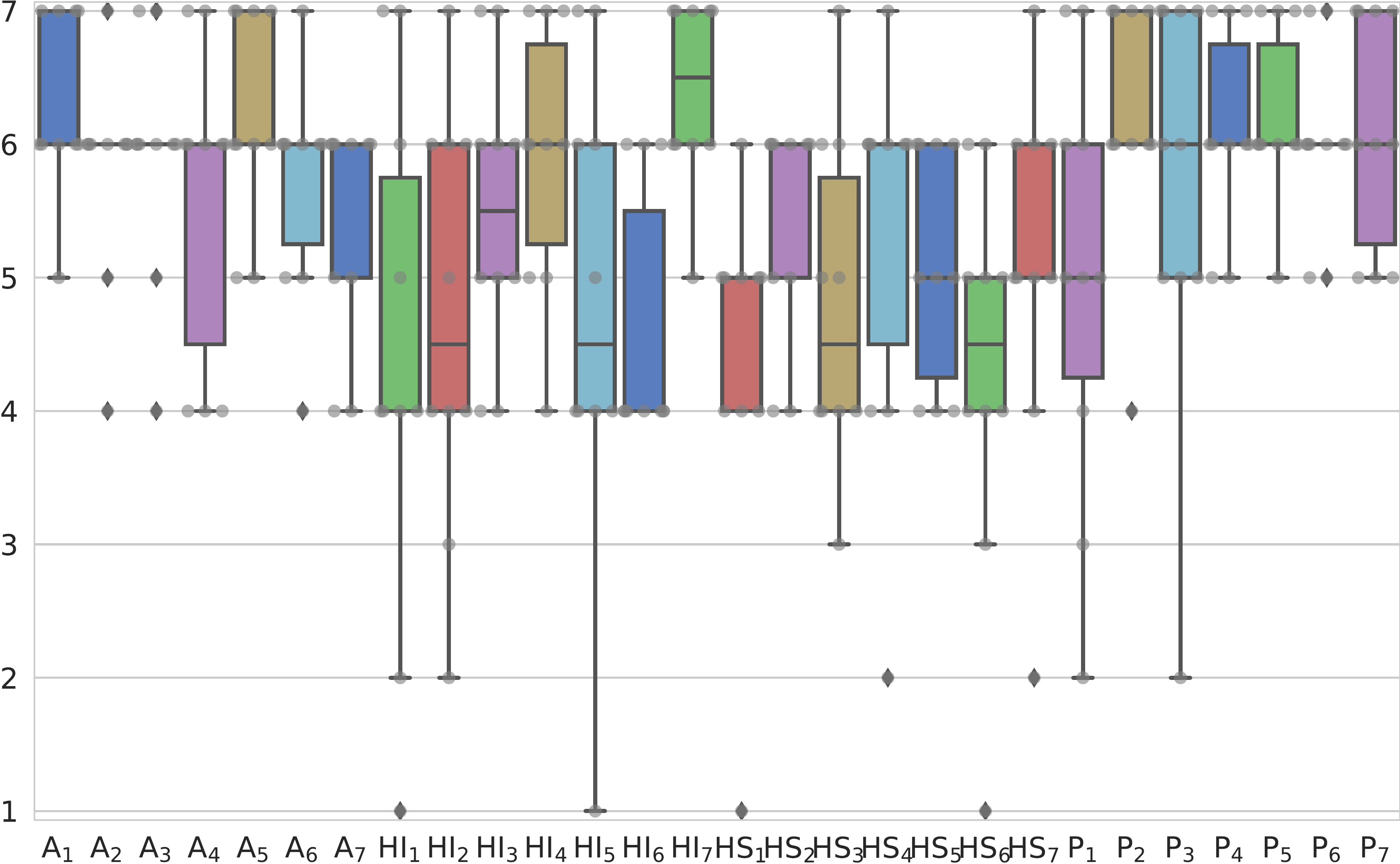} &
			\includegraphics[height=0.2357\textwidth]{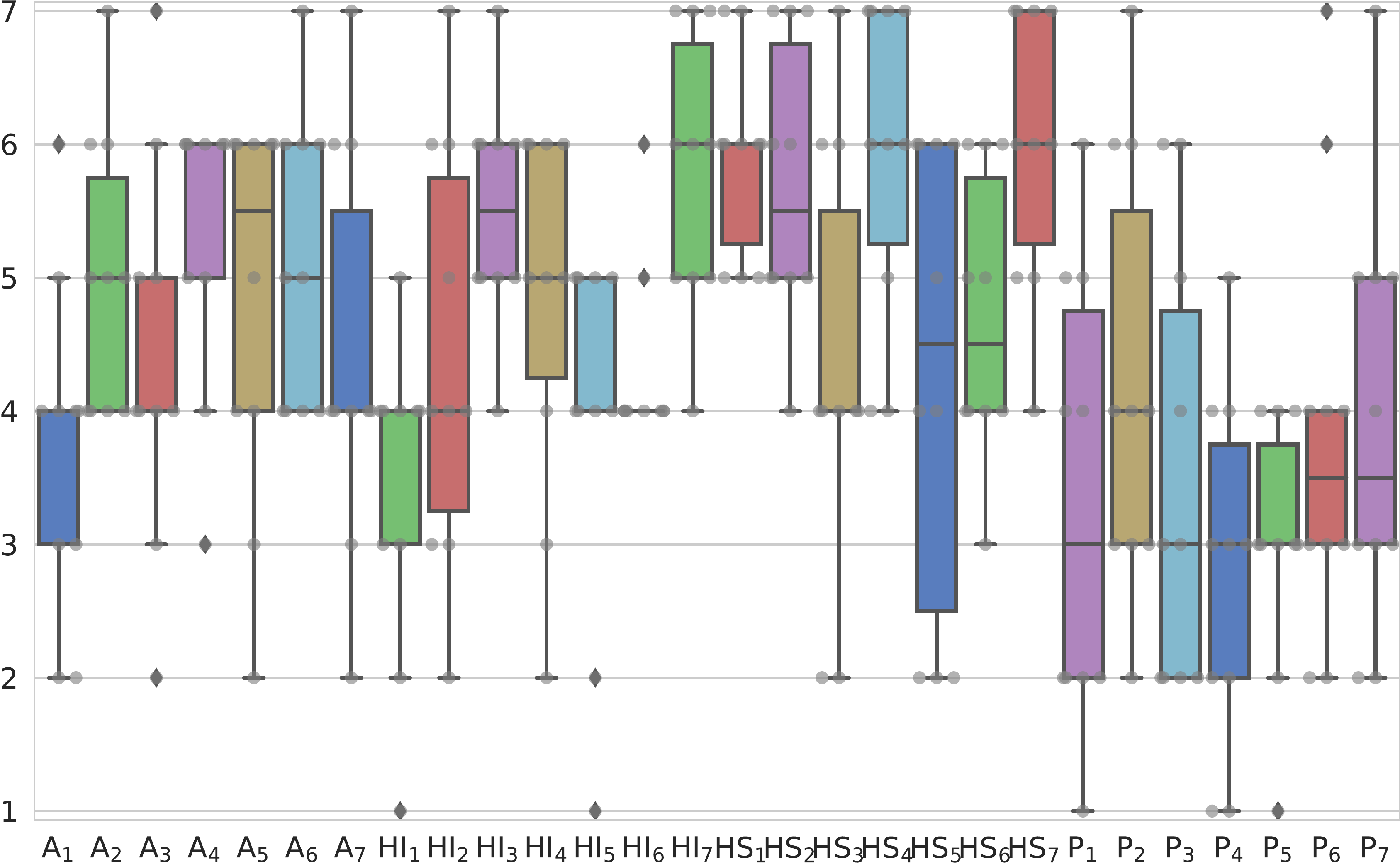} &
			\includegraphics[height=0.2357\textwidth]{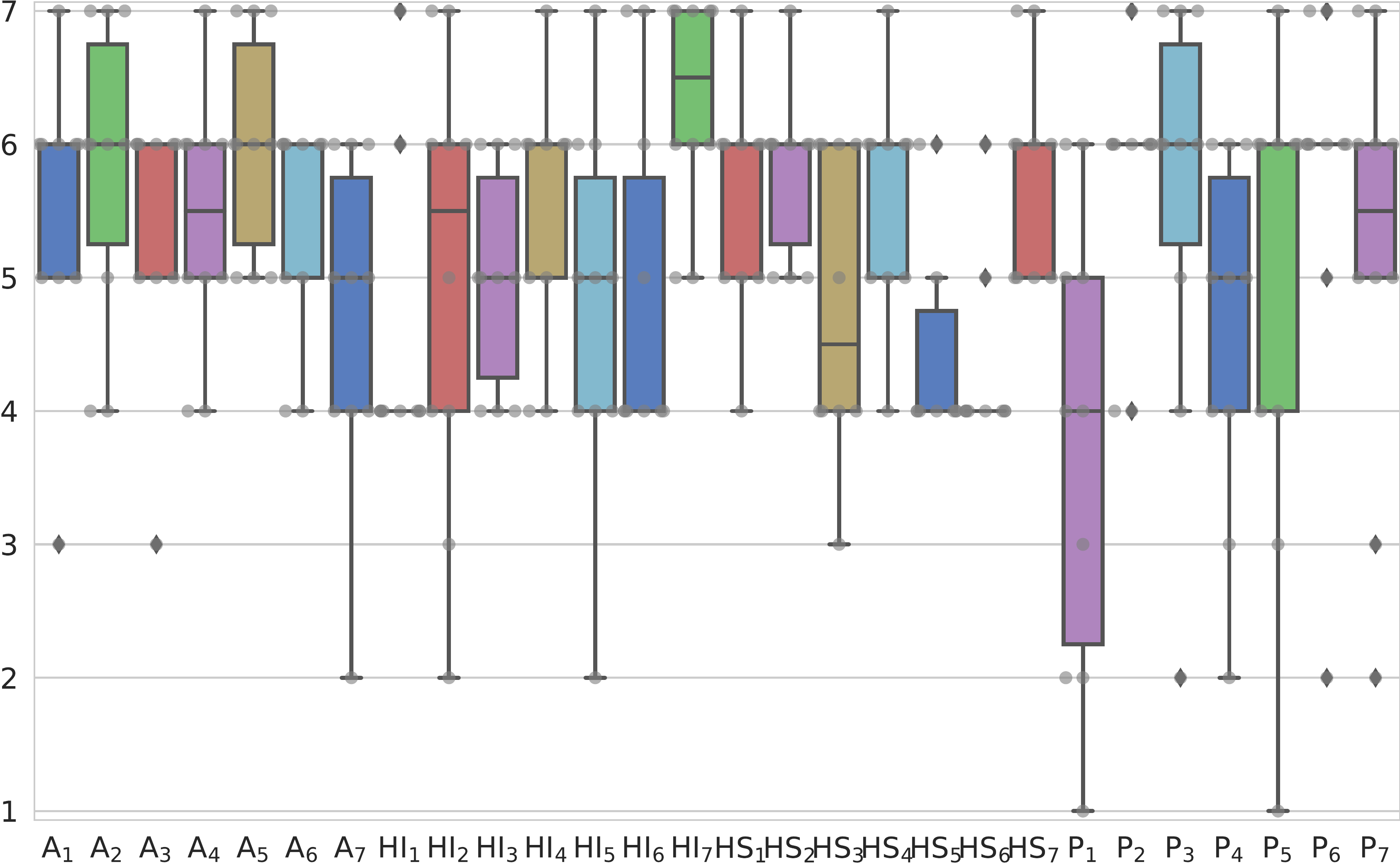} &
			\includegraphics[height=0.2357\textwidth]{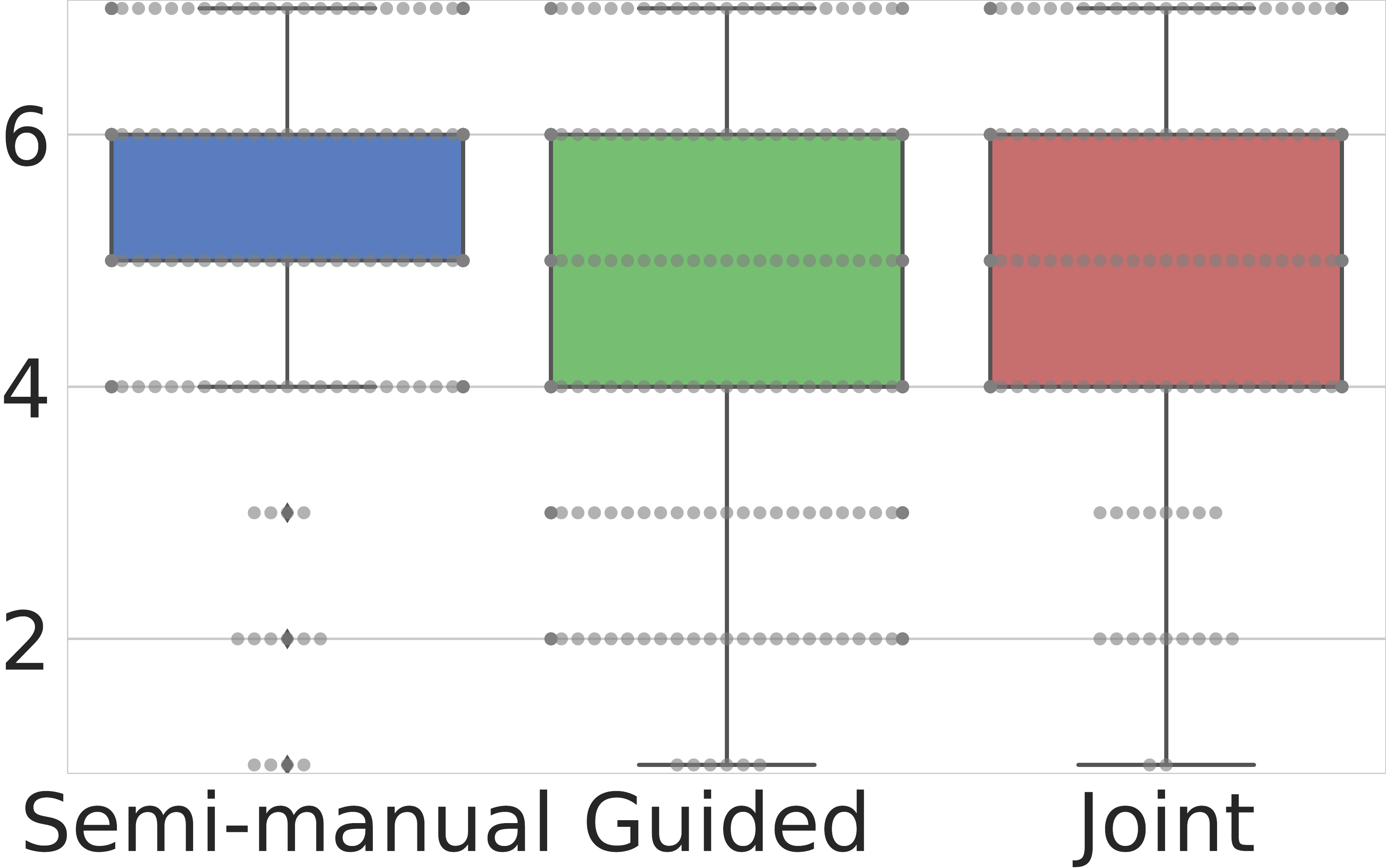} \\
			& \mbox{AttrakDiff-2} \mbox{semi-manual} prototype & \mbox{AttrakDiff-2} guided prototype & \mbox{AttrakDiff-2} joint prototype & \mbox{AttrakDiff-2} overall
		\end{tabular}%
	}
	}%
	\caption{Results of the \mbox{AttrakDiff-2} questionnaires per prototype.
		A value of $7$ is considered the best possible result.
		The \mbox{Semi-manual} prototype's \mbox{AttrakDiff-2} mean is $5.46$, guided prototype's mean is $4.50$, and joint prototype's mean \mbox{AttrakDiff-2} score is $5.22$.
	}
	\label{fig:result_attrakdiff}%
\end{figure*}

The \mbox{PQ} results of the \mbox{AttrakDiff-2} questionnaire are illustrated in \textbf{Fig.}\,\ref{fig:result_attrakdiff}.
The \mbox{PQ} scores for \mbox{semi-manual}, guided, and joint prototypes are $88$\,\%, $50$\,\%, and $74$\,\% of the maximum score, respectively.
Since each of the $95$\,\% confidence intervals are non-overlapping, %
the prototypes' ranking regarding \mbox{PQ} are significant.

\begin{figure*}%
	\resizebox{\textwidth}{!}{%
		{\def\arraystretch{1.0}\tabcolsep=4pt
		\begin{tabular}{cccc}%
			\rotatebox{90}{\Large{$\ $S{\o}rensen-Dice Coefficient}} &
			\includegraphics[height=0.32\textwidth]{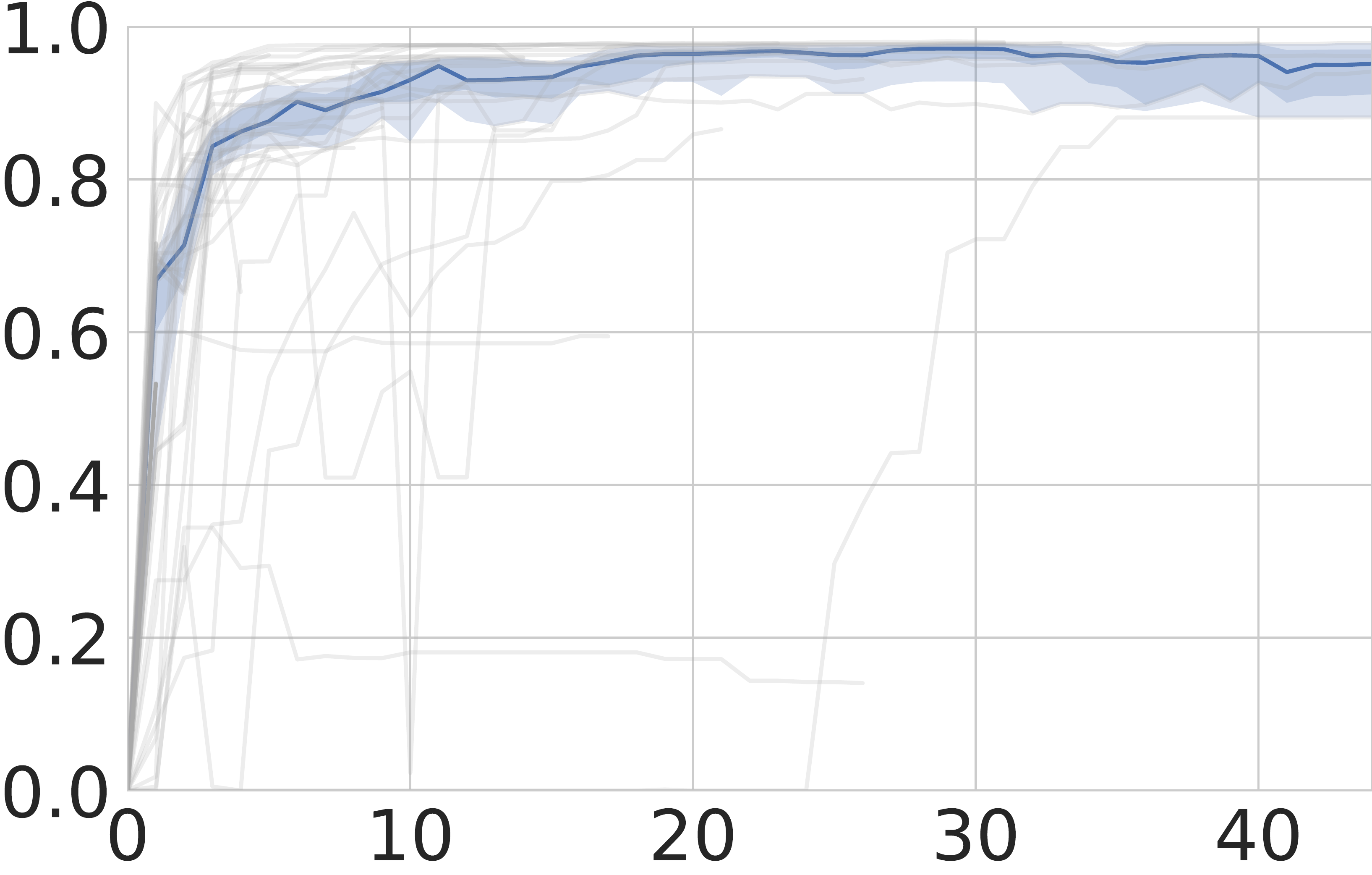} &
			\includegraphics[height=0.32\textwidth]{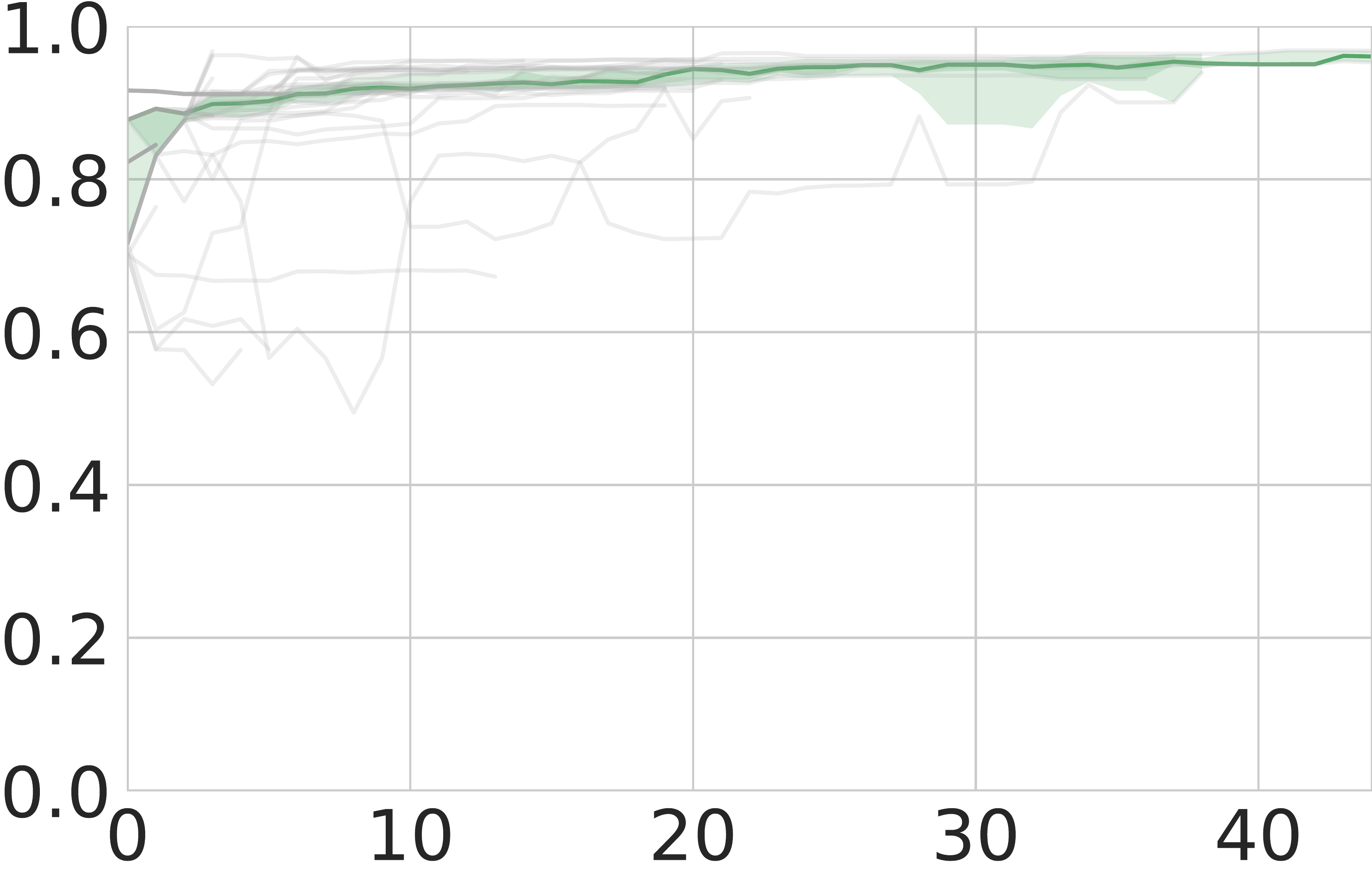} &
			\includegraphics[height=0.32\textwidth]{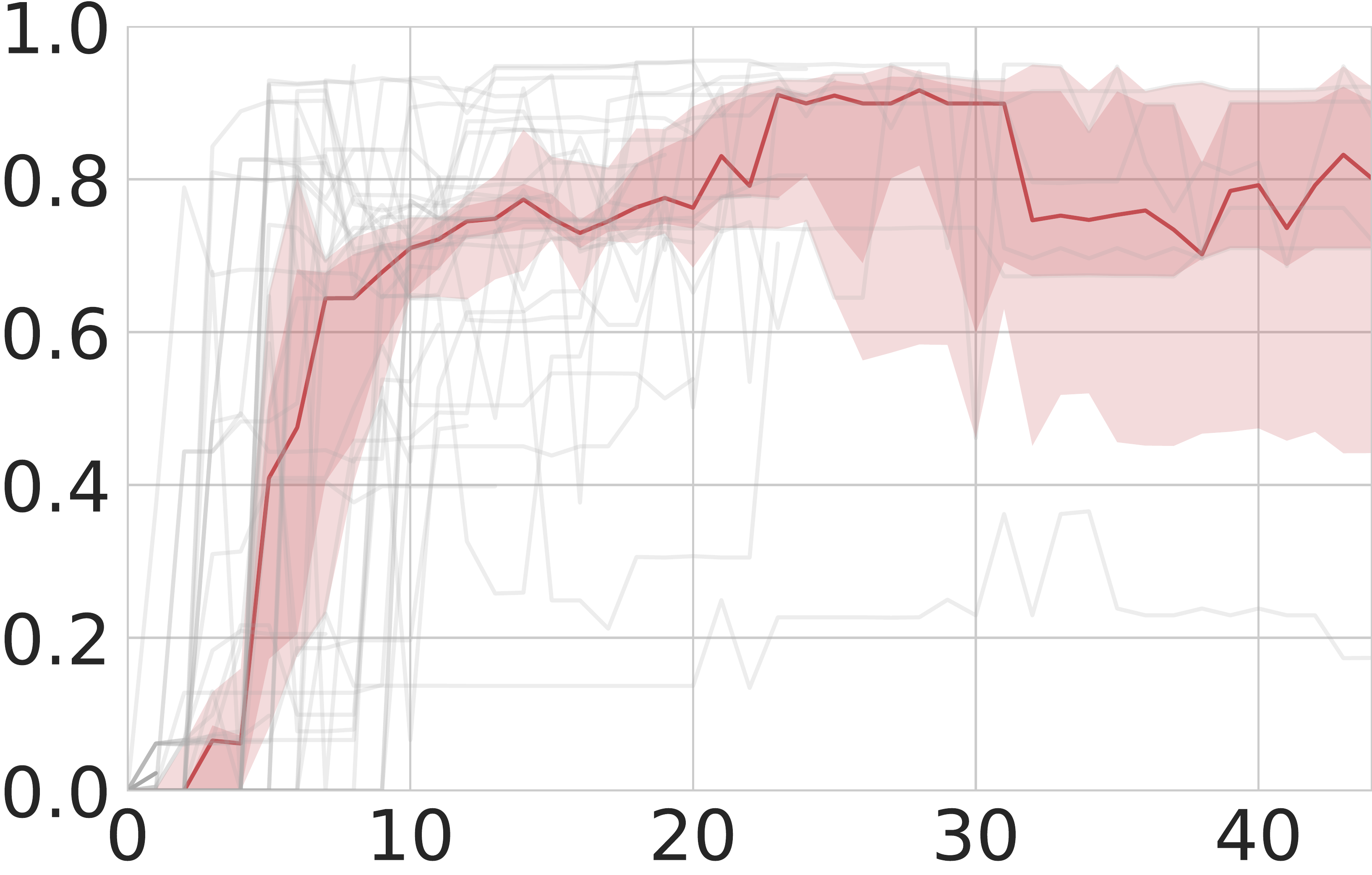} \\ %
			& \Large{\mbox{Semi-manual} Prototype Interactions} & \Large{Guided Prototype Interactions} & \Large{Joint Prototype Interactions} %
		\end{tabular}%
	}
	}%
	\caption{Evaluation of the user interaction data.
		The segmentations' similarity to the ground truth according to the Dice score is depicted per interaction.
		The median Dice rating as well as the $75$\,\% and $95$\,\% confidence intervals are illustrated.
	}
	\label{fig:result_logs}%
\end{figure*}

The quantitative evaluation of recorded interaction data is depicted in \textbf{Fig.}\,\ref{fig:result_logs}.
Dice scores before the first interaction are zero, except for the guided prototype ($0.82\pm0.02$), where few fixed seed points had to be provided to initialize the system.
Utilizing the \mbox{semi-manual} prototype and starting from zero, a similar Dice measure to the guided prototype's initialization is reached after about seven interactions, which takes $13.06\pm2.05$ seconds on average.
The median values of final Dice scores per prototype are $0.95$ (\mbox{semi-manual}), $0.94$ (guided), and $0.82$ (joint). %
The mean overall elapsed wall time in seconds spent for interactive segmentations per prototype are $73\pm11$ (\mbox{semi-manual}), $279\pm36$ (\mbox{guided}), and $214\pm24$ (\mbox{joint}). %
Since segmenting with the guided version takes the longest time and does not yield the highest final Dice scores, the initial advantage from pre-existing seed points does not bias the top ranking of a prototype in this evaluation.

\subsection{Hedonic Quality}\label{sec:results_hedonic}

\subsubsection{Identity and Stimulus}

\begin{figure}%
		\begin{tikzpicture}
		\begin{axis}[width=9cm,height=9cm,grid,xtick={1,3,5,7},xticklabels={1,3,5,7},xmin=1,xmax=7,ytick={1,3,5,7},yticklabels={1,3,5,7},ymin=1,ymax=7,xlabel={Pragmatic Quality (PQ)},ylabel={Hedonic Quality (HQ)}]
			\node[text width=1.5cm,align=center] at (axis cs:2,2) {super-fluous};
			\node[text width=1.5cm,align=center] at (axis cs:2,6) {too self-oriented};
			\node[text width=1.5cm,align=center] at (axis cs:4,4) {neutral};
			\node[text width=1.5cm,align=center] at (axis cs:4,6) {self-oriented};
			\node[text width=1.5cm,align=center] at (axis cs:6,2) {too task-oriented};
			\node[text width=1.5cm,align=center] at (axis cs:6,4) {task-oriented};
			\node[text width=1.5cm,align=center] at (axis cs:6,6) {desired};
			%
			\fill [fill=semi_manual_prot_color,semi_manual_prot_color, fill opacity=0.4] (axis cs:5.64106645752803,4.798427683218617) rectangle (axis cs:6.158933542471971,5.301572316781383);
			\addplot[color=semi_manual_prot_color,mark=otimes*,mark size=5pt,fill=white] coordinates {(5.9,5.05)};
			\fill [fill=guided_prot_color,guided_prot_color, fill opacity=0.4] (axis cs:3.143832023165031,4.6750586928307385) rectangle (axis cs:3.856167976834969,5.167798450026405);
			\addplot[color=guided_prot_color,mark=otimes*,mark size=5pt,fill=white] coordinates {(3.5,4.921428571428572)};
			\fill [fill=joint_prot_color,joint_prot_color, fill opacity=0.4] (axis cs:4.696608735705655,4.953083574359387) rectangle (axis cs:5.446248407151487,5.389773568497756);
			\addplot[color=joint_prot_color,mark=otimes*,mark size=5pt,fill=white] coordinates {(5.071428571428571,5.171428571428572)};
			\addplot[only marks,color=semi_manual_prot_color,mark=otimes*,mark size=1.5pt] coordinates {(6.4285714285714288, 3.8571428571428572)(6.1428571428571432, 6.0)(5.5714285714285712, 4.6428571428571432)(6.4285714285714288, 6.2142857142857144)(5.7142857142857144, 5.2857142857142856)(6.8571428571428568, 5.2857142857142856)(5.1428571428571432, 4.4285714285714288)(6.4285714285714288, 5.1428571428571432)(5.0, 4.6428571428571432)(5.2857142857142856, 5.0)};
			\addplot[only marks,color=guided_prot_color,mark=otimes*,mark size=1.5pt] coordinates {(2.0, 4.2857142857142856)(3.2857142857142856, 5.9285714285714288)(3.8571428571428572, 5.1428571428571432)(4.2857142857142856, 5.5)(2.7142857142857144, 4.7142857142857144)(5.5714285714285712, 5.1428571428571432)(4.7142857142857144, 4.0)(3.2857142857142856, 5.2142857142857144)(3.4285714285714284, 4.9285714285714288)(1.8571428571428572, 4.3571428571428568)};
			\addplot[only marks,color=joint_prot_color,mark=otimes*,mark size=1.5pt] coordinates {(5.1428571428571432, 4.7142857142857144)(4.8571428571428568, 5.0714285714285712)(5.8571428571428568, 5.5714285714285712)(4.4285714285714288, 4.3571428571428568)(6.0, 6.0)(5.2857142857142856, 4.7857142857142856)(5.7142857142857144, 6.3571428571428568)(2.1428571428571428, 4.8571428571428568)(6.4285714285714288, 5.0)(4.8571428571428568, 5.0)};
		\end{axis}
		\end{tikzpicture}
	\caption{\mbox{AttrakDiff-2} portfolio representation, \additioncaption{according to \cite{hassenzahl2008user},} depicting results from the evaluation of the \mbox{semi-manual} segmentation prototype (blue), guided prototype (green), and joint prototype (red).
		The rectangular areas illustrate the $95$\,\% confidence intervals for the mean value in each dimension.
		The mean intervals are $5.5$\,\% for \mbox{PQ} and $4.0$\,\% for \mbox{HQ}. %
	}%
	\label{fig:result_attrakdiff_portfolio}%
\end{figure}
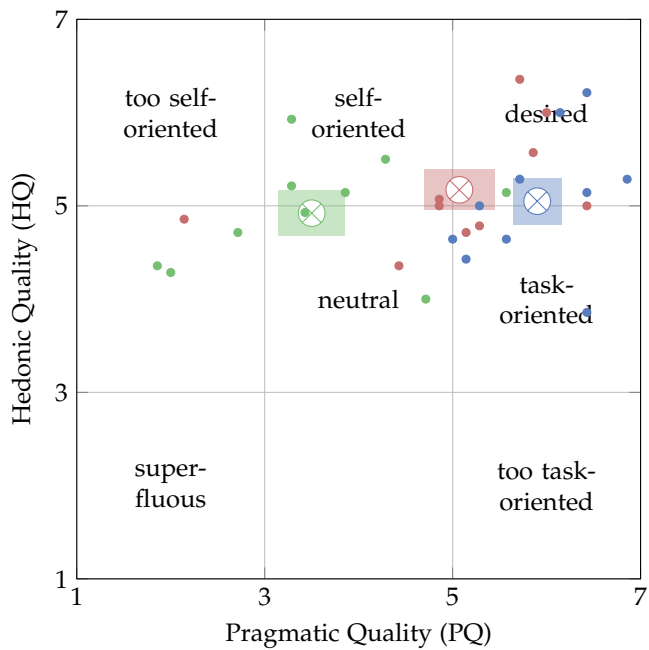

The \mbox{AttrakDiff-2} questionnaire provides a measure for the \mbox{HQ} of identity and stimulus introduced in \textbf{Sec.}\,\ref{sec:questionnaires_attrakdiff}.
The \mbox{HQ} scores for \mbox{semi-manual}, guided, and joint prototypes are $72$\,\%, $70$\,\%, and $77$\,\% of the maximum score, respectively.
Since the $95$\,\% confidence intervals are overlapping for all three prototypes, no system ranks significantly higher than the others. 
An overall evaluation of the \mbox{AttrakDiff-2} results is conducted in the form of a portfolio representation depicted in \textbf{Fig.}\,\ref{fig:result_attrakdiff_portfolio}.

\begin{table}
	\caption{Relative absolute prediction errors for \mbox{AttrakDiff-2} and \mbox{SUS} test set samples. %
		Predictions are computed by six separately trained \changecaption{Stochastic Gradient Boosting Regression Forests (\mbox{GBRFs})}{\mbox{GBRFs}}, one for each figure of merit. Note that each training process only utilizes the interaction log data.
		Results displayed are the median values of $10^4$ randomly initialized training processes.} %
	\label{tab:prediction_results_gbrf}%
	\tabcolsep=5.25pt
	\begin{tabular}{lrrrrrr}
		Relative Error &  ATT &   HQ & HQ-I & HQ-S &   PQ &   SUS  \\[1pt]\hline
		Mean   & 11.5\,\% & 7.4\,\% & 10.5\,\% & 8.0\,\% & 15.7\,\% & 10.4\,\% \\
		Median & 8.9\,\% & 6.3\,\% & 9.4\,\% & 6.2\,\% & 13.7\,\% & 8.8\,\% \\
		Std    & 8.0\,\% & 5.5\,\% & 6.7\,\% & 6.9\,\% & 12.0\,\% & 7.1\,\%
	\end{tabular}
\end{table}

\subsubsection{Qualitative Content Analysis}

A summative qualitative content analysis as described in \textbf{Sec.}\,\ref{sec:qualitative_measures} is conducted on the audio and video data recorded during the study.
After generalization and reduction of given statements, the following user feedback is extracted with respect to three problem statements:
positive usability aspects, 
negative usability aspects, and 
user suggestions concerning existing functions or new functions.

\textbf{Feedback for multiple prototypes}
\begin{enumerate}
	\item Responsiveness: the most common statement concerning the \mbox{semi-manual} and joint version is that the user expected the zoom function to be more responsive and thus more time efficient. %
	\item Visibility: $20$\,\% of the participants had difficulties distinguishing between the segmentation contour line and either the background image or the foreground scribbles in the overlay mask, due to the proximity of their assigned color values.
	\item Feature suggestion: deletion of individual seed points instead of all seeds from last interaction using \emph{undo}.
\end{enumerate}

\textbf{\mbox{Semi-manual} segmentation prototype}
\begin{enumerate}
	\item Mental model: $30$\,\% of test persons suggested clearly visible indication whether the label for the scribble drawn next will be foreground or background.
	\item Visibility: hide previously drawn seed points, in order to prevent confusion with the current contour line and occultation of the underlying image.
\end{enumerate}

\textbf{Guided segmentation prototype}
\begin{enumerate}
	\item Responsiveness: $50$\,\% of test persons suggested an indicator for ongoing computations during their time of waiting.
	\item Control: users would like to influence the location of new seed points, support for manual image zoom, and fine grained control for the \emph{undo} function.
\end{enumerate}

\textbf{Joint prototype}
\begin{enumerate}
	\item Visibility: $64$\,\% of users intuitively found the toggle functionality for seed labels without prior explanation.
	\item Visibility: $64$\,\% of participants suggested visible instructions for manual seed generation.
\end{enumerate}

\subsection{Prediction of Questionnaire Results from Log Data}\label{sec:prediction_of_questionnaire_results_from_log_data} %

\begin{figure}%
	\resizebox{\columnwidth}{!}{%
		\begin{tabular}{rc}%
			\rotatebox{90}{$\qquad\qquad$Feature importance} & 
			\includegraphics[width=\columnwidth]{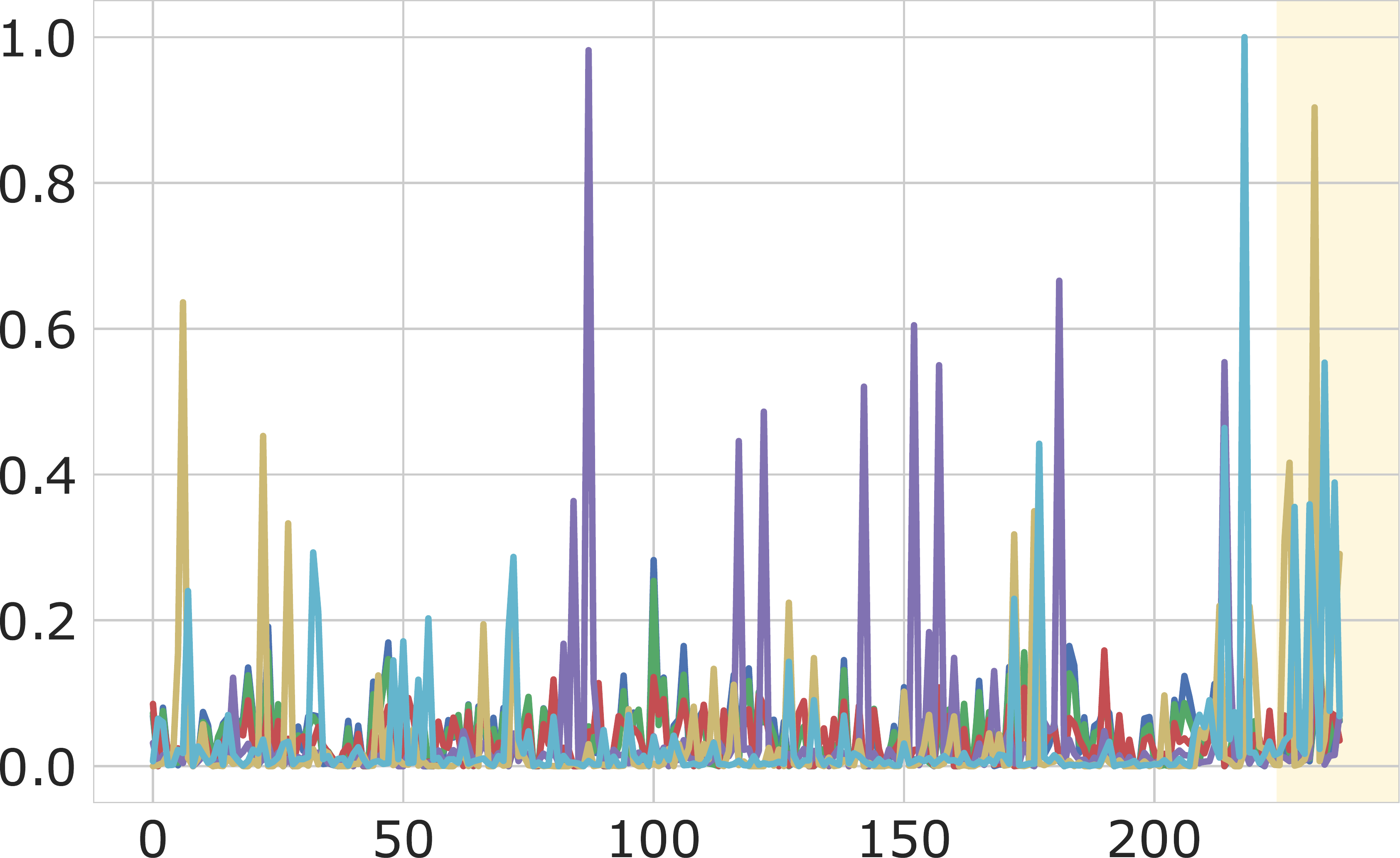} \\
			& Feature Indices\vspace{0.5em} \\
			\rotatebox{90}{$\qquad\qquad$Feature importance} & %
			\includegraphics[width=\columnwidth]{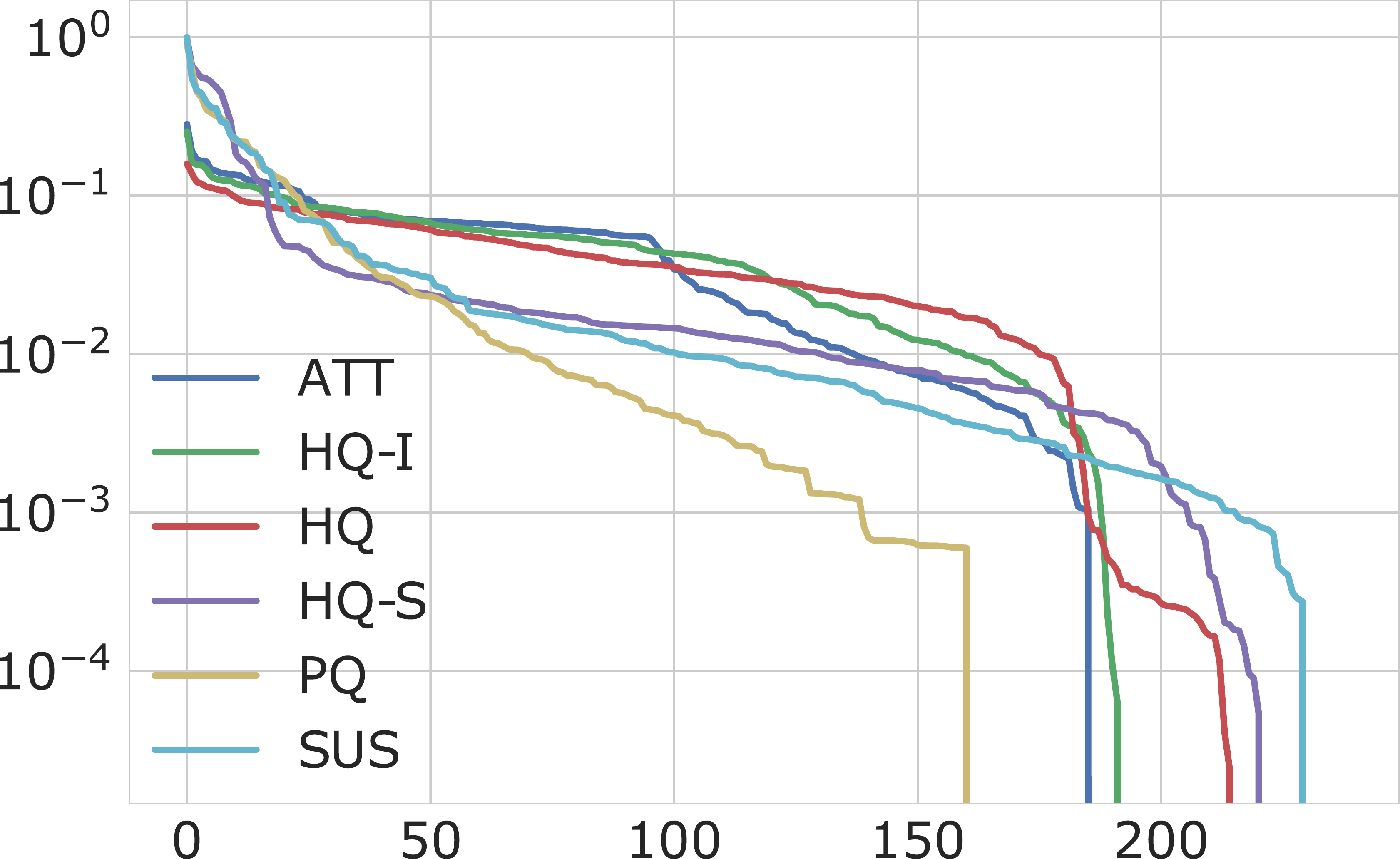} \\
			& Feature indices sorted by importance \\
		\end{tabular}%
	}
	\caption{Relative feature importance measures from $1$\,\% ($205$) of best \mbox{GBRF} estimators from grid search as described in \textbf{Sec.}\,\ref{sec:sus_prediction}.
		The orange rectangle on the \changecaption{top}{upper} right highlights features added via \mbox{PCA} transformation.
		Relative feature importance is depicted on a log scale on the bottom.}
	\label{fig:gbrf_feature_importance}%
\end{figure}

The questionnaires' results are predicted via a regression analysis, based on features extracted from the interaction log data.
A visualization of the feature importances for the regression analysis with respect to the \mbox{GBRF} is depicted in \textbf{Fig.}\,\ref{fig:gbrf_feature_importance}.
An evaluation with the test set is conducted as depicted in \textbf{Tab.}\,\ref{tab:prediction_results_gbrf}.
The mean prediction errors for the questionnaires' results are $15.7$\,\% %
for \mbox{PQ} and $7.4$\,\% %
for \mbox{HQ}.
In both cases, the error of these (first) estimates is larger but close to the average $95$\,\% confidence intervals of $5.5$\,\% (\mbox{PQ}) and $4.0$\,\% (\mbox{HQ}) for the overall questionnaire results in the portfolio representation. %

\begin{figure*}%
	\resizebox{\textwidth}{!}{%
		\includegraphics{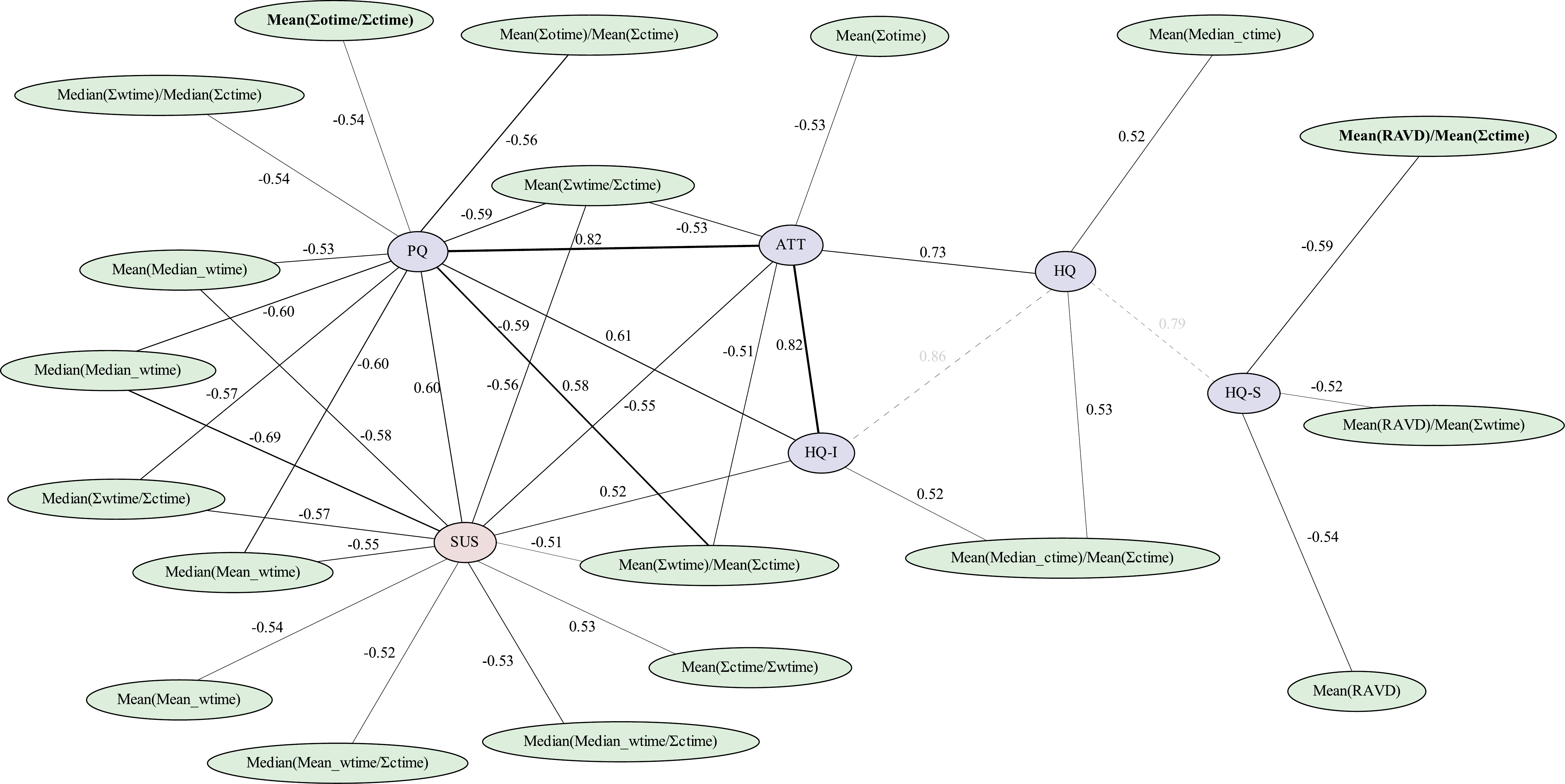}
	}%
	\caption{Features from user interaction logs (green) correlated with \mbox{SUS} (red) and \mbox{AttrakDiff-2} (blue) questionnaire results.
		Bold feature names highlight top five most important features with regards to \mbox{GBRFs}.
		Only relations with a Pearson correlation coefficient $\operatorname{abs}(c) > 0.5$ and $p < 0.05$ are displayed.
		\additioncaption{Note that this visualization is an extension to \textbf{Fig.}\,\ref{fig:result_questionnaire_results_correlation}.}
	}
	\label{fig:feature_correlations_and_feature_importance}%
\end{figure*}

The similarity graph for the acquired usability aspects introduced in \textbf{Fig.}\,\ref{fig:result_questionnaire_results_correlation} can be extended to outline the direct relationship between questionnaire results and recorded features.
Such a graph is depicted in \textbf{Fig.}\,\ref{fig:feature_correlations_and_feature_importance}.
Notably, there is no individual feature, which strongly correlates with one of the questionnaire results.
However, as the results of the regression analysis in \textbf{Tab.}\,\ref{tab:prediction_results_gbrf} depict, there is a noteworthy dependence of the usability aspects measured by the \mbox{SUS} and \mbox{AttrakDiff-2} questionnaires and combinations of the recorded features.
The most important features for the approximation of the questionnaire results are depicted in \textbf{Tab.}\,\ref{tab:most_frequently_used_features}.

\begin{table*}
	\caption{The five most important features per \mbox{GBRF} estimator/label.
		Orange background colors indicate the most frequently used features \additioncaption{in the trained decision trees of the \mbox{GBRFs}}.
		Yellow backgrounds highlight semantically similar feature pairs.
		The abbreviations represent the receiver operating characteristic area under the curve (ROC\_AUC), logistic loss (LOG), and relative absolute area/volume difference (RAVD).
	}
	\label{tab:most_frequently_used_features}
\resizebox{\textwidth}{!}{%
{\def\arraystretch{1.0}\tabcolsep=3pt
\begin{tabular}{r|lllll}
	 & 1.\ & 2.\ & 3.\ & 4.\ & 5.\ \\[1pt]\hline
	ATT & \cellcolor{orange!10}Mean(ROC\_AUC/$\Sigma$wtime) & \cellcolor{orange!5}Mean(Dice)/Mean($\Sigma$wtime) & \cellcolor{orange!5}Mean(LOG)/Mean($\Sigma$ctime) & Med(OBJ\_TPR)/Med($\Sigma$ctime) & Med($\Sigma$ctime) \\
	HQ-I & \cellcolor{orange!10}Mean(ROC\_AUC/$\Sigma$wtime) & \cellcolor{orange!5}PCA\_VAL\_17 & \cellcolor{orange!5}Mean(Dice)/Mean($\Sigma$wtime) & \cellcolor{yellow!5}Med(Med\_ctime)/Med($\Sigma$wtime) & \cellcolor{orange!5}Mean(LOG)/Mean($\Sigma$ctime) \\
	HQ & Med(Jaccard/$\Sigma$ctime) & \cellcolor{orange!5}PCA\_VAL\_17 & 
	\cellcolor{orange!10}Mean(ROC\_AUC/$\Sigma$wtime) & Mean(OBJ\_TPR/$\Sigma$wtime) & \cellcolor{yellow!5}Mean(RAVD/$\Sigma$ctime) \\
	HQ-S & \cellcolor{yellow!5}Mean(RAVD)/Mean($\Sigma$ctime) & Med(Med\_wtime/$\Sigma$wtime) & Med(LOG) & \cellcolor{orange!5}Std(Relative\_Seed\_Coord\_H) & Med(MSE) \\
	PQ & PCA\_VAL\_16 & Mean($\Sigma$otime/$\Sigma$ctime) & Mean(Dice)/Mean($\Sigma$ctime) & PCA\_VAL\_11 & \cellcolor{yellow!5}Med(Med\_ctime/$\Sigma$wtime) \\
	SUS & PCA\_VAL\_2 & PCA\_VAL\_18 & \cellcolor{orange!5}Std(Relative\_Seed\_Coord\_H) & Med(Med\_wtime) & PCA\_VAL\_20
\end{tabular}
}
}%
\end{table*}

\section{Discussion}\label{sec:discussion}

\subsection{Usability Aspects}

Altough the underlying segmentation algorithm is the interactive \mbox{GrowCut} method for all three prototypes tested, the measured user experiences varied significantly.  %
In terms of user stimulus \mbox{HQ-S} a more innovative interaction system like the joint prototype is preferred to a traditional one.
Pragmatic quality aspects, evaluated by \mbox{SUS} as well as \mbox{AttrakDiff-2}'s \mbox{PQ}, clearly outline that the \mbox{semi-manual} approach has an advantage over the other two techniques.
This conclusion also manifests in the {Dice} coefficient values' fast convergence rate towards its maximum for this prototype.
The normalized median \mbox{\emph{$\Sigma$Wall\_time}} %
spent for the overall segmentation of each image are $100$\,\% (\mbox{semi-manual}), $550$\,\% (guided), and $380$\,\% (joint). %
As a result, users prefer the simple, pragmatic interface as well as a substantial degree of freedom to control each iterative step of the segmentation. %
The less cognitively challenging approach is preferred~\cite{ramkumar2016user}. %
The other methods provide more guidance for aspects which the user aims to control themselves.
In order to improve the productivity of an \mbox{ISS}, less guidance should be imposed in these cases, 
while providing more guidance on aspects of the process not apparent to the users' focus of attention~\cite{heron1957perception}.

\subsection{Usability Aspects Approximation}

For \mbox{ATT} and \mbox{HQ-I}, the most discriminative features selected by \mbox{GBRFs} are the %
receiver operating characteristic area under the curve (\mbox{ROC\_AUC}) of the final interactive segmentations over the elapsed real time which passed during segmentation (\mbox{\emph{$\Sigma$Wall\_time}}).
The Jaccard index~\cite{jaccard1912distribution} as well as the relative absolute area/volume difference (\mbox{RAVD}) each divided by the computation time are most relevant for \mbox{HQ}, respective \mbox{HQ-S}.
The pragmatic quality's \mbox{(PQ)} dominant features are composed of final Dice scores and time measurements per segmentation.
The \mbox{SUS} results, quantifying the overall usability of a prototype, is mainly predicted based on %
the features with the highest level of abstraction used.
In the top $10$\,\% ($22$) selected features, $45$\,\% of top \mbox{SUS} features are \mbox{PCA} values, as indicated in \textbf{Tab.}\,\ref{tab:most_frequently_used_features} and \textbf{Fig.}\,\ref{fig:gbrf_feature_importance}(top).
In comparison: \mbox{PQ} $41$\,\%, \mbox{HQ} $36$\,\%, \mbox{HQ-I} $18$\,\%, \mbox{ATT} $14$\,\%, and \mbox{HQ-S} $9$\,\%.

\section{Conclusion}\label{sec:conclusion}

For sufficiently complex tasks like the accurate segmentation of lesions during \mbox{TACE}, fully automated systems are, by their lack of domain knowledge, inherently limited in the achievable quality of their segmentation results.
\mbox{ISS} may supersede fully automated systems in certain niches by cooperating with the human user in order to reach the common goal of an exact segmentation result in a short amount of time.
The evaluation of interactive approaches is more demanding and less automated than the evaluation with other approaches, due to complex human behavior.

However, there are methods like extensive user studies to assess the quality of a given system.
It was shown, that even a suitable approximation of a study's results regarding pragmatic as well as hedonic usability aspects is achievable from a sole analysis of the users' interaction recordings.
Those records are straightforward to acquire during normal (digital) prototype usage and can lead to a good first estimate of the system's usability aspects, without the need to significantly increase the temporal demands on each participant by a mandatory completion of questionnaires after each system usage.

This mapping of quantitative low-level features, which are exclusively based on measurable interactions with the system (like the final Dice score, computation times, or relative seed positions), may allow for a fully automated assessment of an interactive system's quality.

\section{\additioncaption{Outlook}}\label{sec:outlook}

For \change[label=c:c181,ref=c:c18]{the}{this} proposed automation, a rule-based user model (robot user) like~\cite{amrehn2017uinet,amrehn2019interactive} or a learning-based user model could interact with the prototype system instead of a human user.
This evaluation scheme may significantly reduce the amount of resources necessary to investigate each variation of a prototype's \mbox{UI} features and segmentation methodologies.
\addition[label=c:a181,ref=c:c18]{An estimate of a system's usability can therefore be acquired fully automatically with dependence only on the chosen user model.}
\addition[label=c:a182,ref=c:c18]{In addition, the suitable approximation of a usability study's result can be used as a descriptor, i.\,e.\ feature vector, for a user.
These features can be utilized for a clustering of users, which is a necessary step for the application of a personalized segmentation system.
Such an interactive segmentation system might benefit from prior knowledge about a user's preferences and input patterns in order to achieve accurate segmentations from less interactions.}

\section*{Disclaimer}
The concept and software presented in this paper are based on research and are not commercially available.
Due to regulatory reasons its future availability cannot be guaranteed.

{\color{addedmarkupcolor}
\section*{Conflicts of Interest}
The authors declare that there are no conflicts of interest regarding the publication of this paper.
}

\section*{Acknowledgment}
Thanks to Christian Kisker and Carina Lehle for their hard work with the data collection.


\ifCLASSOPTIONcaptionsoff
  \newpage
\fi



\bibliographystyle{IEEEtran}
\bibliography{references}

\appendix

\section*{Example for \mbox{SUS} Evaluation (\textbf{Eq.}~\ref{eq:sus_score})}

\color{addedmarkupcolor} 

The result of the \mbox{SUS} survey is a single scalar value, in the range of zero to $100$ as a composite measure of the overall usability.
The score is computed according to \textbf{Eq.}\,\ref{eq:sus_score}, as outlined in~\cite{brooke1996sus}, 
given $S$ participants, where $\mathbf{x}^\text{SUS}_{s,i}$ is the response to the statement $i$ by subject $s$.
%
\begin{equation*}
\operatorname{sus}(\mathbf{x}) = \frac{2.5}{S} \sum_{s}\left[\, \sum_{\text{odd } i} \mathbf{x}^\text{SUS}_{s,i} + \sum_{\text{even } i} (4 - \mathbf{x}^\text{SUS}_{s,i})\, \right]
\end{equation*}

\noindent Let $S=3$ participants answer the $10$ questions (listed in \mbox{\textbf{Sec.}\,\ref{sec:questionnaires_sus})} of the SUS questionnaire as follows: %

\[ \mathbf{x}^\text{SUS} = \left| \begin{array}{c}
\mathbf{x}^\text{SUS}_0 \\
\mathbf{x}^\text{SUS}_1 \\
\mathbf{x}^\text{SUS}_2 \end{array} \right| = \left| \begin{array}{cccccccccc}
 0 & 1 & 2 & 3 & 4 & 0 & 1 & 2 & 3 & 4 \\
 1 & 2 & 3 & 4 & 0 & 1 & 2 & 3 & 4 & 0 \\
 2 & 3 & 4 & 0 & 1 & 2 & 3 & 4 & 0 & 1 \end{array} \right|,\]

\noindent where $\mathbf{x}^\text{SUS}_s$ are rows in matrix $\mathbf{x}^\text{SUS}$. Then: %

\begin{align*}%
\operatorname{sus}(\mathbf{x}) = \frac{2.5}{3} \cdot (&(0 + 3 + 2 + 1 + 4 + 4 + 1 + 2 + 3 + 0) + \\
 & (1 + 2 + 3 + 0 + 0 + 3 + 2 + 1 + 4 + 4) + \\
 & (2 + 1 + 4 + 4 + 1 + 2 + 3 + 0 + 0 + 3))
\end{align*}%

\addition[label=c:a152,ref=c:c15]{ %
\noindent In this case, $\operatorname{sus}(\mathbf{x}) = 50$.
Note that the factor $2.5$ in \mbox{\textbf{Eq.}\,\ref{eq:sus_score}} normalizes the \mbox{SUS} score to a value $0 \le \operatorname{sus}(.) \le 100$. %
}

\color{black} 

\section*{Example for \mbox{AttrakDiff} Evaluation (\textbf{Eq.}~\ref{eq:attrakdiff_score})}

\color{addedmarkupcolor} 

For the questionnaire's evaluation for subject \mbox{$s \in \left\{0, 1, \dots, S-1\right\}$}, each of the seven adjective pairs \mbox{$i \in \left\{0, 1, \dots, 6\right\}$} per group 
\mbox{$g \in \{\text{PQ}, \,\text{ATT}, \,\text{HQ-I}, \,\text{HQ-S}\}$} is assigned a score \mbox{$\mathbf{x}^g_{s,i} \in \left\{1, 2, \dots, 7\right\}$} by each participant, reflecting their tendency towards the positive of the two adjectives.
The overall ratings per group are defined in \cite{hassenzahl2003attrakdiff} as the mean scores computed over all subjects $s$ and statements $i$, as depicted in \textbf{Eq.}\,\ref{eq:attrakdiff_score}.
Here, $S$ is the number of participants in the survey. %
%
\begin{equation*}
\operatorname{attrakdiff}(\mathbf{x}, \,g) = \frac{1}{7 \cdot S} \sum_{s} \sum_{i} \mathbf{x}^g_{s,i}
\end{equation*}
%
\noindent Let $S=3$ participants fill in the $28$ choices (listed in \mbox{\textbf{Tab.}\,\ref{tab:attrakdiff_statements})} of the \mbox{AttrakDiff-2} questionnaire as follows, where $\mathbf{x}^g_s$ are rows in matrix $\mathbf{x^g}$:

Group PQ: 

\[ \mathbf{x}^\text{PQ} = \left| \begin{array}{c}
\mathbf{x}^\text{PQ}_0 \\
\mathbf{x}^\text{PQ}_1 \\
\mathbf{x}^\text{PQ}_2 \end{array} \right| = \left| \begin{array}{ccccccc}
1 & 2 & 3 & 4 & 5 & 6 & 7 \\
2 & 3 & 4 & 5 & 6 & 7 & 7 \\
3 & 4 & 5 & 6 & 7 & 7 & 7 \end{array} \right|\]

Group ATT: 

\[ \mathbf{x}^\text{ATT} = \left| \begin{array}{c}
\mathbf{x}^\text{ATT}_0 \\
\mathbf{x}^\text{ATT}_1 \\
\mathbf{x}^\text{ATT}_2 \end{array} \right| = \left| \begin{array}{ccccccc}
2 & 3 & 4 & 5 & 6 & 7 & 7 \\
3 & 4 & 5 & 6 & 7 & 7 & 7 \\
4 & 5 & 6 & 7 & 7 & 7 & 7 \end{array} \right|\]

Group HQ-I: 

\[ \mathbf{x}^\text{HQ-I} = \left| \begin{array}{c}
\mathbf{x}^\text{HQ-I}_0 \\
\mathbf{x}^\text{HQ-I}_1 \\
\mathbf{x}^\text{HQ-I}_2 \end{array} \right| = \left| \begin{array}{ccccccc}
3 & 4 & 5 & 6 & 7 & 7 & 7 \\
4 & 5 & 6 & 7 & 7 & 7 & 7 \\
5 & 6 & 7 & 7 & 7 & 7 & 7 \end{array} \right|\]

Group HQ-S: 

\[ \mathbf{x}^\text{HQ-S} = \left| \begin{array}{c}
\mathbf{x}^\text{HQ-S}_0 \\
\mathbf{x}^\text{HQ-S}_1 \\
\mathbf{x}^\text{HQ-S}_2 \end{array} \right| = \left| \begin{array}{ccccccc}
4 & 5 & 6 & 7 & 7 & 7 & 7 \\
5 & 6 & 7 & 7 & 7 & 7 & 7 \\
6 & 7 & 7 & 7 & 7 & 7 & 7 \end{array} \right|\]

After evaluation via \textbf{Eq.}\,\ref{eq:attrakdiff_score}: 

\begin{align*}%
\operatorname{attrakdiff}(\mathbf{x}, \text{\makebox[2.6em][r]{PQ}}) = (&(1 + 2 + 3 + 4 + 5 + 6 + 7) + \\
& (2 + 3 + 4 + 5 + 6 + 2 \cdot 7) + \\
& (3 + 4 + 5 + 6 + 3 \cdot 7)) \,/\, 21 \\
\operatorname{attrakdiff}(\mathbf{x}, \text{\makebox[2.6em][r]{ATT}}) = (&(2 + 3 + 4 + 5 + 6 + 2 \cdot 7) + \\
& (3 + 4 + 5 + 6 + 3 \cdot 7) + \\
& (4 + 5 + 6 + 4 \cdot 7)) \,/\, 21 \\
\operatorname{attrakdiff}(\mathbf{x}, \text{\makebox[2.6em][r]{HQ-I}}) = (&(3 + 4 + 5 + 6 + 3 \cdot 7) + \\
& (4 + 5 + 6 + 4 \cdot 7) + \\
& (5 + 6 + 5 \cdot 7)) \,/\, 21 \\
\operatorname{attrakdiff}(\mathbf{x}, \text{\makebox[2.6em][r]{HQ-S}}) = (&(4 + 5 + 6 + 4 \cdot 7) + \\
& (5 + 6 + 5 \cdot 7) + \\
& (6 + 6 \cdot 7)) \,/\, 21 \\
\end{align*}%

\noindent In this case, 
\mbox{$\operatorname{attrakdiff}(\mathbf{x},\text{PQ}) = 4.81$},  
\mbox{$\operatorname{attrakdiff}(\mathbf{x},\text{ATT}) = 5.52$},  
\mbox{$\operatorname{attrakdiff}(\mathbf{x},\text{HQ-I}) = 6.10$}, and  
\mbox{$\operatorname{attrakdiff}(\mathbf{x},\text{HQ-S}) = 6.52$}.  

\noindent The confidence intervals $\operatorname{conf}(.)$ can then be extracted via the percent point function $\operatorname{ppf}(.)$ (also called quantile function or inverse cumulative distribution function) for the selected $95$\,\% confidence interval. 
\begin{align*} 
z &= \operatorname{ppf}(0.95 \cdot 0.5) = 1.95996 \\
\operatorname{conf}(\mathbf{x}, \,g) &= \operatorname{mean}(\mathbf{x}^g) \pm z \cdot \frac{\operatorname{std}(\mathbf{x}^g)}{\sqrt{7 \cdot S}} \\
\end{align*}

\addition[label=c:a171,ref=c:c17]{ %
\noindent Note that $\operatorname{mean}(.)$ and $\operatorname{std}(.)$ flatten the input matrix to a vector first, such that mean and standard deviation are computed from a list of values and the outcome is one scalar value per function.
The confidence intervals for the example data are 
\mbox{$\operatorname{conf}(\mathbf{x},\text{PQ}) = 4.81 \pm 0.81$},  
\mbox{$\operatorname{conf}(\mathbf{x},\text{ATT}) = 5.52 \pm 0.68$},  
\mbox{$\operatorname{conf}(\mathbf{x},\text{HQ-I}) = 6.10 \pm 0.53$}, and  
\mbox{$\operatorname{conf}(\mathbf{x},\text{HQ-S}) = 6.52 \pm 0.36$}.  
} %

\end{document}